\documentclass[preprint,12pt, 3p]{elsarticle}

\usepackage{amsmath}
\usepackage{amssymb}
\usepackage{mathrsfs}
\usepackage{hyperref}
\usepackage{bm}
\usepackage{accents}
\usepackage{xcolor}
\usepackage{graphicx}
\usepackage{wrapfig}
\usepackage[table]{xcolor}
\usepackage{color,soul}

\usepackage{float}
\usepackage{placeins}
\usepackage{algorithm}
\usepackage{algpseudocode}
\usepackage{tcolorbox}
\usepackage{algorithm}
\usepackage{algpseudocode}
\tcbuselibrary{breakable}

\newtcolorbox{codeboxraw}{%
  breakable,
  colback=gray!10,
  colframe=gray!10,
  boxrule=0pt,
  left=6pt,
  right=6pt,
  top=6pt,
  bottom=6pt
}

\usepackage{caption}

\newdefinition{rmk}{Remark}

\journal{arXiv}

\begin{document}

\begin{frontmatter}

\title{Instabilities and Phase Transformations in Architected Metamaterials: a Gradient-Enhanced Continuum Approach}

\author[1]{Sarvesh Joshi}
\author[1]{S. Mohammad Mousavi}
\author[2]{Craig M. Hamel}
\author[3]{Stavros Gaitanaros}
\author[4]{Prashant K. Purohit}
\author[2]{Ryan Alberdi}
\author[1,5]{Nikolaos Bouklas\corref{cor1}}
\ead {nbouklas@cornell.edu}

\cortext[cor1]{Corresponding Author}

\address[1]{Sibley School of Mechanical and Aerospace Engineering, Cornell University, Ithaca, NY, USA\fnref{label1}}
\address[2]{Sandia National Laboratories, Albuquerque, NM, USA\fnref{label2}}
\address[3]{DTU Engineering Technology, Technical University of Denmark, Ballerup, 2750 Denmark\fnref{label3}}
\address[4]{Department Mechanical Engineering and Applied Mechanics, University of Pennsylvania, Philadelphia, PA, USA\fnref{label4}}
\address[5]{Pasteur Labs, Brooklyn, NY, USA\fnref{label5}}






\begin{abstract}

Architected metamaterials such as foams and lattices exhibit a wide range of properties governed by microstructural instabilities and emerging phase transformations. Their macroscopic response--including energy dissipation during impact, large recoverable deformations, morphing between configurations, and auxetic behavior--remains difficult to capture with conventional continuum models, which often rely on discrete approaches that limit scalability. We propose a nonlocal continuum formulation that captures both stable and unstable responses of elastic architected metamaterials. The framework extends anisotropic hyperelasticity by introducing nonlocal variables and internal length scales reflective of microstructural features. Local polyconvex free-energy models are systematically augmented with two families of non-(poly)convex energies, enabling both metastable and bistable responses. Implementation in a finite element framework enables solution using a hybrid monolithic--staggered strategy. Simulations capture densification fronts, forward and reverse transformations, hysteresis loops, imperfection sensitivity, and globally coordinated auxetic modes. Overall, this framework provides a robust foundation for accelerated modeling of instability-driven phenomena in architected materials, while enabling extensions to anisotropic, dissipative, and active systems as well as integration with data-driven and machine learning approaches.

\end{abstract}


\end{frontmatter}

\section{Introduction}

Architected metamaterials --materials with carefully tailored low relative density microstructures that exhibit non-standard responses-- have garnered significant attention due to their ability to exhibit extraordinary mechanical behaviors that transcend those of conventional materials. These behaviors, including auxeticity, recoverable densification, programmable stiffness, and multistability \cite{yang_review_2004, faure_design_2017, kaminakis_design_2015}, emerge not just from the composition of the material but mainly due to its micro-architecture~\cite{bertoldi_negative_2010, lakes_deformation_1991, deshpande_isotropic_2000, scarpa_dynamic_2002}. As a result, architected metamaterials have found application in a wide range of fields, such as energy absorption for impacts \cite{gaitanaros2012crushing},  soft robotics~\cite{estrin_architectured_2019}, biomedical scaffolds~\cite{kim_foam-like_2016, liang2017phase,chen20243d}, morphing structures~\cite{ament_autonomous_2021}, and lightweight deployable structures in aerospace~\cite{hopkins_design_2016, korner_systematic_2015}.

Among the many emergent behaviors in these architected systems, auxeticity, in which a material contracts laterally under axial compression, is particularly striking. While classically achieved through re-entrant or rotating unit cell geometries~\cite{lakes_foam_1987, prall_properties_1997}, more recent studies highlight the role of long-range cooperative interactions in enabling auxetic response~\cite{bertoldi_negative_2010-1, hughes_auxetic_2010, ganghoffer_frontiers_2023}. These deformation modes are highly sensitive to imperfections, boundary conditions, and spatial constraints~\cite{liu_spatially_2024, pasternak_materials_2012, kotani_materials_2016}.
The unconventional responses of architected metamaterials are rooted in microstructural instabilities and non-affine micro-deformation mechanisms that govern the redistribution of strain and stress across the structure~\cite{alderson_elastic_2010, dirrenberger_effective_2013}. The multiscale cascading effect of these microstructural instabilities, manifests through the emergence of apparent phase transformations (e.g. a transition between a rarefied and densified phase). Consequently, the (macro)structural response can become unstable, leading to collapse and failure, but also enabling strategies for fast morphing.

However, classical (local\footnote{Within the manuscript, local, refers to classical continuum theories that do not consider nonlocal effects.}) continuum formulations for perfectly elastic materials are fundamentally limited in their ability to capture such phase transitions that induce a spatially heterogeneous material behavior~\cite{truskinovsky1996ericksen}. Such features are crucial for the development of predictive models without having to resort to resolving the complete microstructural details, a practice that can quickly lead to significant computational cost. Despite their limitations, local continuum models have been used to model instabilities, often incorporating free energy densities that permit multiple local minima in the energy landscape~\cite{gao2008multiple, gao_multiple_2008}.  Building on Ericksen’s foundational work on non-monotonic stress–strain relations~\cite{ericksen1975equilibrium}, models for phase transformations in solids have also been proposed using kinetic laws for tracking the transformation front~\cite{abeyaratne2006evolution,purohit2019compression} but these are mostly restricted to a 1D setting.

Although standard computational homogenization approaches, where a representative volume element (RVE) is utilized to obtain a local model, are commonly utilized for problems in architected metamaterials~\cite{geers2010multi}, the resulting models are also local continuum models, and thus suffer from the same deficiencies. Additionally, if the RVE localizes it alludes to multiple localizations in the microscale due to periodicity assumptions, and as such, the RVE that has to be considered is the infinite RVE to capture the energy landscape for the phase transition. There have been some successful models that capture complex deformation mechanisms at the RVE level and upscale them to continuum models~\cite{khajehtourian2021continuum,khajehtourian2021soft} but these do not aim to capture sharp transitions, nor can generalize to other types of materials.

\begin{figure}[ht]
    \centering
    \includegraphics[width=\linewidth]{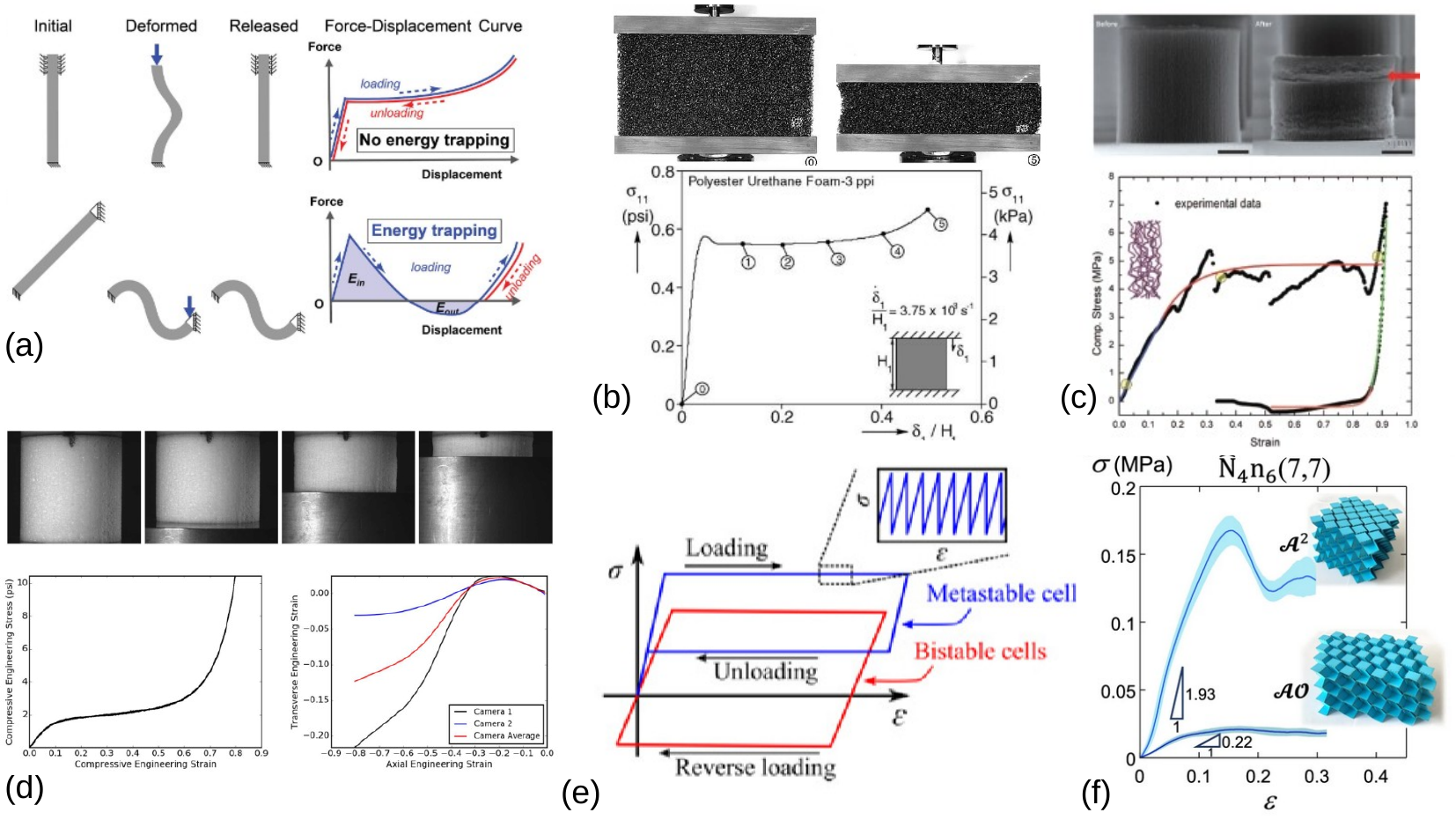}
    \caption{Microscopic and macroscopic instabilities and emerging phase transformations in architected metamaterials: .(a)unstable force displacement responses for microscopic members of a lattice-type metamaterial (adapted from \cite{shan2015multistable}), (b) response of a polyester urethane foam under compression (adapted from  \cite{gong2005compressive}), (c) cyclic response of cylindrical CNT forest specimen under compression exhibiting localization (adapted  from \cite{purohit2019compression}),  (d) compression of cylindrical polyurethane foam specimens, stress strain response and axial vs. transverse strain response (adapted from  \cite{long_2020}), (e) cyclic response for bistable and metastable response of architected metamaterials (adapted from  \cite{restrepo_phase_2015}), (f)Stress strain response of rigidly flat-foldable  lockable origami-inspired metamaterials with topological stiff states (adapted  from \cite{jamalimehr2022rigidly}).}
    \label{fig:main.pdf}
\end{figure}

Early attention on modeling of metallic foams focused on the effects of yielding and hardening developing local constitutive models~\cite{deshpande_isotropic_2000,deshpande2000high,deshpande2001multi}, but the importance of accounting for material length scales in this type of continuum models was also identified early~\cite{chen2002size}, inspired from previously introduced strain gradient plasticity models\cite{aifantis1984microstructural,fleck1997strain}. These works, in turn, followed early approaches by Toupin and Mindlin to introduce characteristic material length scales and regularize localization in elasticity~\cite{toupin1962elastic,mindlin1963microstructure,mindlin1965second}. Gradient-enhanced models --where additional internal variables are used to introduce energetic gradient penalties-- have also been developed~\cite{sagiyama_numerical_2018, bacigalupo_homogenization_2014, dalaq_finite_2016, iltchev_computational_2015,mousavi2025chain}. Phase field models can also be viewed as a sub-class of these methods, and are often used to model phase transformations, introducing an order parameter and utilizing free energy formulations that allow for multiple local minima~\cite{chen2002phase,borden2012phase,li2020variational}. Micromorphic and generalized continuum models also followed that direction~\cite{forest_continuum_2005, combescure_generalized_nodate}, and while their complexity often is limiting for the development of interpretable models, there have been recent efforts in this direction related to metamaterials~\cite{sperling2023enriched,guo2025reduced,maraghechi2024harvesting,van2020newton}.  Even though there have been significant developments, a general formulation that can enable capturing the macroscopic response but also the emerging phase transitions that occur in architected metamaterials is still lacking.

On the design front, advances in homogenization~\cite{brechet_architectured_2013, dirrenberger_homogenization_2011}, concurrent multiscale modeling~\cite{xu_concurrent_2016, vicente_concurrent_2016, allaire_shape_2002}, and topology optimization~\cite{guest_optimizing_2006, xu_design_2016, andreassen_design_2014, ghaedizadeh_tuning_2016, osanov_topology_2016, challis_design_2008} have opened pathways for embedding instabilities into functional materials. Data-driven and neural-network-based methods offer accelerated design tools~\cite{kumar2020inverse,zheng2023unifying,thakolkaran2025experiment,kalina_neural_2024, huang_optimized_2024}, and learning approaches~\cite{jones2025differentiable,fuhg2024review}, though challenges remain in generalizing these approaches across diverse loading conditions, geometries, and ensuring thermodynamic admissibility.

In this work, we develop a thermodynamically consistent, gradient-enhanced finite element framework for modeling phase transitions, along with the corresponding stable and unstable structural responses that arise in architected metamaterials. Relevant response at the microstructural and macroscopic level for various types of architected metamaterials --from foams, to carbon nanotube forests (CNT) and origami-type structures-- are shown in Fig. \ref{fig:main.pdf}. To construct the nonlocal model, we enhance local polyconvex models with two families of non-(poly)convex energies which additionally incorporate length scales relevant to the microstructure. This allows the model to exhibit metastable and bistable responses in constrained compression experiments. Spatial regularization is achieved through gradient penalties, and artificial viscosity is introduced. The framework is implemented via finite elements and a hybrid solution strategy is implemented utilizing both monolithic and staggered approaches. A model is specialized for phase transitions between a densified and a rarefied phase, and numerical simulations demonstrate the emergence of densification fronts, forward and reverse transformations,  hysteresis loops, tunable imperfection sensitivity, and globally coordinated auxetic modes. 

The remainder of the paper is organized as follows. In Section~\ref{Section:LocalDescription}, we introduce a framework for anisotropic hyperelasticity, further non-(poly)convex free energy formulations are introduced to connect to phase transformations, and finally, these models are specialized to deliver auxetic response. Section~\ref{Section:NonLocalDescription} develops the nonlocal extension using a gradient-enhanced model with internal variables related to mixed invariants introduced for anisotropic hyperelasticity. Further, the model is specialized for an isotropic case to account for instabilities that are governed by the volume ratio. Section~\ref{Section:FEMImplementation} outlines the finite element implementation, including discretization, nonlinear solution strategies, and mixed space construction. Section~\ref{Section:Results} presents simulations to examine densification fronts, hysteresis, graded imperfection sensitivity, and emergent auxeticity. Conclusions and future directions are provided in Section~\ref{Section:Conclusion}, with supplementary details in the Appendix.


\section{A Local Description of Instabilities and Phase Transitions}\label{Section:LocalDescription}
\subsection{A framework for anisotropic hyperelasticity}
Consider an elastic body occupying the reference configuration \( \Omega_0 \subset \mathbb{R}^3 \). Let the boundary of the body $\partial \Omega_0$ be composed of two parts $\partial \Omega_0^{t}$ and $\partial \Omega_0^{u}$ such that $\partial \Omega_0 = \partial \Omega_0^{t} \cup \partial \Omega_0^{u}$.
Here,  $\partial \Omega_0^{u}$ and  $\partial \Omega_0^{t}$ describe the boundary sections that displacement and traction boundary conditions are prescribed.
The motion between the referential position $\mathbf{X}$ and the current position $\mathbf{x}$ can be defined by
\begin{equation}
\mathbf{x} = \mathbf{\varphi}(\mathbf{X},t) = \mathbf{X} + \mathbf{u}(\mathbf{X},t)
\end{equation} 
where $\mathbf{u}$ describes the time-dependent displacement field and $\mathbf{\varphi}(\mathbf{X},t)$ denotes the motion of the body. 
This allows to define the deformation gradient
\begin{equation}
\mathbf{F} = \nabla \mathbf{\varphi}(\mathbf{X})
\end{equation}

In the hyperelastic framework, existence of the free energy density function $\Psi$ is postulated, which is assumed to be defined per unit reference volume \citep{holzapfel2002nonlinear}.
The formulation of a free energy density function is dependent on the symmetry group that complies with the symmetries that correspond to a specific material. 
A symmetry group of a material is a set of transformations that allow for material symmetry to be preserved. If a scalar-valued tensor function, such as the strain energy density function, is invariant under a rotation it may be written in terms of the invariants ($I_{1}, \ldots, I_{\mathrm{n}}$) of its arguments as $\Psi(I_{1}, \ldots, I_{\mathrm{n}}) $ \cite{holzapfel2002nonlinear}. 
For purely mechanical processes of perfectly elastic materials, the second law of thermodynamics requires that the Clausius-Planck inequality turns into an equality, leading to
\begin{equation}\label{eq::StressTens}
\begin{aligned}
\mathbf{S} &=  2\frac{\partial \Psi(I_{1}, \ldots, I_{\mathrm{n}})}{\partial \mathbf{C}} = 2 \sum_{i=1}^{\mathrm{n}} \frac{\partial \Psi}{\partial I_{i}} \frac{\partial I_{i}}{\partial \mathbf{C}}.
\end{aligned}
\end{equation}
where $\mathbf{S}$ is the second Piola-Kirchhoff stress tensor, $\mathbf{C}=\mathbf{F}^T\mathbf{F}$ is the right Cauchy-Green deformation tensor, ($I_{1}, \ldots, I_{\mathrm{3}}$) are its principal invariants, and ($I_{4}, \ldots, I_{\mathrm{n}}$) are the so-called pseudo-invariants.

Commonly, hyperelastic constitutive laws are constructed so that the reference configuration corresponds to a stress-free state with zero energy. This can be achieved by requiring that
\begin{equation}
    \Psi(\mathbf{I}, \bullet) = 0
\end{equation}
in the reference configuration where $\mathbf{F} = \mathbf{C} =\mathbf{I}$. This condition and the physically sound assumption that the strain energy function increases under deformation, i.e. $ \Psi(\mathbf{C}, \bullet) \geq 0$, ensures that the stress in the reference configuration is zero, i.e.
\begin{equation}
    \mathbf{S}(\mathbf{I}, \bullet) = \mathbf{0}
\end{equation}
Additionally, the growth condition ensures coercivity of the strain energy density and is typically formulated by requiring that
\begin{equation}
    \Psi(\mathbf{C}, \bullet) \to \infty \quad \text{as} \quad \det(\mathbf{C}) \to 0 \quad \text{or} \quad \|\mathbf{C}\| \to \infty,
\end{equation}
which guarantees that the material response remains physically meaningful under extreme volumetric deformations. 
Finally, Ball \cite{ball1976convexity} introduced the concept of \emph{polyconvexity} as a sufficient condition for ensuring the existence of minimizers in nonlinear elasticity. A function $\Psi(\mathbf{F})$ is said to be polyconvex if it can be expressed as a function $h$, that is individually convex with respect to its arguments $\mathbf{F}$, $\det \mathbf{F}$, and $\operatorname{cof} \mathbf{F}$. That is,
\begin{equation}
    \Psi(\mathbf{F}) = h(\mathbf{F}, \operatorname{cof} \mathbf{F}, \det \mathbf{F}).
\end{equation}
Polyconvexity also implies rank-one convexity which in turn ensures that ellipticity is preserved, preventing short-wavelength instabilities

Related to the cascade of instabilities this work focuses in, loss of ellipticity~\cite{knowles1978failure}, is a common indicator for instabilities, that can be triggered by various deformation mechanisms, including purely elastic deformations in composites~\cite{lopez2007homogenization}, but also plasticity~\cite{agoras2025effect}. These instabilities can often also lead to the emergence of phase transformations~\cite{furer2018macroscopic,iordanidis2025effect}, triggered from various mechanisms, including shear deformation.

\subsection{Non-(poly)convex models}
Extensive theoretical investigations have been dedicated to understanding phase transitions and twinning phenomena in solid materials~\cite{ball1989fine}. Central to this effort is the formulation of energy functionals incorporating \textit{non-(poly)convex} free energy densities, which serve as prototypical models for capturing the emergence of multiple stable or metastable states within the material~\cite{gao2008multiple}. 
Pioneering work by Ericksen~\cite{ericksen1975equilibrium} elucidated the implications of non-(poly)convex energy landscapes by analyzing an elastic bar governed by a non-monotonic stress-strain relation. His results revealed the existence of multiple non-unique minimizers, corresponding to distinct equilibrium configurations separated by sharp discontinuities. These discontinuities are interpreted as \textit{phase interfaces}—transition zones across which the material switches between coexisting phases. The presence of such interfaces is crucial for interpreting experimental manifestations of hysteresis, as material points become arrested in metastable states during loading-unloading cycles.

Building upon the foundational framework developed by Abeyaratne and Knowles~\cite{abeyaratne2006evolution} in the context of hyperelasticity, a nonlinearly elastic material can be characterized through the Helmholtz free energy density function \( \Psi(\gamma) \), representing the strain energy per unit reference volume as a function of the strain measure \( \gamma \). The associated stress response is given by:
\begin{equation}
    \sigma = \widehat{\sigma}(\gamma) = \frac{d\Psi}{d\gamma}.
\end{equation}
On the other hand, the Gibbs free energy (referred to as potential energy in \cite{abeyaratne2006evolution}) per unit reference volume, incorporating the work of external stress, is expressed as:
\begin{equation}
    \widehat{\Psi}(\gamma, \sigma) = \Psi(\gamma) - \sigma\gamma.
\end{equation}
and is simply a Legendre transform of the Helmholtz free energy, moving from a displacement-control to a force-control setting.
The identification of extrema in \( \widehat{\Psi}(\cdot, \sigma) \) is therefore central to predicting phase transitions, as each local minimum represents a metastable or stable phase configuration. When multiple such minima exist, with distinct energy barriers, the system can exhibit spatial phase segregation, which are hallmarks of materials with rich internal energy landscapes.

Such a scenario necessitates a non-(poly)convex form of \( \Psi(\gamma) \), resulting in a stress-strain relationship that is inherently non-monotonic. This behavior underlies the phase coexistence and transition mechanisms observed in numerous soft and crystalline materials. In the broader context of architected and biological materials, these insights form the theoretical foundation for modeling metastability, structural reorganization, and hysteresis as emergent phenomena driven by energy minimization in non-(poly)convex energy landscapes. This framework not only explains mechanical instabilities but also provides a rigorous basis for tracking the evolution of interfaces between competing phases through variational methods. 

\subsection{Instabilities and auxetic effects} \label{sec::local}

In architected metamaterials, auxetic response is often encountered and is typically the result of microstructural deformation mechanisms that trigger a cascade that spans microscopic and macroscopic length scales~\cite{chen20243d}. In such auxetic responses, significant densification is observed at the microstructure level (the spatial density of the material is increasing). In an idealized case in isotropic hyperelasticity we can consider that the transition from a rarefied to a densified phase and vice versa, is controlled just by the dependence of the free energy on the volume ratio $J=det(\mathbf{F})$ (as it relates to the third principal invariant of the right Cauchy-Green deformation tensor $\mathbf{C}$). 

The base constitutive model adopted in this work is the compressible Neo-Hookean formulation. This simple model captures nonlinearities associated with soft solids and enables the representation of finite compressibility effects. To account for unstable auxetic behavior, we build upon the formulation proposed by Gao and Ogden~\cite{gao2008multiple} (with a change of variable from $\gamma$ to $J$) and also the double-well formulation often used in phase-field theory. The corresponding local form of the proposed free energy density is defined via an additive decomposition:
\begin{equation}\label{Eq: LocalForm_Psi}
    \Psi = \Psi_{\textrm{C}}+\Psi_{\textrm{NC}} 
\end{equation}
where the first term \( \Psi_{\textrm{C}} \) is the compressible Neo-Hookean contribution, and the second term \( \Psi_{\textrm{NC}} \) represents a non-(poly)convex component, chosen either as the Gao–Ogden model \( \Psi_{\textrm{GO}} \) or a double-well form \( \Psi_{\textrm{DW}} \). These are given by:
\begin{equation}\label{Eq:NeoHookean}
    \Psi_{\textrm{C}} = \frac{\mu}{2}\left(I_1 - 3 - 2\ln J\right) + \frac{\kappa}{2}\left(\ln J\right)^2
\end{equation}
\begin{equation}\label{Eq:Non-Convex}
\Psi_{\textrm{NC}} =
\begin{cases}
\displaystyle \Psi_{\textrm{GO}} = \frac{\alpha}{2}\left(\frac{(1 - J)^2}{2} - \beta(1 - J)\right)^2 \\
\displaystyle \Psi_{\textrm{DW}} = \zeta(J - K)^2(J - 1)^2
\end{cases}
\end{equation}
Here, \( \mu \) and \( \kappa \) denote the shear and bulk moduli, \( \alpha \) and \( \beta \) are tuning parameters controlling non-convexity in the Gao–Ogden model, and \( \zeta \), \( K \) are the corresponding parameters in the double-well model. 

\begin{figure}[ht]
    \centering
    \includegraphics[width=\linewidth]{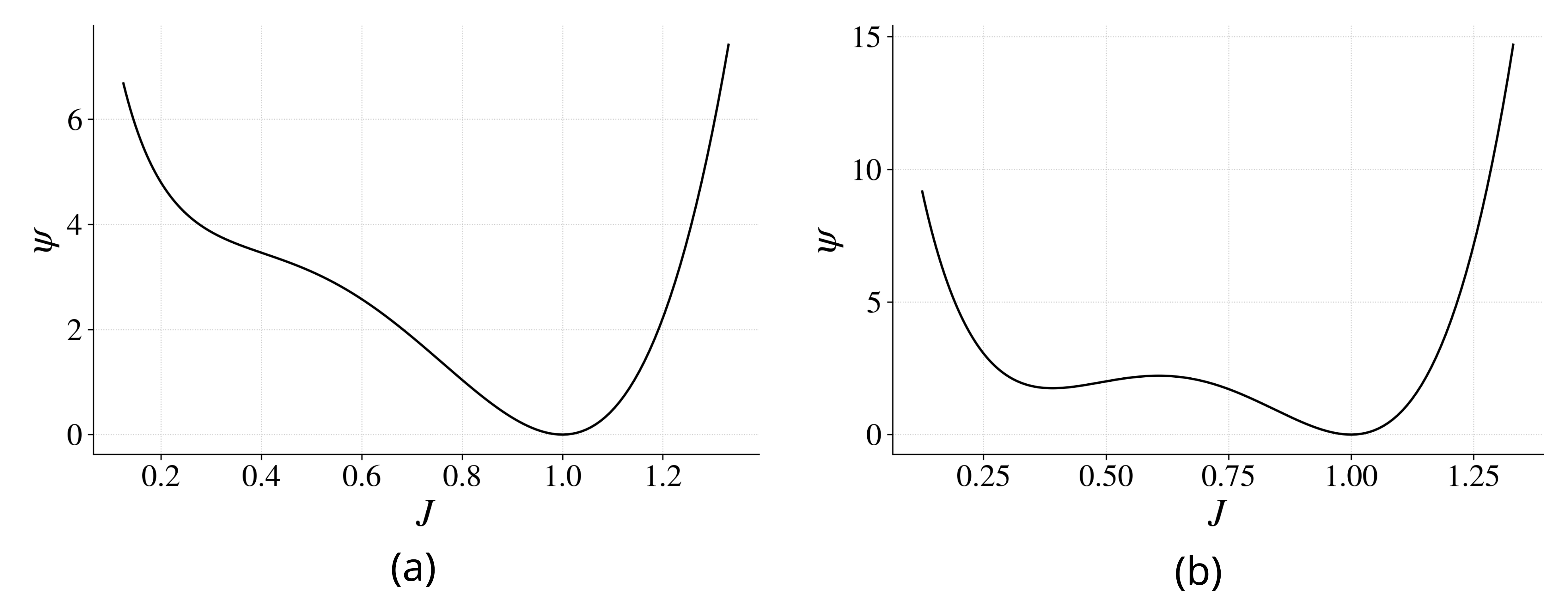}
    \caption{Variation of the proposed free energy density \( \Psi \) with respect to the Jacobian determinant \( J \), under hydrostatic deformation. (a): non-(poly)convex Helmholtz free energy density with a single minimum corresponding to \( \alpha = 300.0 \), \( \beta = 0.5 \); (b): non-(poly)convex Helmholtz free energy density with two local minima for \( \alpha = 1000.0 \), \( \beta = 0.35 \). In both cases, \( \mu = 2.0 \), \( \kappa = 2.0 \).}
    \label{fig:combined_Psi_vs_J}
\end{figure}

To examine the energy landscape defined by Eq.~\ref{Eq: LocalForm_Psi} and Eq.~\ref{Eq:Non-Convex}a for the Gao-Ogden model, we consider a hydrostatic deformation with principal stretch \( \lambda \), so that \( J = \lambda^3 \) and \( I_1 = 3\lambda^2 \). Further, all results are presented in a non-dimensional setting. The resulting energy landscape for the hydrostatic deformation is plotted in Fig.~\ref{fig:combined_Psi_vs_J}, by varying $J$ about the undeformed state at $J=1$. Fig.~\ref{fig:combined_Psi_vs_J}(a) and (b) illustrate the effect of modulating parameters \( \alpha \) and \( \beta \) while keeping other parameters fixed, transitioning from a non-(poly)convex landscape with a single minimum at the undeformed state, to a landscape with two local minima (one at the undeformed configuration $J=1$, and one at high compression $J\approx0.35$).

\begin{figure}[ht]
    \centering
    \includegraphics[width=\linewidth]{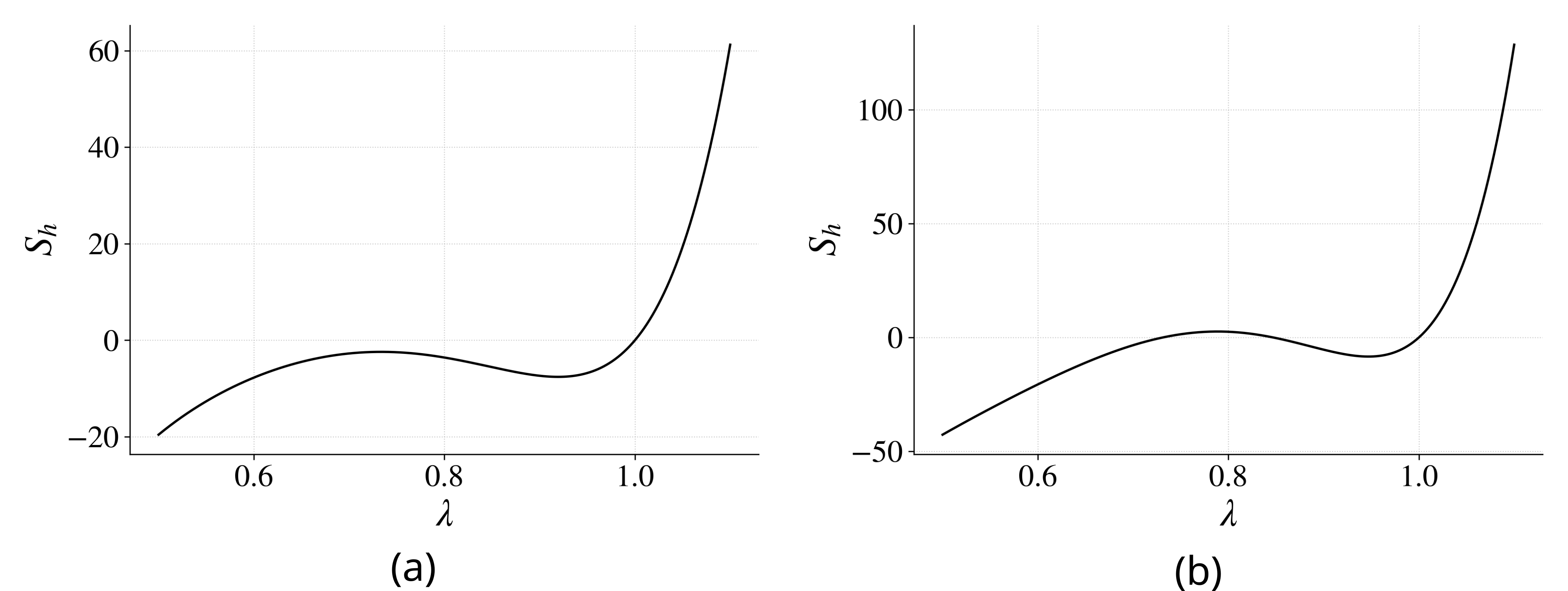}
    \caption{Stress–stretch response under hydrostatic loading for energies in Fig.~\ref{fig:combined_Psi_vs_J}. The hydrostatic second Piola–Kirchhoff stress component \( S_{h} \) is plotted against the stretch \( \lambda \). (a): single-well Helmholtz free energy density for \( \alpha = 300.0 \), \( \beta = 0.5 \); (b): bistable energy density with two local minima for \( \alpha = 1000.0 \), \( \beta = 0.35 \). }
    \label{fig:combined_S_vs_E_plot}
\end{figure}

For a hydrostatic deformation the second Piola–Kirchhoff stress tensor can be obtained as:
\begin{equation}\label{Eq: lmda_based_S}
    \mathbf{S}(\lambda) = \mu\left(\mathbf{I} - \lambda^2\mathbf{I}\right) + 3\kappa\lambda^2\ln\lambda\mathbf{I} + \alpha\left(\frac{(1 - \lambda^3)^2}{2} - \beta(1 - \lambda^3)\right)(\lambda^3 - 1 + \beta)\lambda^5\mathbf{I}
\end{equation}
reducing to \( \mathbf{S} = S_h \mathbf{I} \), allowing a single scalar component to characterize the stress–stretch behavior. Fig.~\ref{fig:combined_S_vs_E_plot} shows the resulting hydrostatic stress–strain response for both the sets of parameters from Fig.~\ref{fig:combined_Psi_vs_J}. As illustrated in the figure, the non-monotonicity in the stress response in compression is indicative of an unstable response and rapid transition from a rarefied to a densified phase.

\begin{figure}[ht]
    \centering
    \includegraphics[width=\linewidth]{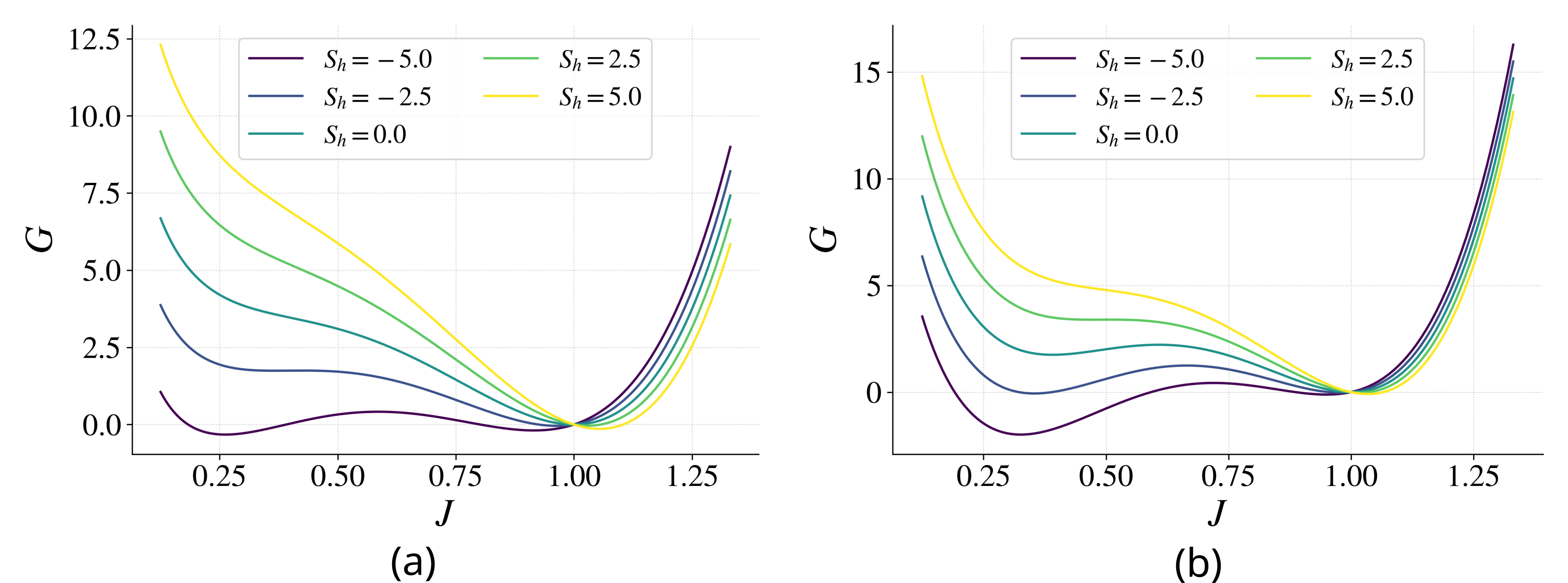}
    \caption{Gibbs free energy landscape \( \widehat{\Psi}(J) \) under hydrostatic deformation for varying stress levels. (a): metastable case with \( \alpha = 300.0 \), \( \beta = 0.5 \); (b): bistable case with \( \alpha = 1000.0 \), \( \beta = 0.35 \).}
    \label{fig:combined_G_vs_J}
\end{figure}

Further, we examine the associated Gibbs free energy density \( \widehat{\Psi} \), which through a Legendre transform describes the material in a force-control setting:
\begin{equation}\label{Eq: Pot_Energy}
    \widehat{\Psi} = \Psi - \mathbf{S} : \mathbf{E}.
\end{equation}
This transformation enables evaluation of \( \widehat{\Psi} \) under various prescribed levels of hydrostatic stress. As shown in Fig.~\ref{fig:combined_G_vs_J}, the model for the first set of parameters seen in Fig.~\ref{fig:combined_G_vs_J}(a) exhibits a single minimum at zero stress, but develops multiple minima under compression indicative of a metastable response. In contrast, the second set in Fig.~\ref{fig:combined_G_vs_J}(b) retains two stable configurations across all stress levels, indicative of a bistable response.

Additional supporting analyses are provided in Appendix~\ref{appendix: Parametric_Analytical_Plots}, which presents parameter sweeps of both Helmholtz free energy $\Psi(J)$ and corresponding stress-stretch response as a function of $\alpha$, $\beta$, and $\kappa$, along with Gibbs energy $G(J)$ landscapes under varying $\kappa$.

While the local formulation captures the key characteristics of phase transitions and densification under hydrostatic compression, it overlooks spatial interactions and the role of microstructural gradients. However, in many physical systems the mechanical response is governed by inherently nonlocal phenomena. To account for these effects, we extend the framework to a nonlocal formulation in a gradient-enhanced setting. This regularization not only enables the resolution of diffuse phase boundaries but also mitigates unphysical localization, yielding a more stable, accurate, and physically meaningful representation of material behavior.

\section{A Nonlocal Description for Architected Materials}\label{Section:NonLocalDescription}
 This aims to allow for the ability to resolve emerging phase transformations and multistability in the context of anisotropic hyperelasticity. Even though our construction for a non-(poly)convex free energy density in Section \ref{sec::local} focused on the volumetric response, here a more general framework will be presented that allows for an extension to modeling anisotropic architected materials. To build a physically meaningful connection to the underlying microstructural features, we will extend to a nonlocal formulation. This nonlocality allows the model to capture long-range spatial interactions that are essential for resolving the emerging phase-boundary topology at a macroscopic level~\cite{tekog2011size}. 
In particular, instabilities arising due to the non-(poly)convex local formulation presented in Section \ref{sec::local} could be recast in terms of a nonlocal counterpart of the Jacobian determinant, denoted $\tilde{J}$. Equivalently, the nonlocal counterparts for the mixed invariants could also be considered as ($\tilde{I}_{1}, \ldots, \tilde{I}_{\mathrm{n}}$).

\subsection{Thermodynamic considerations}
We begin by identifying the relevant state variables governing the system. The primary kinematic variable is the displacement field \( \mathbf{u} \), which defines the deformation of the material. To capture nonlocal effects and regularize the phase transitions, we augment the state space with a nonlocal mixed invariants ($\tilde{I}_{1}, \ldots, \tilde{I}_{\mathrm{n}}$). Together, $\mathbf{u}$ and ($\tilde{I}_{1}, \ldots, \tilde{I}_{\mathrm{n}}$) provide a comprehensive set of state variables capable of capturing both the macroscopic deformation and the microstructural evolution associated with phase transformation in anisotropic architected materials. One can physically interpret these nonlocal quantities as corresponding mixed invariants averaged through a representative volume relevant to the micro-architecture of the material. With these state variables defined, we employ the principle of virtual power to establish the governing equations. 

The internal mechanical power, $P_{\text{int}}$, in the reference configuration $\Omega_0$, is expressed as
\begin{equation}\label{Eq: IntPower}
    P_{\text{int}}=\int_{\Omega_0} \left[\mathbf{P} \colon \nabla\dot{\mathbf{u}} + \sum_{i=1}^{\mathrm{n}} \left(f_{i}\dot{\tilde{I}}_{i} + \boldsymbol{\xi}_{\tilde{i}} \cdot \nabla\dot{\tilde{I}}_{i} \right) \right]\, dV,
\end{equation}
where $\mathbf{P}$ denotes the first Piola-Kirchhoff stress tensor, while $f_{\tilde{J}}$ and $\boldsymbol{\xi}_{\tilde{J}}$ represent the generalized microforces conjugate to the nonlocal mixed invariants ($\tilde{I}_{1}, \ldots, \tilde{I}_{\mathrm{n}}$). Their material gradients are $\nabla\tilde{I}_{1}$, respectively ($\nabla(\,)=\partial(\,)/\partial\mathbf{X}$), and $dV$ denotes the infinitesimal volume element in $\Omega_0$.

Neglecting body forces, the external mechanical power \( P_{\text{ext}} \) over the reference configuration \( \Omega_0 \) can be expressed as
\begin{equation}
    P_{\text{ext}} = \int_{\partial \Omega_0} \mathbf{T} \cdot \dot{\mathbf{u}} \, dS,
\end{equation}
where \( \mathbf{T} \) denotes the prescribed surface traction vector and \( \dot{\mathbf{u}} \) is the material velocity. The surface element \( dS \) is measured with respect to the reference configuration.
Invoking the principle of virtual power, which equates internal and external virtual power contributions for all admissible variations, we obtain $\delta P_{\text{int}} = \delta P_{\text{ext}}$. This variational identity encapsulates the first law of thermodynamics in mechanical systems. Applying the divergence theorem to convert boundary integrals into volume integrals, the following field equations and boundary conditions emerge:

\begin{itemize}
    \item \textbf{Mechanical equilibrium and boundary conditions:}
    \begin{subequations}
        \begin{align}
            \nabla \cdot \mathbf{P} &= \mathbf{0} \quad \text{in } \Omega_0, \\
            \mathbf{u} &= \check{\mathbf{u}} \quad \text{on } \partial \Omega_0^u, \\
            \mathbf{P} \cdot \mathbf{N} &= \mathbf{T} \quad \text{on } \partial \Omega_0^t,
        \end{align}
    \end{subequations}

    \item \textbf{Microforce balance and boundary conditions for each nonlocal mixed invariant \( \tilde{I}_i \):}
    \begin{subequations}
        \begin{align}
            f_{\tilde{I}_i} + \nabla \cdot \boldsymbol{\xi}_{\tilde{I}_i} &= 0 \quad \text{in } \Omega_0, \\
            \tilde{I}_i &= \check{I}_i \quad \text{on } \partial \Omega_0^{I_i}, \\
            \boldsymbol{\xi}_{\tilde{I}_i} \cdot \mathbf{N} &= \check{\iota}_i \quad \text{on } \partial \Omega_0^{\xi_i},
        \end{align}
    \end{subequations}
\end{itemize}
Here, \( \mathbf{N} \) denotes the outward unit normal to the boundary \( \partial \Omega_0 \), and \( \check{\iota}_i \) are prescribed microtractions. These field equations collectively govern the macroscopic deformation and internal microstructural evolution within the nonlocal formulation.

To ensure thermodynamic admissibility, the rate of change of the Helmholtz free energy density \( \Psi = \Psi(\mathbf{F}, \lbrace \tilde{I}_i \rbrace, \lbrace \nabla \tilde{I}_i \rbrace) \) is expressed using the chain rule as
\begin{equation}\label{eq:RatePsiCorrected}
    \frac{d\Psi}{dt} = \frac{\partial \Psi}{\partial \mathbf{F}} \colon \dot{\mathbf{F}} + \sum_i \left( \frac{\partial \Psi}{\partial \tilde{I}_i} \dot{\tilde{I}}_i + \frac{\partial \Psi}{\partial \nabla \tilde{I}_i} \cdot \nabla \dot{\tilde{I}}_i \right).
\end{equation}
Combining this with the internal power expression and invoking the second law of thermodynamics via the Clausius–Duhem inequality, we obtain the dissipation inequality:
\begin{equation}
    \mathcal{D} = \left( \mathbf{P} - \frac{\partial \Psi}{\partial \mathbf{F}} \right) \colon \dot{\mathbf{F}} + \sum_i \left( f_{\tilde{I}_i} - \frac{\partial \Psi}{\partial \tilde{I}_i} \right) \dot{\tilde{I}}_i + \sum_i \left( \boldsymbol{\xi}_{\tilde{I}_i} - \frac{\partial \Psi}{\partial \nabla \tilde{I}_i} \right) \cdot \nabla \dot{\tilde{I}}_i \geq 0.
\end{equation}

To guarantee that the inequality is satisfied for all admissible processes, the following constitutive relations are adopted:
\begin{equation}\label{Eq: ConstitutiveRelations}
    \mathbf{P} = \frac{\partial \Psi}{\partial \mathbf{F}}, \quad
    f_{\tilde{I}_i} = \frac{\partial \Psi}{\partial \tilde{I}_i}, \quad
    \boldsymbol{\xi}_{\tilde{I}_i} = \frac{\partial \Psi}{\partial \nabla \tilde{I}_i}, \quad \text{for all } i.
\end{equation}
These relations reflect the thermodynamically consistent evolution of the system and serve as the basis for implementing regularized phase transformation models within the framework of architected materials. 

\subsection{A specific material model}\label{Subsection: NL_Psi}
To incorporate nonlocal effects into the constitutive framework, we introduce a nonlocal volume ratio \( \tilde{J} \), which serves as a regularized approximation of the volumetric deformation measure \( J \). This nonlocal mixed invariant captures the essential features of volumetric behavior while enabling spatial interactions to be embedded within the formulation.

Building upon the local expression defined in Eq.~\ref{Eq: LocalForm_Psi}, we reformulate the non-(poly)convex energy contribution from the Gao-Ogden model \( \Psi_{\text{GO}} \) in terms of \( \tilde{J} \), yielding:
\begin{equation}
    \Psi_{\text{GO}} = \frac{\alpha}{2} \left( \frac{(1 - \tilde{J})^2}{2} - \beta(1 - \tilde{J}) \right)^2,
\end{equation}
where the parameters \( \alpha \) and \( \beta \) control the shape and depth of the resulting potential landscape, thereby enabling the modeling of phase coexistence and bistable or metastable behavior.

We include an energetic penalty term for the discrepancies between \( \tilde{J} \) and \( J \)
\begin{equation}
    \Psi^{\text{coup}} = c(J - \tilde{J})^2 ,
\end{equation}
where $c$ is the modulus of that penalty. We also introduce a gradient regularization term that penalizes the unbounded growth of the gradient of the nonlocal volume ratio as
\begin{equation}
    \Psi^{\text{grad}} = dl^2 \lvert \nabla \tilde{J} \rvert^2
\end{equation}
where $d$ is the modulus of the gradient regularization and $l$ is a material nonlocal length scale relevant to the micro-architecture. As we will see later, controlling the length scale can promote or deter imperfection sensitivity in the response. 

By combining the Neo-Hookean volumetric response, the non-(poly)convex energetic contribution, and the local/nonlocal coupling and the gradient regularization, we arrive at the full expression for the Helmholtz free energy density:
\begin{equation}\label{Eq: Psi_NL_GO}
    \Psi = \frac{\mu}{2}\left(I_1 - 3 - 2 \ln J\right) 
         + \frac{\kappa}{2}\left(\ln J\right)^2 
         + \frac{\alpha}{2} \left( \frac{(1 - \tilde{J})^2}{2} - \beta(1 - \tilde{J}) \right)^2 
         + c(J - \tilde{J})^2 
         + dl^2 \lvert \nabla \tilde{J} \rvert^2,
\end{equation}
where \( \mu \) and \( \kappa \) denote the shear and bulk moduli, respectively, and \( I_1 \) is the first invariant of the right Cauchy--Green deformation tensor.

Alternatively, a second variant of the nonlocal formulation can be constructed utilizing the double-well \( \Psi_{\text{DW}} \) in terms of the nonlocal volume ratio. This yields:
\begin{equation}\label{Eq: Psi_NL_DW}
    \Psi = \frac{\mu}{2}\left(I_1 - 3 - 2 \ln J\right) 
         + \frac{\kappa}{2}\left(\ln J\right)^2 
         + \zeta(\tilde{J} - K)^2(\tilde{J} - 1)^2 
         + c(J - \tilde{J})^2 
         + dl^2 \lvert \nabla \tilde{J} \rvert^2.
\end{equation}

Both formulations establish a thermodynamically consistent, gradient-enhanced framework for modeling volumetric phase transformations and interfacial phenomena in architected soft materials. The flexibility to interchange between non-(poly)convex energy landscapes allows for tailored representations of metastability, phase topology, and energy barriers across a range of material systems.

\subsection{Constitutive relations}

In this section, we identify the power conjugate quantities associated with the primary state variables: the displacement field \( \mathbf{u} \), the nonlocal volume ratio \( \tilde{J} \), and its spatial gradient \( \nabla \tilde{J} \). These conjugates emerge naturally from the expression of the internal mechanical power and are determined via the constitutive relations introduced in Eq.~\ref{Eq: ConstitutiveRelations}. Specifically, the power conjugates correspond to the first Piola--Kirchhoff stress tensor \( \mathbf{P} \), and the microforce conjugates, \( f_{\tilde{J}} \) and \( \boldsymbol{\xi}_{\tilde{J}} \) respectively.

For the nonlocal formulation outlined in Section.~\ref{Subsection: NL_Psi}, this yields:
\begin{equation}\label{Eq: 1stPK_Stress}
    \mathbf{P} = \mu\left(\mathbf{F} - \mathbf{F}^{-T}\right) + \kappa \ln J \, \mathbf{F}^{-T} + 2c(J - \tilde{J}) J \mathbf{F}^{-T}.
\end{equation}
The first two terms correspond to the standard compressible Neo-Hookean model, while the third term arises from the coupling energy \( \Psi^{\text{coup}} \).

The microforce conjugate is:
\begin{equation}\label{Eq: boldXi}
    \boldsymbol{\xi}_{\tilde{J}}= 2dl^2 \nabla \tilde{J}.
\end{equation}
This term governs the spatial smoothness of the nonlocal volume ratio \( \tilde{J} \), effectively penalizing sharp phase boundaries and promoting diffuse transitions relative to $l$.

Finally, \( f_{\tilde{J}} \) is derived, and it is noted that its expression depends on the specific form of the non-(poly)convex potential employed. For the Gao--Ogden formulation \( \Psi_{\text{GO}} \) and the alternative double-well form \( \Psi_{\text{DW}} \), the corresponding expressions are:
\begin{subequations}\label{Eq: fJVariants}
    \begin{align}
        f_{\tilde{J}}^{\text{GO}} &= \alpha\left( \frac{(1 - \tilde{J})^2}{2} - \beta(1 - \tilde{J}) \right)(\tilde{J} - 1 + \beta) - 2c(J - \tilde{J}), \\
        f_{\tilde{J}}^{\text{DW}} &= 2\zeta(\tilde{J} - K)(\tilde{J} - 1)\left(2\tilde{J} - (1 + K)\right) - 2c(J - \tilde{J}).
    \end{align}
\end{subequations}
In both formulations, the final term \( -2c(J - \tilde{J}) \) stems from the coupling energy \( \Psi^{\text{coup}} \).

Together, these expressions define the thermodynamic driving forces associated with each state variable and serve as the foundation for deriving the weak form and numerical implementation of the nonlocal model.

\subsection{Strong and weak forms}

Similar to the phase field model for fracture ~\cite{miehe2010phase, ambati2015review}, in this work we aim to capture unstable transitions of a materials system. Anticipating the potential convergence issues of the numerical solution, we incorporate an artificial viscosity term that introduces a controlled dissipation mechanism by penalizing the rapid temporal evolution of nonlocal variables (in this case the nonlocal volume ratio), thereby enhancing the stability and robustness of the numerical solution. This damping mechanism is artificial, but it is often taken to be representative with dissipation related to the rate of the micro-mechanism in consideration. Below, we revert to the general notation for our formulation.


\vspace{1em}
\noindent \textbf{Strong Form.} The governing equations, modified to include viscous damping, are:
\begin{itemize}
    \item \textbf{Field Equations:}
    \begin{subequations}
        \begin{align}
            \nabla \cdot \mathbf{P} &= \mathbf{0} \quad \text{in } \Omega_0, \\
            f_{\tilde{I}_i} + \nabla \cdot \boldsymbol{\xi}_{\tilde{I}_i} + \eta_i \, \dot{\tilde{I}}_i &= 0 \quad \text{in } \Omega_0,
        \end{align}
    \end{subequations}
    
    \item \textbf{Boundary Conditions:}
    \begin{subequations}
        \begin{align}
            \mathbf{u} &= \check{\mathbf{u}} \quad \text{on } \partial\Omega_0^{u}, \\
            \mathbf{P} \cdot \mathbf{N} &= \mathbf{T} \quad \text{on } \partial\Omega_0^{t}, \\
            \tilde{I}_i &= \check{I}_i \quad \text{on } \partial\Omega_0^{I_i}, \\
            \boldsymbol{\xi}_{\tilde{I}_i} \cdot \mathbf{N} &= \check{\iota}_i \quad \text{on } \partial\Omega_0^{\xi_i}.
        \end{align}
    \end{subequations}
\end{itemize}

Unless stated otherwise, we prescribe homogeneous natural microforce boundary conditions, i.e., zero microtraction on the entire boundary: \( \check{\iota} = 0 \), hence \( \boldsymbol{\xi}_{\tilde{I}_i} \cdot \mathbf{N} = 0 \) on \( \partial \Omega_0 \).

With, \( \eta_i > 0 \) being the artificial viscosity coefficient associated with the \( i \)-\text{th} nonlocal mixed invariant, and \( \dot{\tilde{I}}_i \) denotes its rate of change over time. The introduction of artificial viscosity serves a dual purpose: it improves numerical stability in the presence of steep gradients in the internal field, \( \tilde{I}_i \) and provides a consistent means of regulating the incremental evolution of nonlocal mixed invariants through a dissipative-like mechanism.

To transition to the weak form, we define appropriate function spaces for the trial and test functions. These are expressed as:
\begin{subequations}\label{Eq: FunctionSpaces}
    \begin{align}
        \mathbb{U} &= \left\{ \mathbf{u} \in [H^1(\Omega_0)]^d \ \big| \ \mathbf{u} = \check{\mathbf{u}} \text{ on } \partial\Omega_0^u \right\}, \\
        \mathbb{I}_i &= \left\{ \tilde{I}_i \in H^1(\Omega_0) \ \big| \ \tilde{I}_i = \check{I}_i \text{ on } \partial\Omega_0^{I_i} \right\},
    \end{align}
\end{subequations}
where \( d \) indicates the number of spatial dimensions (e.g., \( d = 2 \) for planar problems). 

\vspace{1em}
\noindent \textbf{Weak Form.} Let \( (\delta \mathbf{u}, \delta \tilde{I}_i) \in \mathbb{U}_0 \times \mathbb{I}_{i,0} \) be admissible test functions vanishing on the corresponding Dirichlet boundaries. The weak form of the system is then to find \( (\mathbf{u}, \{ \tilde{I}_i \}) \in \mathbb{U} \times \prod_i \mathbb{I}_i \) such that:
\begin{subequations}
    \begin{align}
        \int_{\Omega_0} \mathbf{P} : \nabla \delta \mathbf{u} \, dV &= \int_{\partial\Omega_0^t} \mathbf{T} \cdot \delta \mathbf{u} \, dS, \\
        \int_{\Omega_0} f_{\tilde{I}_i} \, \delta \tilde{I}_i \, dV + \int_{\Omega_0} \boldsymbol{\xi}_{\tilde{I}_i} \cdot \nabla \delta \tilde{I}_i \, dV + \int_{\Omega_0} \eta_i \, \dot{\tilde{I}}_i \, \delta \tilde{I}_i \, dV &= \int_{\partial\Omega_0^{\xi_i}} \check{\iota}_i \, \delta \tilde{I}_i \, dS \quad \forall i.
    \end{align}
\end{subequations}
This generalized formulation offers a robust framework for modeling complex boundary value problems.

For the case of a single scalar nonlocal variable, the volume ratio \( \tilde{J} \) as we previously utilized in the specialization of the Helmholtz free energy, the weak form simplifies accordingly. Find \( (\mathbf{u}, \tilde{J}) \in \mathbb{U} \times \mathbb{J} \) such that:
\begin{subequations}
    \begin{align}
        \int_{\Omega_0} \mathbf{P} : \nabla \delta \mathbf{u} \, dV &= \int_{\partial\Omega_0^t} \mathbf{T} \cdot \delta \mathbf{u} \, dS, \\
        \int_{\Omega_0} f_{\tilde{J}} \, \delta \tilde{J} \, dV + \int_{\Omega_0} \boldsymbol{\xi}_{\tilde{J}} \cdot \nabla \delta \tilde{J} \, dV + \int_{\Omega_0} \eta \, \dot{\tilde{J}} \, \delta \tilde{J} \, dV &= \int_{\partial\Omega_0^{\xi}} \check{\iota} \, \delta \tilde{J} \, dS.
    \end{align}
\end{subequations}

\section{Finite Element Implementation}\label{Section:FEMImplementation}

The proposed computational framework is implemented using the open-source platform \texttt{FEniCS} \cite{logg2012automated, FEniCSAlnaes2015}, which facilitates the concise definition and efficient evaluation of variational forms. Automated differentiation via the Unified Form Language (UFL) \cite{UFL}, coupled with \texttt{PETS}c-based solver backends, enables consistent linearization and robust handling of nonlinear systems. All simulation scripts are made available in a public GitHub repository.\footnote{GitHub link to be provided upon publication.}

The governing equations are posed over the reference domain \( \Omega_0 \subset \mathbb{R}^d \) and involve two primary fields: the displacement \( \mathbf{u} : \Omega_0 \to \mathbb{R}^d \), and nonlocal volume ratio \( \tilde{J} : \Omega_0 \to \mathbb{R} \), which captures volumetric softening and gradient-regularized phase transitions. The total potential is defined via a non-(poly)convex Helmholtz free energy functional incorporating both classical strain energy and gradient-penalized microforces.

\subsection{Discretization and function Spaces}
A mixed finite element strategy is employed to resolve the coupled field problem. We choose a Taylor-Hood space, where the displacement \( \mathbf{u} \) is discretized using quadratic Lagrange elements, while \( \tilde{J} \) is approximated with linear elements.

\subsection{Solution strategy}
The discrete system is solved via a Newton–Raphson procedure applied to the coupled residual.
Artificial viscosity is introduced to regularize the evolution of \( \tilde{J} \) in pseudo-time. This regularization is discretized using a backward Euler scheme, an implicit first-order time integration method. At each pseudo-time step, the evolution of \( \tilde{J} \) is computed using its current value and its value from the previous step, treating the update as fully implicit. 
The backward Euler method is particularly well-suited for these problems, as it suppresses high-frequency numerical oscillations and enables stable convergence across a wide range of parameter regimes.

Nonlinear systems are solved using \texttt{PETS}c’s \texttt{SNES} interface with line search globalization and \texttt{MUMPS} direct solver backends, augmented by Hypre algebraic multigrid (AMG) preconditioners. Convergence is monitored via relative and absolute residual norms, with tolerances set to \( 10^{-9} \), and a maximum of 50 iterations per step. These criteria enable accurate resolution of sharp transitions while maintaining computational efficiency.

Further, a hybrid scheme is proposed. When convergence is not attained in the coupled system, the algorithm reverts to a staggered scheme~\footnote{It is noted that this is opposite from the strategy often followed in phase field fracture problems, where first a staggered scheme is followed and the solution scheme reverts to monolithic upon satisfactory reduction of the residual.}. In this alternate strategy, the displacement \( \mathbf{u} \) and nonlocal volume ratio \( \tilde{J} \) are solved sequentially: first, \( \mathbf{u} \) is updated for fixed \( \tilde{J} \), and subsequently \( \tilde{J} \) is evolved using the current configuration of \( \mathbf{u} \). This alternating minimization proceeds until the update norm satisfies
\[
\| \tilde{J}^{(k+1)} - \tilde{J}^{(k)} \|_{L^2(\Omega_0)} < \epsilon_{\text{stag}},
\]
with \( \epsilon_{\text{stag}} = 10^{-3} \). The staggered approach significantly improves robustness in non-(poly)convex regimes, allowing the simulation to proceed through regions of severe instability while maintaining fidelity to the governing equations.

\begin{codeboxraw}
\begin{algorithmic}[1]

\State Generate mesh on \( \Omega_0 \); define function spaces \( \mathbb{U}, \mathbb{J} \), and mixed space \( \mathbb{V} = \mathbb{U} \times \mathbb{J} \)
\State Initialize \( \mathbf{u}_0, \tilde{J}_0 \)

\For{each load step \( i = 1, \dots, N \)}
    \State Apply increment \( u^{(i)}_{\mathrm{inc}} \)
    \State Solve \( \mathcal{R}(\mathbf{u}, \tilde{J}) = 0 \) via monolithic Newton method (\texttt{SNES})
    \If{converged}
        \State Record solution; compute derived quantities (stress, \( J \), \( \Psi \), etc.)
    \Else
        \State Initialize staggered fields: \( \mathbf{u}^{(0)} \leftarrow \mathbf{u}_{i-1}, \ \tilde{J}^{(0)} \leftarrow \tilde{J}_{i-1} \)
        \Repeat
            \State Solve for \( \mathbf{u}^{(k+1)} \) with fixed \( \tilde{J}^{(k)} \)
            \State Solve for \( \tilde{J}^{(k+1)} \) with fixed \( \mathbf{u}^{(k+1)} \)
        \Until{ \( \|\tilde{J}^{(k+1)} - \tilde{J}^{(k)}\|_{L^2(\Omega_0)} < \epsilon_{\mathrm{stag}} \) }
        \State Set \( \mathbf{u}_i \leftarrow \mathbf{u}^{(k+1)}, \ \tilde{J}_i \leftarrow \tilde{J}^{(k+1)} \)
    \EndIf
\EndFor

\end{algorithmic}
\end{codeboxraw}

\vspace{4pt}
{\noindent\small
\textbf{Box 1.} Pseudocode summary of the finite element implementation. The procedure integrates a monolithic Newton solver with a staggered approach.
}

\section{Results and Discussion} \label{Section:Results}

This section presents a detailed assessment of the proposed nonlocal finite element framework through a series of numerical simulations designed to evaluate its physical fidelity and computational robustness. The simulations investigate the mechanical response of architected solids under displacement-controlled monotonic and cyclic loading, with a focus on capturing emerging phase transformations and corresponding structural responses.

\begin{figure}[h!]
    \centering
    \includegraphics[width=0.85\linewidth]{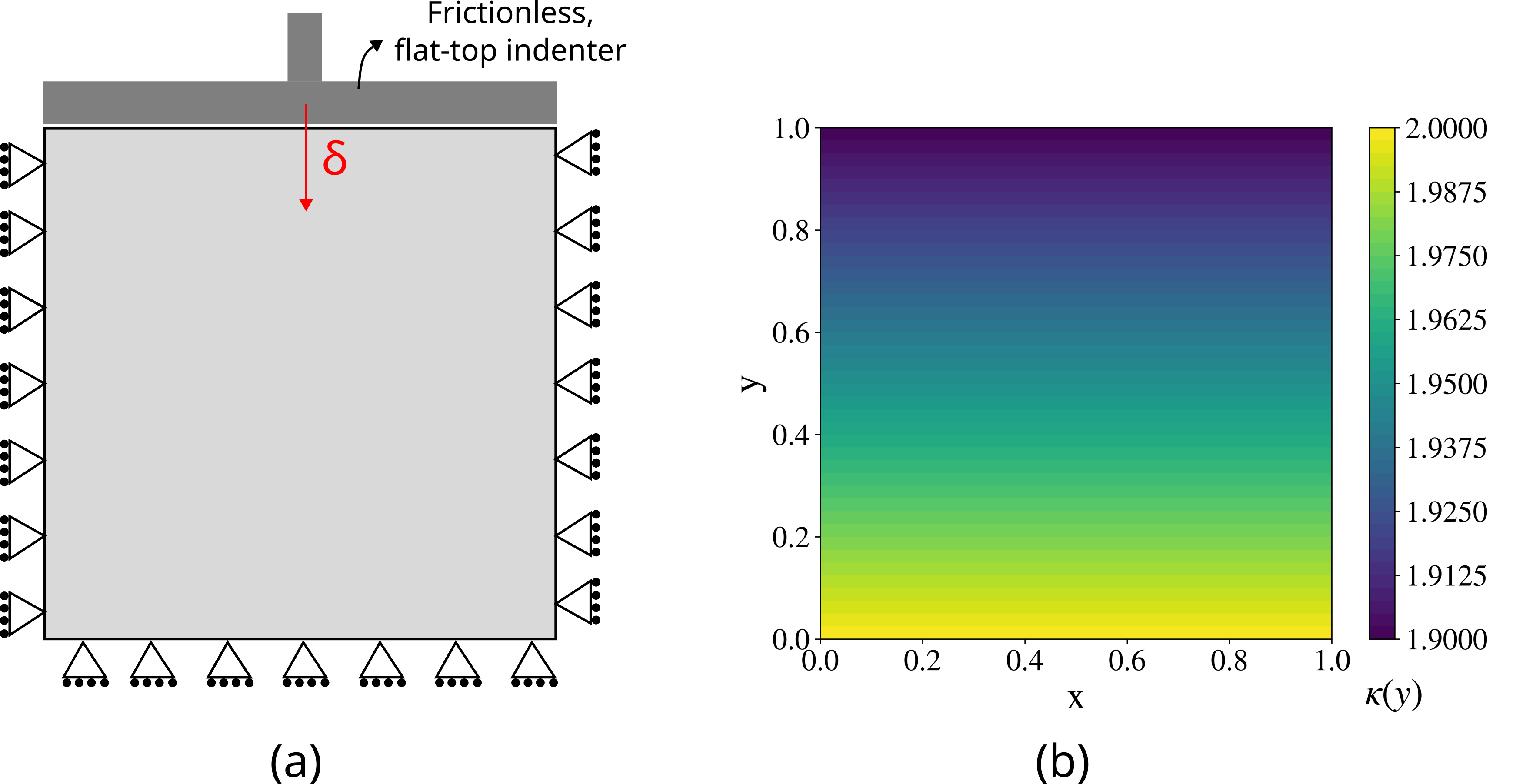}
    \caption{Simulation setup and imposed inhomogeneity. (a) Loading configuration for both monotonic and cyclic cases. (b) Spatial variation of the bulk modulus \( \kappa(y) \), shown for a 5\% linear grading. }
    \label{fig:loading_schematics_combined}
\end{figure}

Fig.~\ref{fig:loading_schematics_combined} outlines the simulation setup. Fig.~\ref{fig:loading_schematics_combined}(a) corresponds to a unit square domain that undergoes confined compression in plane strain; this setup is utilized for the majority of the results. A flat, frictionless indenter imposes vertical displacement on the top surface, while roller boundary conditions along the lateral and bottom edges are applied. In cyclic simulations, the indenter is not imposed through a prescribed displacement but through a contact constraint; this is crucial for identifying potential equilibrium states during the unloading. Fig.~\ref{fig:loading_schematics_combined}(b) illustrates the spatial variation of the bulk modulus \( \kappa(\mathbf{X}) \) that is utilized for some simulation. It is implemented as a linear gradient along the domain height $y$ ($y=0$ at the bottom face) with a prescribed percentage deviation around a maximum value (e.g., 5\% about \( \kappa = 2.0 \)). 

Two representative material models are considered to construct a metastable and a bistable case.  In both cases (all quantities are dimensionless), the shear modulus \( \mu \) and the maximum bulk modulus \( \kappa \) are set to 2.0. The metastable case uses \( \alpha = 300.0 \), \( \beta = 0.5 \), \( c = 230.0 \), \( d = 1.0 \), producing a shallow single-well energy landscape that allows for reversible phase transformations. The bistable case adopts \( \alpha = 1000.0 \), \( \beta = 0.35 \), \( c = 700.0 \), \( d = 1.0 \), yielding a double-well profile with a secondary energy minimum. The reaction force \( \bar{F} \) is normalized by the shear modulus \( \mu \), and the vertical displacement \( \delta \) by the domain height. The nonlocal length scale that we select, normalized by the height of the domain is $l/H=0.017$, and is considered to be relevant to the representative volume element (RVE) or statistical volume element (SVE) size.

The subsections that follow investigate key modeling features in depth: solver verification and effects of artificial viscosity, imperfection sensitivity under stiffness grading, spatial evolution of local and nonlocal volumetric fields, cyclic response with reversibility and retention, artificial viscosity-driven softening behavior, and the emergence of auxetic patterns under strong nonlocal coupling. Additional supporting analyses are provided in the appendices, including stress evolution in cyclic loading (Appendix~\ref{appendix: Cyclic_Loading}), viscous regularization effects (Appendix~\ref{appendix: viscous_micro_buckling}), and a convergence study on length scale and mesh refinement (Appendix~\ref{section:length_scale_study}).


\subsection{Performance of the hybrid solution scheme}

\begin{figure}[H]
    \centering
    \includegraphics[width=\linewidth]{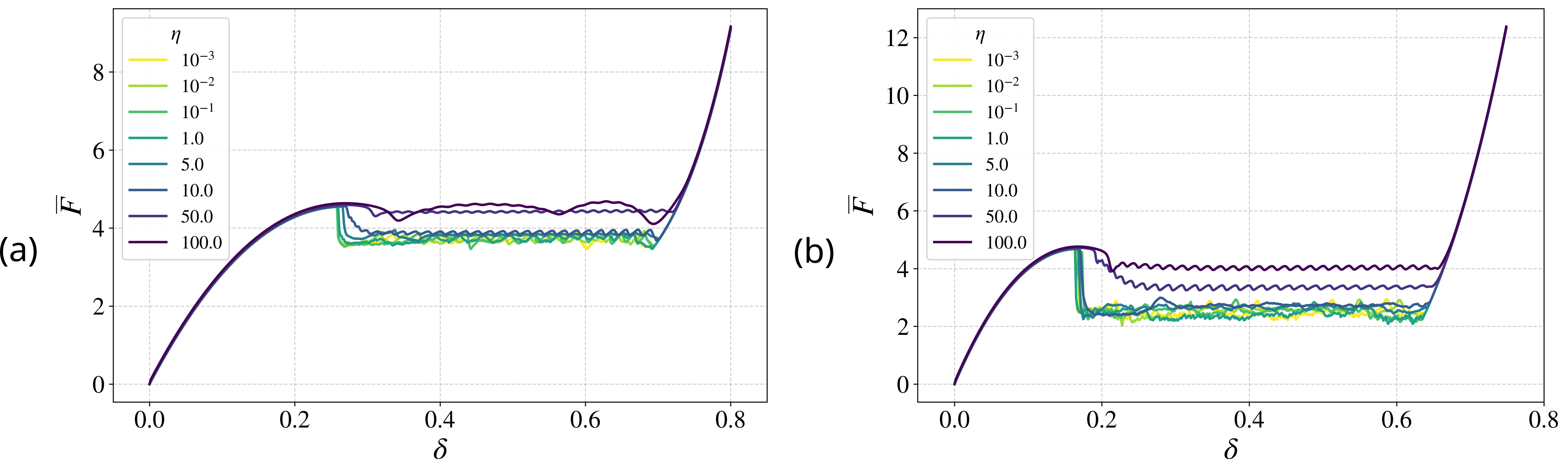}
    \caption{Effect of artificial viscosity \( \eta \) on the force-displacement response for (a) metastable and (b) bistable energy landscapes. Simulations correspond to variation of bulk modulus \( \kappa \) by 5.0\% along the height of the domain with \( \kappa \) smallest at the top.}
    \label{fig:viscosity_study}
\end{figure}

\begin{figure}[h!]
    \centering
    \includegraphics[width=\linewidth]{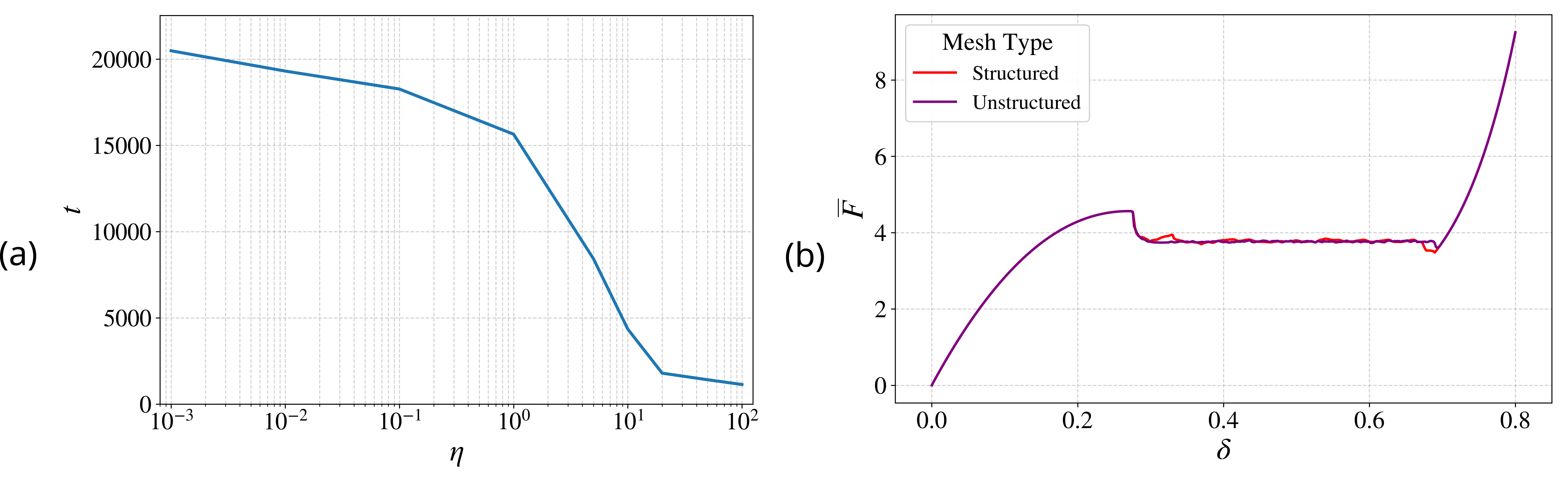}
    \caption{(a) Total wall–clock time $t$, in seconds versus artificial viscosity $\eta$ for identical meshes, loading protocol, and solver tolerances, with variation of bulk modulus $\kappa$ by $5.0\%$. Increasing $\eta$ consistently accelerates convergence, producing up to an order–of–magnitude runtime reduction across the range tested. Timings were obtained on an AMD Ryzen Threadripper 1950X (16 cores / 32 threads, SMT enabled; max 3.4\,GHz; dual–NUMA; 32\,MiB L3). (b) Comparison of force-displacement responses obtained using structured and unstructured meshes. Results correspond to a bulk modulus variation of \( \kappa \) by 1.0\% along the domain height and viscosity \( \eta = 5.0 \).}
    \label{fig:eta_time_&_mesh_effect}
\end{figure}

We begin by examining the influence of the artificial viscosity parameter \( \eta \) using a fixed structured mesh. Fig.~\ref{fig:viscosity_study} presents the force–displacement response for both selected cases for the metastable and bistable models under varying values of \( \eta \), with a fixed vertical bulk modulus variation of 5.0\%. Overall the response is separated in three stages i) elastic loading, ii) plateau, and iii) densification. For artificial viscosity $\eta<1$ the hybrid solver resorts to the staggered scheme to reach convergence, whereas for $\eta\geq 1$ convergence is attained through a monolithic scheme. Higher values of artificial viscosity lead to significantly faster convergence (Fig.~\ref{fig:eta_time_&_mesh_effect}(a); total wall–clock time $t$ in seconds, measured on an AMD Ryzen Threadripper 1950X, 16 cores / 32 threads). During elastic loading a limit load is reached, followed by a plateau with prominent oscillations indicative of instabilities, and eventually a stable densification path. This response is indicative of the elastic response of  low relative density materials with a high degree of regularity. Foams also show similar responses (with three distinct regimes) but without such strong signatures of localization and a corresponding force drop. For low artificial viscosity, the solutions exhibits a sharp drop after the limit load, which is smoothened for higher values. To maintain the important features of the solution profile and take advantage of the acceleration provided when adopting higher values of artificial viscosity, we utilize an intermediate value of \( \eta = 5.0 \) was found to provide an effective balance between physical fidelity and acceleration. It is important to note that artificial viscosity is often taken to mimic micro-level dissipation associated with rate effects for the evolution of the nonlocal variables. 

Further, we investigate the sensitivity of the solution to mesh topology, comparing structured and unstructured triangular meshes for a fixed viscosity \( \eta = 5.0 \) and a vertical bulk modulus variation of 1.0\%.

Fig.~\ref{fig:eta_time_&_mesh_effect}(b) shows that both mesh types successfully capture key mechanical features, including stiffness softening and onset of internal transitions. Minor variations in the unstable plateau stage are attributed to mesh-induced heterogeneity of the solution. One can note that the unstructured mesh leads to a smoother plateau compared to the structured case. The increased heterogeneity and approximation error expected in the unstructured case lead to local response-type (instead of limit load-type) behavior during the plateau. In reality, imperfection sensitivity in these materials is critical (as will be further discussed), and in homogeneous numerical cases, these imperfections lie in the spatial fluctuations due to numerical approximation error.

These findings collectively underscore the effectiveness of the proposed formulation in delivering stable and physically consistent results across a range of discretization schemes and regularization parameters.  Building on this, we now investigate how spatial heterogeneity of material properties --introduced through bulk modulus grading-- can further influence the onset and progression of localized phase transitions.

\subsection{Imperfection sensitivity and graded structural properties}

We introduce a controlled vertical gradient in the bulk modulus \( \kappa(\mathbf{X}) \) across the height of the domain. This gradient is designed such that \( \kappa(\mathbf{X}) \) increases from the top to the bottom of the specimen, thereby having lower compressibility at the top as indicated by the bulk to shear ratio $\kappa/\mu$ corresponding to the Neo-Hookean model; we note that these are not the true linearized bulk modulus of the proposed material models, but the comparison to the base Neo-Hookean model is useful for the stable elastic loading segment of the response. Additionally, in this study, we have not perturbed any of the parameters of the non-(poly)convex energy contribution $\Psi_\textrm{NC}$. This spatial variation serves as an intentional imperfection that biases the mechanical response and encourages preferential zones for phase transformation.

\begin{figure}[h!]
    \centering
    \includegraphics[width=\linewidth]{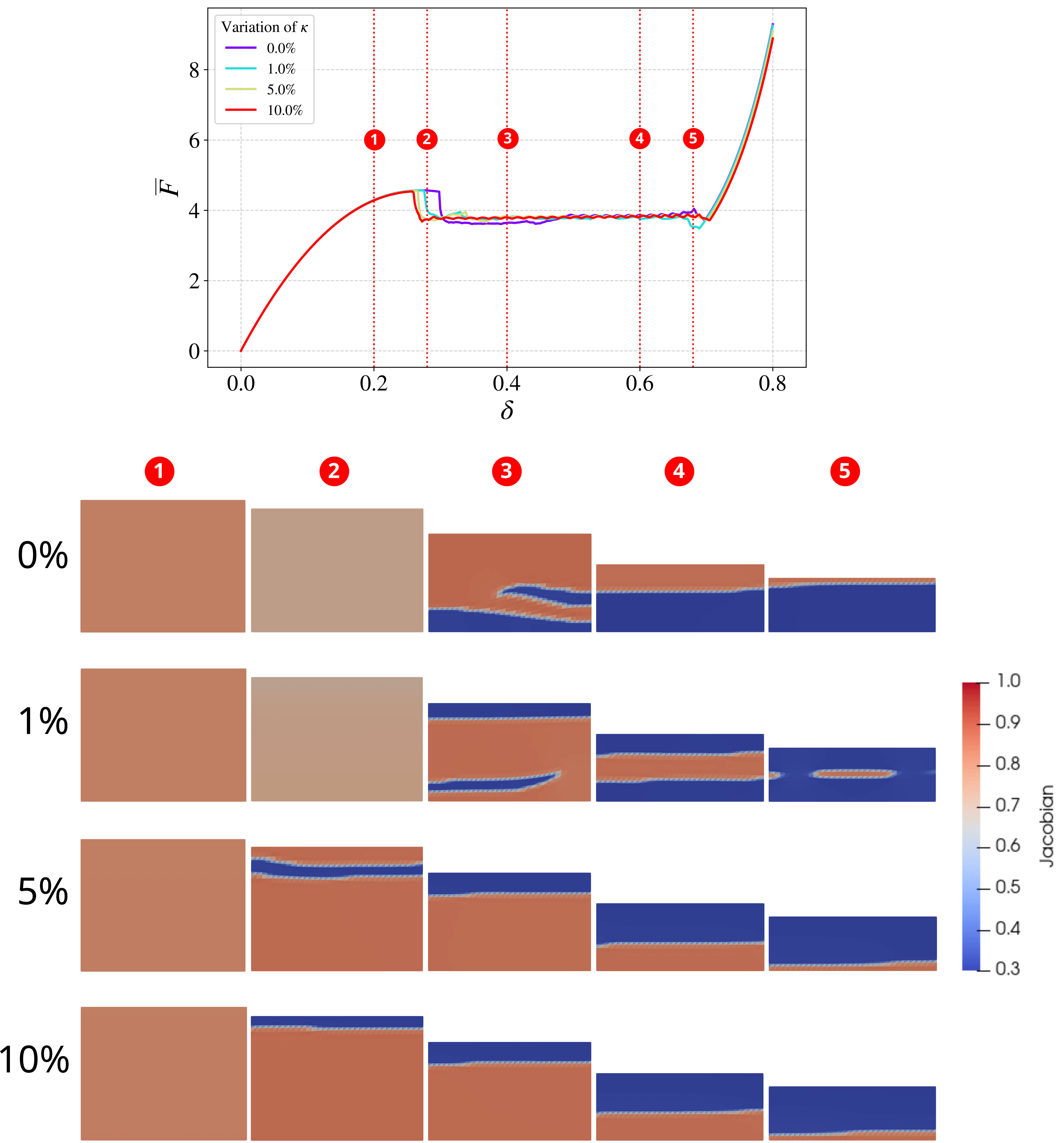}
    \caption{Effect of vertical grading in bulk modulus \( \kappa \) on the force-displacement response and the evolution of the Jacobian field \( J \) for a metastable material. The graded profiles correspond to 0.0\%, 1.0\%, 5.0\%, and 10.0\% variation in \( \kappa \) over the domain height. Snapshots at five loading stages illustrate the impact of heterogeneity of material properties on densification.}
    \label{fig:kappa_variation}
\end{figure}

Fig.~\ref{fig:kappa_variation} illustrates the force–displacement behavior alongside snapshots of the Jacobian field \( J \) at five distinct loading stages, for four different levels of vertical grading: 0.0\%, 1.0\%, 5.0\%, and 10.0\%. We focus on 5 snapshots of the loading up to the end of plateau. When the material properties are homogeneous throughout the domain (0.0\% variation), the system exhibits homogeneous deformation during elastic loading with spontaneous densification in two distinct regions, which later merge, and ultimately the entirety of the domain is densified. The location of the initial densified regions is purely due to imperfection sensitivity to the noise due to the approximation error. The introduction of a modest gradient (e.g., 1.0\%) causes the densification front to nucleate earlier and more distinctly near the top region, even though there is still a secondary densification region in the lower part of the domain. In this case, two fronts propagate and ultimately merge at the center of the domain.
As the magnitude of the gradient increases, this behavior becomes more pronounced and localized. Higher variation in \( \kappa \) (e.g., 5.0\% and 10.0\%) results in sharper front formation, earlier observation of densification. In all cases, the emergence of densification matched the drop in the force-displacement plot transitioning from the elastic loading stage to the plateau. In the plateau stage, there is coexistence of a highly densified phase ($J\approx0.3$) and a rarefied phase ($J\approx0.9$), indicating $70\%$ and $10\%$ volumetric compression accordingly. During this apparent phase transformation the force required is almost constant and only starts increasing during the densification stage.

\subsection{Local and nonlocal volume ratio evolution}

\begin{figure}[h!]
    \centering
    \includegraphics[width=\linewidth]{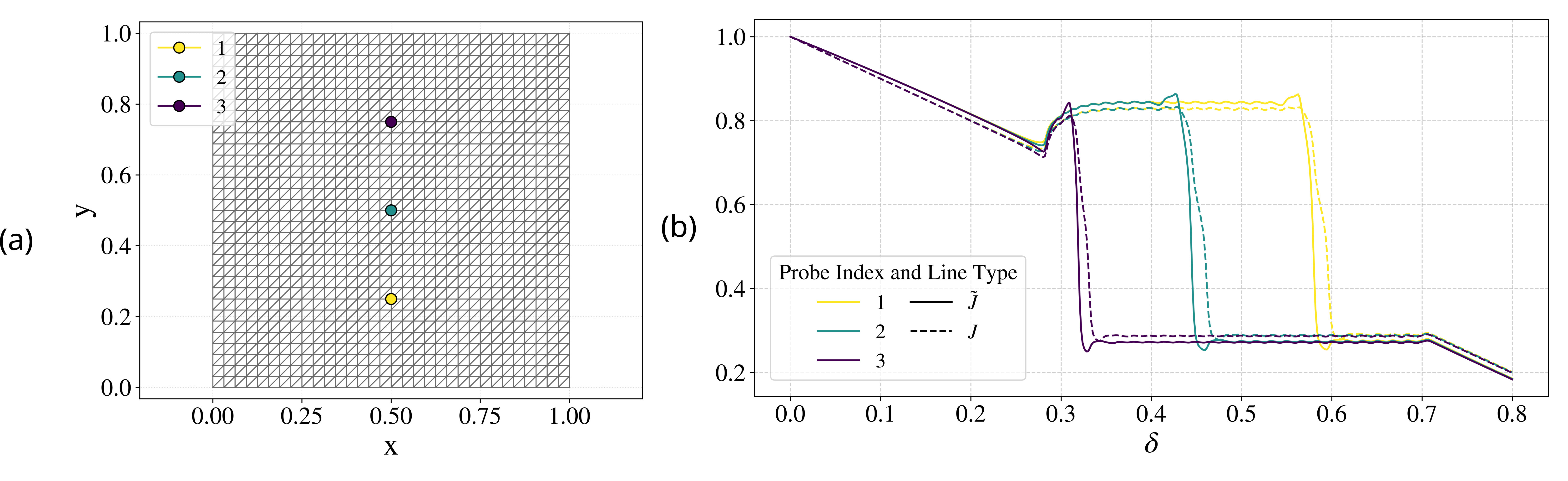}
    \caption{Evolution of local volume ratio \( J \) and nonlocal volume ratio \( \tilde{J} \) at three vertical probe points for a metastable material with artificial viscosity \( \eta = 20.0 \). Spatial probe locations are shown in (a), with corresponding variable trajectories shown in (b). A slight rise in response is observed prior to the collapse, indicating relaxation of the underlying material near the probe point as the densification front passes.}
    \label{fig:evolution}
\end{figure}

To investigate the evolution of local and nonlocal volume ratio during loading, we track the local volumetric deformation \( J \) and its nonlocal counterpart \( \tilde{J} \) at three vertical probe points positioned at increasing heights along the specimen’s centerline, as shown in Fig.~\ref{fig:evolution}(a). These probes are chosen to monitor the densification through the domain under increasing compressive loading.

The plot in Fig.~\ref{fig:evolution}(b) shows the evolution of \( J \) and \( \tilde{J} \) as functions of the applied displacement \( \delta \). Initially, all probes exhibit a uniform, nearly linear reduction in volume due to elastic compression. As loading continues, a sudden drop in \( J \) is observed at each probe, indicating the onset of local collapse as the densification front reaches that region. Just before this drop, however, a slight increase in both \( J \) and \( \tilde{J} \) is noticeable. This jump reflects the relaxation of the material below the probe as the region above densifies, causing local unloading that momentarily counteracts the ongoing compression.

The nonlocal variable \( \tilde{J} \) evolves more smoothly than its local counterpart due to spatial averaging and possibly the influence of artificial viscosity. This smoothing effect becomes particularly evident near the instability threshold, where \( J \) undergoes sharp transitions, while \( \tilde{J} \) changes more gradually. The temporal ordering of these transitions across the probes also reflects the downward propagation of the densification front.

\subsection{Cyclic loading response}

To examine whether our proposed models can capture the path-dependent behavior of architected solids under repeated loading, we perform cyclic compression tests on both metastable and bistable material systems using a flat-top indenter. Specifically, for the both the metastable and the bistable case we have in mind the corresponding responses recorded by Restrepo {\it{et al.}} (2015)~\cite{restrepo_phase_2015} for metastable and bistable architected materials and we aim to qualitatively reproduce them while also pushing the response into the densification stage. 

The indenter imposes displacement-controlled loading from the top surface, with unilateral contact constraints governing its interaction with the deformable body. This contact-based approach is utilized such that the indenter can detach from the body in case a secondary stable equilibrium state is attained. This setup facilitates realistic simulation of compression and partial recovery, capturing how deformation fronts form, propagate, and evolve under cyclic loading conditions, with responses shaped by the underlying energy landscape and material history.

\begin{figure}[h!]
    \centering
    \includegraphics[width=\linewidth]{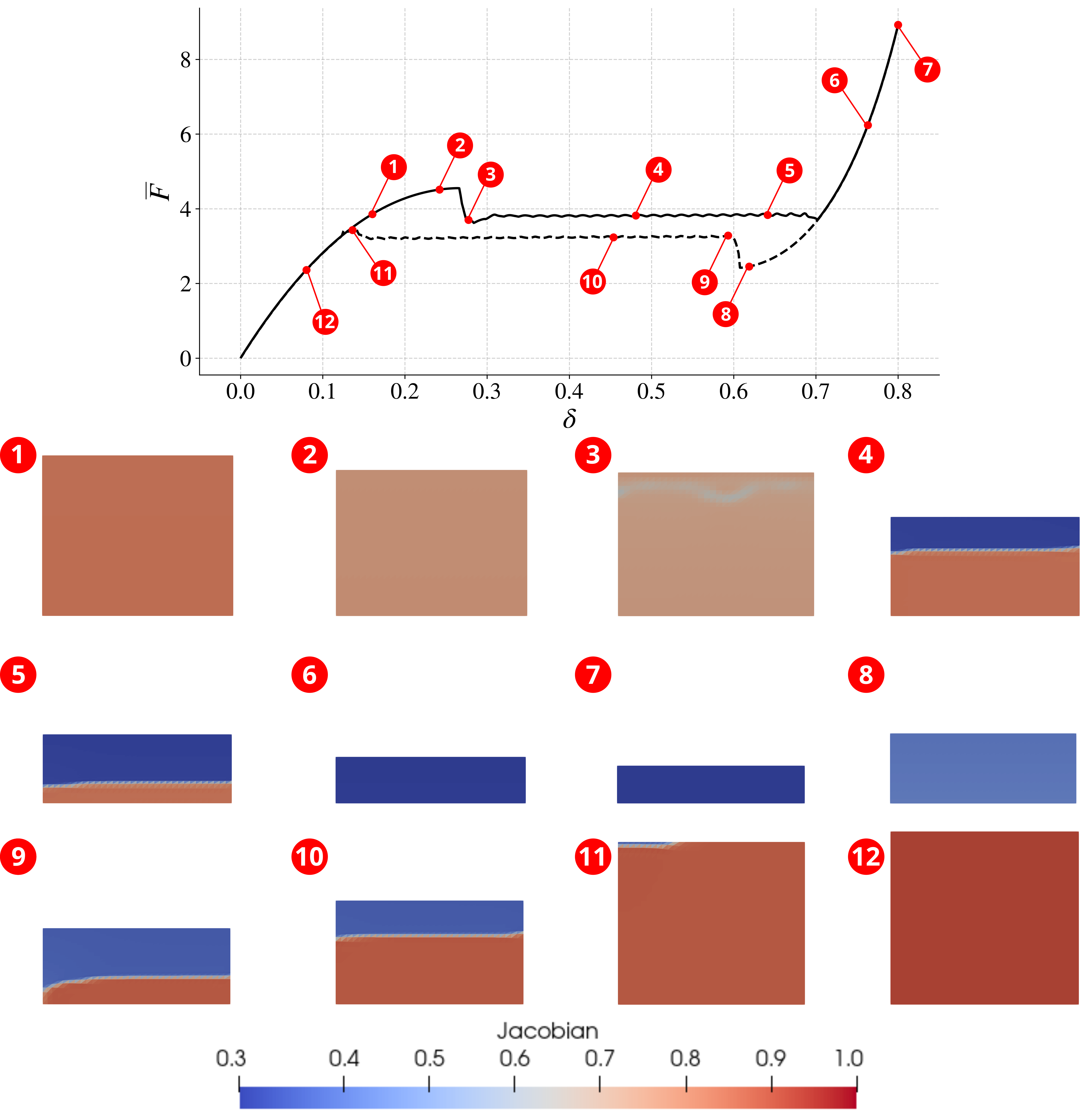}
    \caption{Cyclic response for a metastable material system with 5.0\% vertical grading in bulk modulus \( \kappa \) and artificial viscosity \( \eta = 5.0 \). Displacement-controlled loading and unloading are applied via a flat-top indenter in contact with the top surface. The top plot shows the force–displacement trajectory over one full cycle; the numbered snapshots correspond to key loading stages and show the evolution of the Jacobian field \( J \).}
    \label{fig:metastable_cyclic}
\end{figure}

Fig.~\ref{fig:metastable_cyclic} illustrates the cyclic response for the metastable case. The force–displacement curve exhibits a nearly reversible trajectory with mild hysteresis. Snapshots at 12 key points highlight the evolution of the local volume ratio \( J \) during loading (points 1–6) and unloading (points 7–12). As the indenter compresses the body, a densification front emerges near the top and progresses downward in response to increasing deformation. Upon unloading, partial recovery occurs as the front recedes until a full recovery to the initial state is achieved. This reversible behavior is characteristic of metastable systems with moderate energy barriers.

\begin{figure}[h!]
    \centering
    \includegraphics[width=\linewidth]{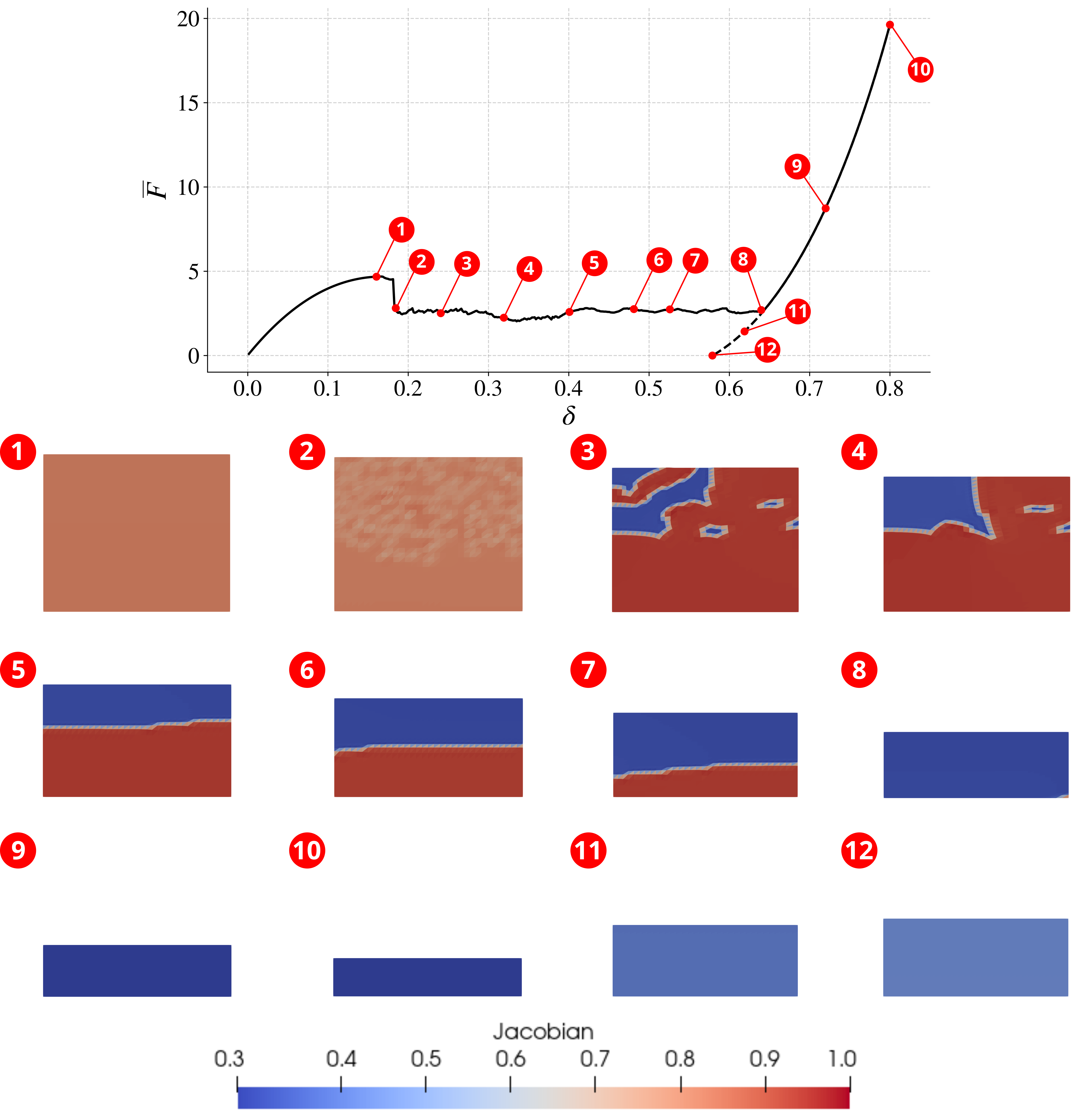}
    \caption{Cyclic response for a bistable material system with 5.0\% vertical grading in bulk modulus \( \kappa \) and viscosity \( \eta = 5.0 \). Loading and unloading are applied through a flat-top indenter in contact. The top plot shows the force–displacement hysteresis, and snapshots of \( J \) reveal permanent deformation and residual fronts.}
    \label{fig:bistable_cyclic}
\end{figure}

In contrast, the bistable case shown in Fig.~\ref{fig:bistable_cyclic} displays a markedly different cyclic response. The force–displacement curve reveals strong hysteresis and significant irreversibility, indicating that the material undergoes a phase transition into a deep energy well from which it cannot fully recover upon unloading. As the flat indenter compresses the structure (points 1–9), multiple densified zones nucleate and ultimately merge to a single densified region. This indicates that the induced $5\%$ grading was not strong enough to bias the densification strictly to the top region. In architected materials this is usually a sign of heterogeneity/polydispersity. During unloading (points 10–12), the reverse transformation does not initiate and the material reaches a secondary stable equilibrium state, indicative of the bistable response.

Detailed spatial stress contour plots corresponding to these loading stages are provided in Appendix`\ref{appendix: Cyclic_Loading}, offering further insight into the evolution of internal stresses during this cyclic process.

Overall, these simulations demonstrate that the proposed nonlocal framework, when coupled with contact mechanics, can accurately capture hysteresis, energy dissipation, and recovery patterns in cyclic loading scenarios, as well as recapitulate experimentally observed phase transitions as well as metastable and bistable responses for architected materials. The contrast between metastable and bistable materials highlights the role of energy landscape topology in governing reversibility. To better approximate irreversible behaviors seen in real-world architected materials, we next examine how artificial viscosity can be leveraged to mimic rate-effects at the microstructure level.

\FloatBarrier

\subsection{Approximating microstructural deformation rate effect}

Even though the materials examined here are purely elastic, the extreme deformations that they undergo at the microstructural level to accommodate densification and phase transformations, can lead to rate-dependent dissipation associated with the non-affine microstructural deformation mechanisms. To explore how artificial viscosity can emulate the aforementioned rate-effects in architected materials, we perform comparative simulations for both metastable and bistable systems under cyclic compression. Specifically, we vary the artificial viscosity parameter \( \eta \) to assess its influence on the loading–unloading response and energy dissipation characteristics.

\begin{figure}[htbp]
    \centering
    \includegraphics[width=\linewidth]{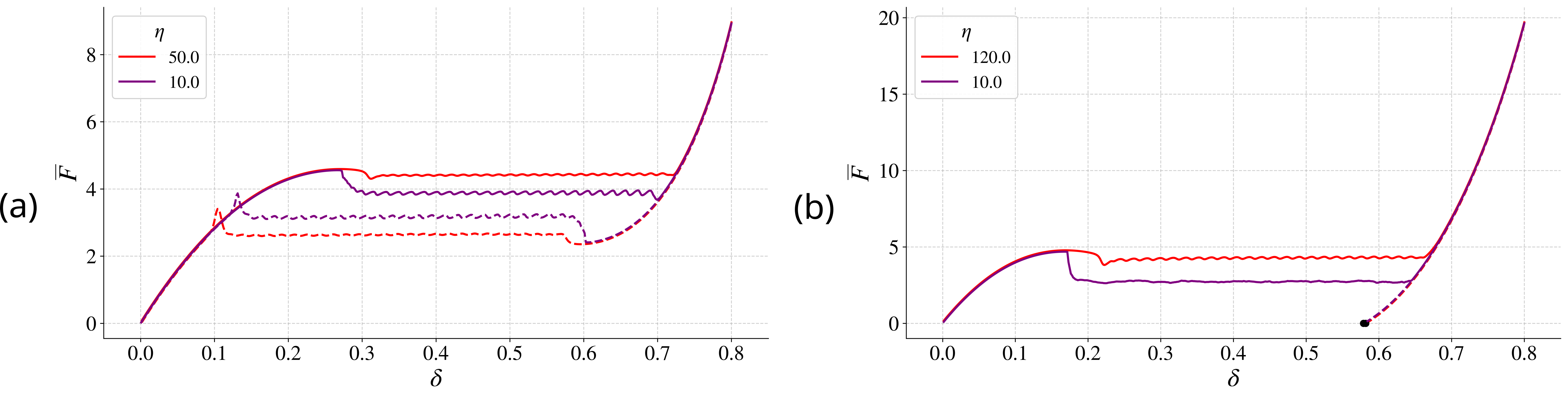}
    \caption{Illustrating the approximation of a viscous microbuckling effect in (a) metastable and (b) bistable materials. Force–displacement curves are shown for increasing values of artificial viscosity \( \eta \). Solid and dashed lines correspond to loading and unloading phases, respectively.}
    \label{fig:viscous_LnU}
\end{figure}

Fig.~\ref{fig:viscous_LnU} presents the force–displacement responses for two representative values of \( \eta \), highlighting distinct hysteresis behavior in both metastable Fig.~\ref{fig:viscous_LnU}(a) and bistable Fig.~\ref{fig:viscous_LnU}(b) systems. The solid curves represent the loading path, while dashed lines denote the unloading phase. For lower viscosity (\( \eta = 10.0 \)), both material systems exhibit sharper softening transitions during loading and faster recovery during unloading, resulting in narrower hysteresis loops. However, as viscosity increases (\( \eta = 50.0 \) and \( \eta = 120.0 \)), the system begins to exhibit a more gradual force decay during loading and significant residual deformation upon unloading—characteristics analogous to what a viscous micro-deformation mechanism would induce. The stable segments of the path (elastic loading and densification), remain essentially unchanged by this modification of artificial viscosity maintaining fidelity to the elastic solution.

The metastable case demonstrates a pronounced plateau and delayed re-stiffening in the high-viscosity regime, while the bistable case features a more abrupt force drop followed by a long plateau. Notably, the increased hysteresis area at higher \( \eta \) reflects greater energy dissipation. Complementary contour plots of the Jacobian field illustrating the spatial evolution of densification fronts with varying $\eta$ are provided in Appendix~\ref{appendix: viscous_micro_buckling}.

These observations suggest that artificial viscosity can be used not only as a regularization mechanism but also as a phenomenological means of emulating microstructural rate-effects. By tuning $\eta$, one can modulate hysteresis and residual strain—offering new pathways to model complex metamaterials at the structural scale, and not limited to small RVEs.

\subsection{Imperfection insensitive structures and auxetic effects}

Metamaterials like foams and lattices show varying degrees of imperfection sensitivity~\cite{BOLINTINEANU2021,long_2020,long_2022,luan2022microscopic}. Our proposed model was shown to effectively capture such effects. On the other hand, there are architected metamaterials that have a more ordered/collective behavior. This is often the case when a well designed mechanism for the microstructural deformation is at play governed by response-type and not limit load-type behavior~\cite{bertoldi2017flexible}. To investigate the emergence of imperfection-insensitive mechanical response and auxetic behavior, we take inspiration from the use of gradient-enhanced damage models for fracture, and specifically towards capturing imperfection sensitivity. In those cases tuning the nonlocal length scale $l$ can effectively filter out defects/imperfection with a characteristic length below that length scale\cite{mousavi2025capturing}. 

We simulate the compression of a structured metastable system subjected to uniform vertical loading through a frictionless flat punch. In this configuration, no geometric or material imperfections are introduced, allowing us to isolate the intrinsic response of the architecture. A key distinction in this case is the choice of the nonlocal length scale, which is set equal to the full height of the domain $l=H$. This large interaction length promotes filtering imperfections induced by numerical error and penalizing large gradients of the nonlocal volume ratio. This effectively suppresses the sensitivity to localized heterogeneities or numerical noise, inducing an order collective response insensitive to imperfections.

\begin{figure}[H]
    \centering
    \includegraphics[width=\linewidth]{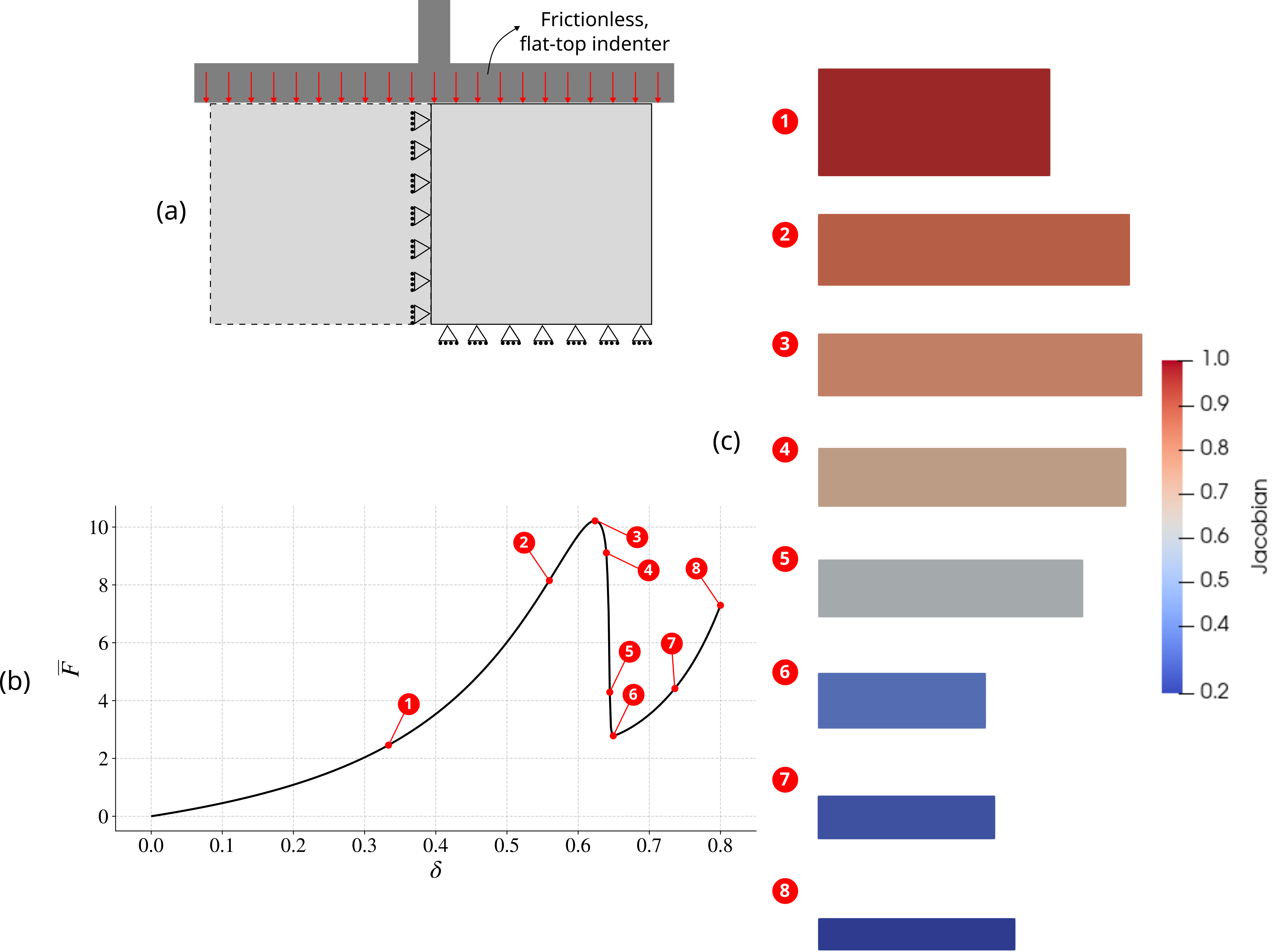}
    \caption{Auxetic response in a metastable architected structure under uniform compression, with its schematic illustrated in (a). Force–displacement curve, showcased in (b), is annotated with numbered markers corresponding to snapshots, presented in (c), showing the evolution of the Jacobian field \( J \). The system uses a large nonlocal length scale equal to the domain height, promoting imperfection-insensitive and coordinated deformation.}
    \label{fig:auxetic_response}
\end{figure}

Fig.~\ref{fig:auxetic_response} illustrates the normalized force–displacement response alongside eight deformation snapshots that capture the evolution of the Jacobian determinant \( J \). The structure exhibits a smooth, progressive densification pattern without the emergence of a phase transformation front. The force is increasing up to a point of an instability accommodating an auxetic effect.
The Poisson effects through this unconfined compression experiment are non-monotonic. Initially, lateral expansion is observed during stable elastic loading, followed by a transition into an auxetic deformation mode where lateral contraction is observed, induced by the instability and collective phase transformation . 

This study underscores the ability of long nonlocal interaction lengths to suppress imperfection-driven instabilities and facilitate coordinated, auxetic deformation modes. The resulting behavior is robust, and less sensitive to geometric or material noise, making it well-suited for adaptive or morphing applications. These insights round out the comprehensive evaluation of the framework, illustrating its versatility in capturing both localized and global deformation mechanisms in architected materials.

\section{Conclusion}\label{Section:Conclusion}

In this study, we proposed and validated a nonlocal finite element framework tailored to simulate the complex loading response of elastic architected metamaterials. The framework was specialized to capture metastable and bistable behaviors associated with volumetric phase transformations between a densified and a rarefied phase. The formulation introduces a nonlocal volume ratio and incorporates gradient-based regularization through penalization of the gradients of the nonlocal volume ratio, enabling the resolution of sharp internal transitions such as nucleation of densified phase and propagation of densified front under compressive loading.
Through a comprehensive series of numerical experiments, we demonstrated the robustness, accuracy, and adaptability of the framework across a range of material parameters and mesh topologies. Artificial viscosity was utilized as a tunable parameter to emulate rate-dependent micro-deformation phenomena that control dissipation. The model was shown to recapitulate phase transformations that emerge in architected materials following the macroscopic response and appropriate signatures of the microstructural response, while being able to capture sharp phase transformation fronts and hysteresis loops for metastable and bistable models.

By introducing heterogeneity of the materials' properties, we showcased how grading of material properties can guide densification. 
Finally, simulations with a large nonlocal length scale relative to the domain size, demonstrated the emergence of globally coordinated deformation and auxetic response. Notably, this framework is the first to introduce a structured and scalable phenomenological construction to capture macroscopic and microscopic behavior for architected metamaterials. Once an appropriate energy is constructed and relevant nonlocal variables and length scales are identified, the model can perform predictive calculation in a structural scale without having to resort to discrete simulations. Additionally, it overcomes barriers encountered by local homogenization methods towards capturing the emergence of localized phase transformation fronts opening the door for the development of new theories and models. 

Overall, this work establishes a versatile computational foundation for capturing a broad spectrum of mechanical phenomena in architected solids and metamaterials, particularly under loading regimes where phase transitions, localization, and nonlocal interactions play a central role. These insights lay the foundation for future efforts in incorporating material heterogeneity, dynamic as well as inelastic effects into advanced continuum models for architected materials. Additionally, incorporating machine learning approaches in the transitions from discrete RVE or experimental data to the identification of nonlocal continuum models for accelerating and automating the model development process is a task that we are planning to undertake in future studies.

\section{Acknowledgements}\label{Section:Conclusion}
S.J., C.M.H., R.A. and N.B. were supported by the Laboratory Directed Research and Development program at Sandia National Laboratories, a multimission laboratory managed and operated by National Technology and Engineering Solutions of Sandia, LLC, a wholly owned subsidiary of Honeywell International, Inc., for the U.S. Department of Energy's National Nuclear Security Administration under contract DE-NA-0003525. 
This paper describes objective technical results and analysis. Any subjective views or opinions that might be expressed in the paper do not necessarily represent the views of the U.S. Department of Energy or the United States Government. P.K.P acknowledges support from the University Research Foundation at the University of Pennsylvania.

\newpage

\appendix
\gdef\thesection{\Alph{section}}
\makeatletter
\renewcommand\@seccntformat[1]{Appendix \csname the#1\endcsname.\hspace{0.5em}}
\makeatother

\section{Parametric Study of Energy Landscape under Hydrostatic Deformation} \label{appendix: Parametric_Analytical_Plots}

This appendix investigates how the constitutive parameters \( \alpha \), \( \beta \), and \( \kappa \) influence the volumetric energy function \( \Psi(J) \) introduced in Section~2.2, and the corresponding mechanical response. These parametric plots provide insight into the qualitative regimes explored in the simulations, including transitions between metastable and bistable behavior.

\begin{figure}[h!]
    \centering
    \includegraphics[width=0.95\linewidth]{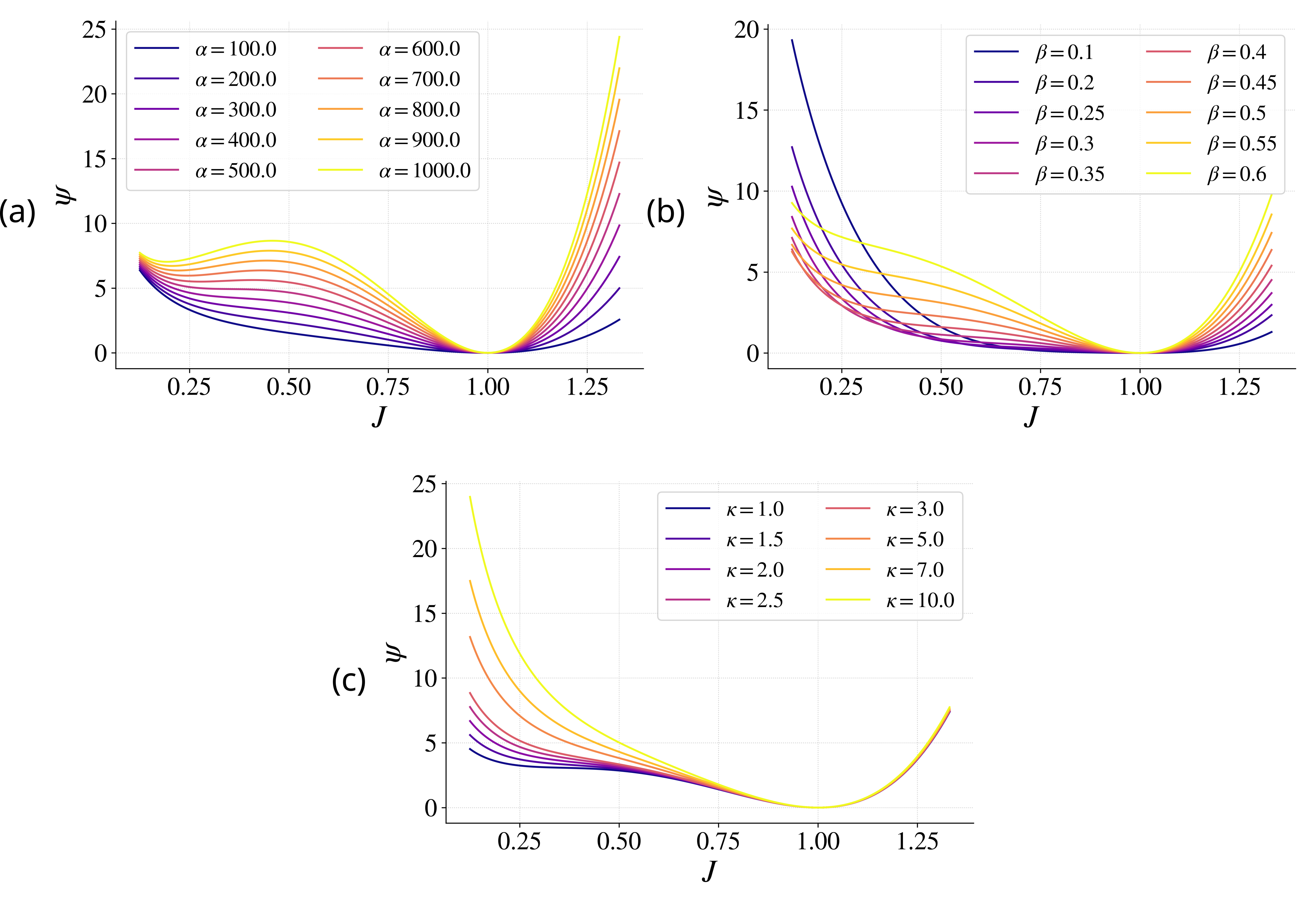}
    \caption{Parametric variation of the volumetric energy density \( \Psi(J) \) for different values of (a) \( \alpha \), (b) \( \beta \), and (c) \( \kappa \).}
    \label{fig:Parametric_Psi_vs_J}
\end{figure}

Fig.~\ref{fig:Parametric_Psi_vs_J} shows how changes in each parameter affect the shape and convexity of the energy landscape. Increasing \( \alpha \) deepens the non-convex well and promotes bistability, while higher \( \beta \) values sharpen the well and skew its shape. The bulk modulus \( \kappa \), meanwhile, acts as a stiffness modulator—larger \( \kappa \) values lead to stronger energy penalties for volumetric changes, particularly near \( J = 1 \).

\begin{figure}[h!]
    \centering
    \includegraphics[width=0.95\linewidth]{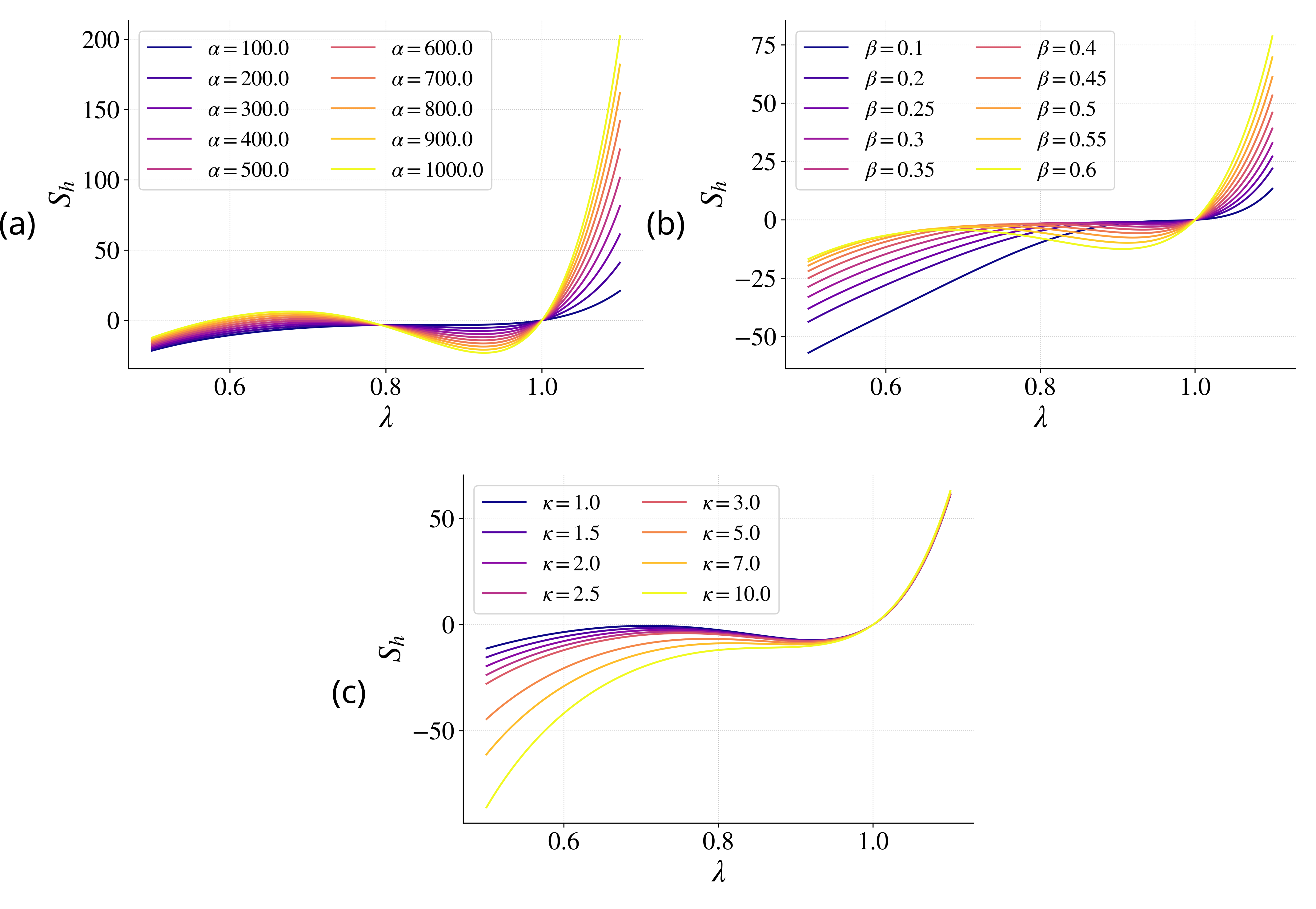}
    \caption{Stress–stretch behavior derived from the volumetric free energy, showing the effect of (a) \( \alpha \), (b) \( \beta \), and (c) \( \kappa \) on hydrostatic second Piola–Kirchhoff stress, \( S_{h} \) as a function of stretch \( \lambda \).}
    \label{fig:Parametric_S_vs_lambda}
\end{figure}

Fig.~\ref{fig:Parametric_S_vs_lambda} illustrates the resulting second Piola–Kirchhoff stress under uniaxial deformation. In Fig.\ref{fig:Parametric_S_vs_lambda}(a), increasing \( \alpha \) induces pronounced softening followed by strong re-stiffening, indicating bistable behavior. In Fig.\ref{fig:Parametric_S_vs_lambda}(b), varying \( \beta \) alters the symmetry and smoothness of the stress transition. Fig.\ref{fig:Parametric_S_vs_lambda}(c) shows that larger \( \kappa \) enhances compressive stiffness, consistent with a greater resistance to volumetric collapse.

\begin{figure}[h!]
    \centering
    \includegraphics[width=0.95\linewidth]{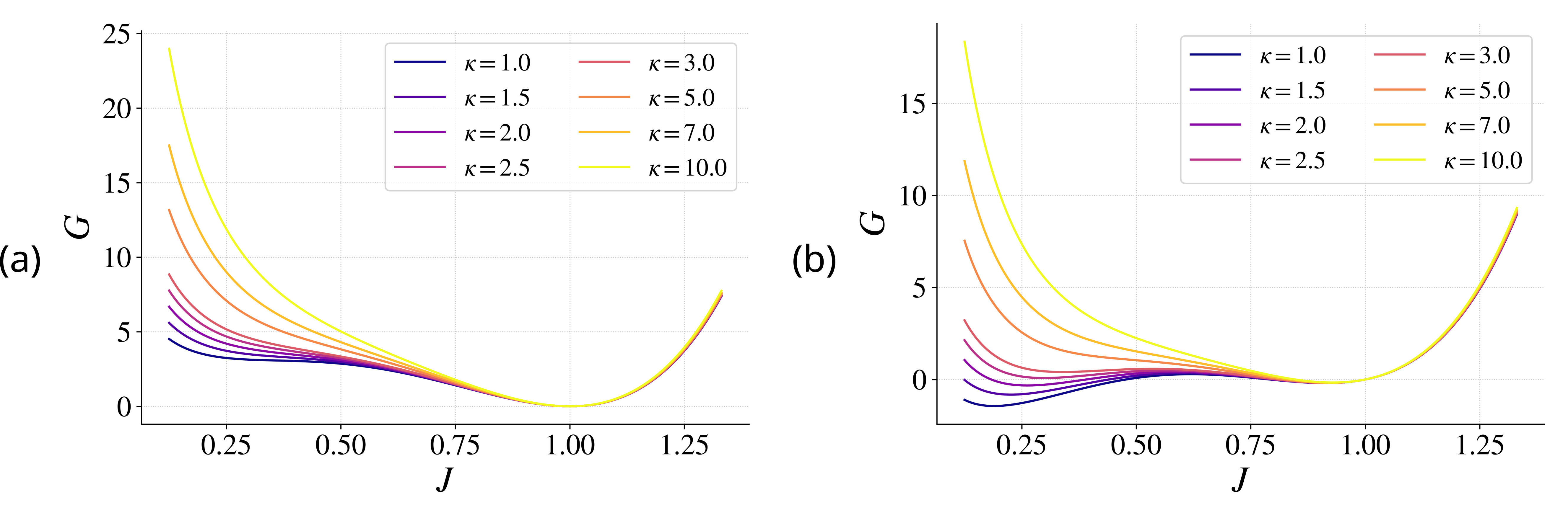}
    \caption{Gibbs free energy \( G(J) \) plotted for various \( \kappa \) values under (a) zero load (\( S_{h} = 0.0 \)) and (b) applied compressive load (\( S_{h} = -5.0 \)).}
    \label{fig:Combined_Kappa_Sweep_G_vs_J}
\end{figure}

Finally, Fig.~\ref{fig:Combined_Kappa_Sweep_G_vs_J} shows the Gibbs free energy \( G(J) \) for different \( \kappa \) values under two loading conditions. In the absence of load, the minima coincide near \( J = 1 \), but as compressive stress is applied, the energy well shifts and deepens asymmetrically. The variation of \( \kappa \) strongly affects the depth and curvature of these minima, directly modulating the system's compressibility and stability under external loading.
These plots clarify how material parameters influence energy landscapes and mechanical response, forming the simulations discussed in the main text.


\section{Stress Contours for Cyclic Loading} \label{appendix: Cyclic_Loading}

To complement the force–displacement and Jacobian field observations presented in the main text (Fig.~\ref{fig:bistable_cyclic}), we now provide detailed snapshots of the stress components during cyclic loading for the bistable material system. These plots help visualize the spatial evolution of internal stress fields and provide further insight into the irreversibility and path-dependence associated with bistable responses.

\begin{figure}
    \centering
    \includegraphics[width=\linewidth]{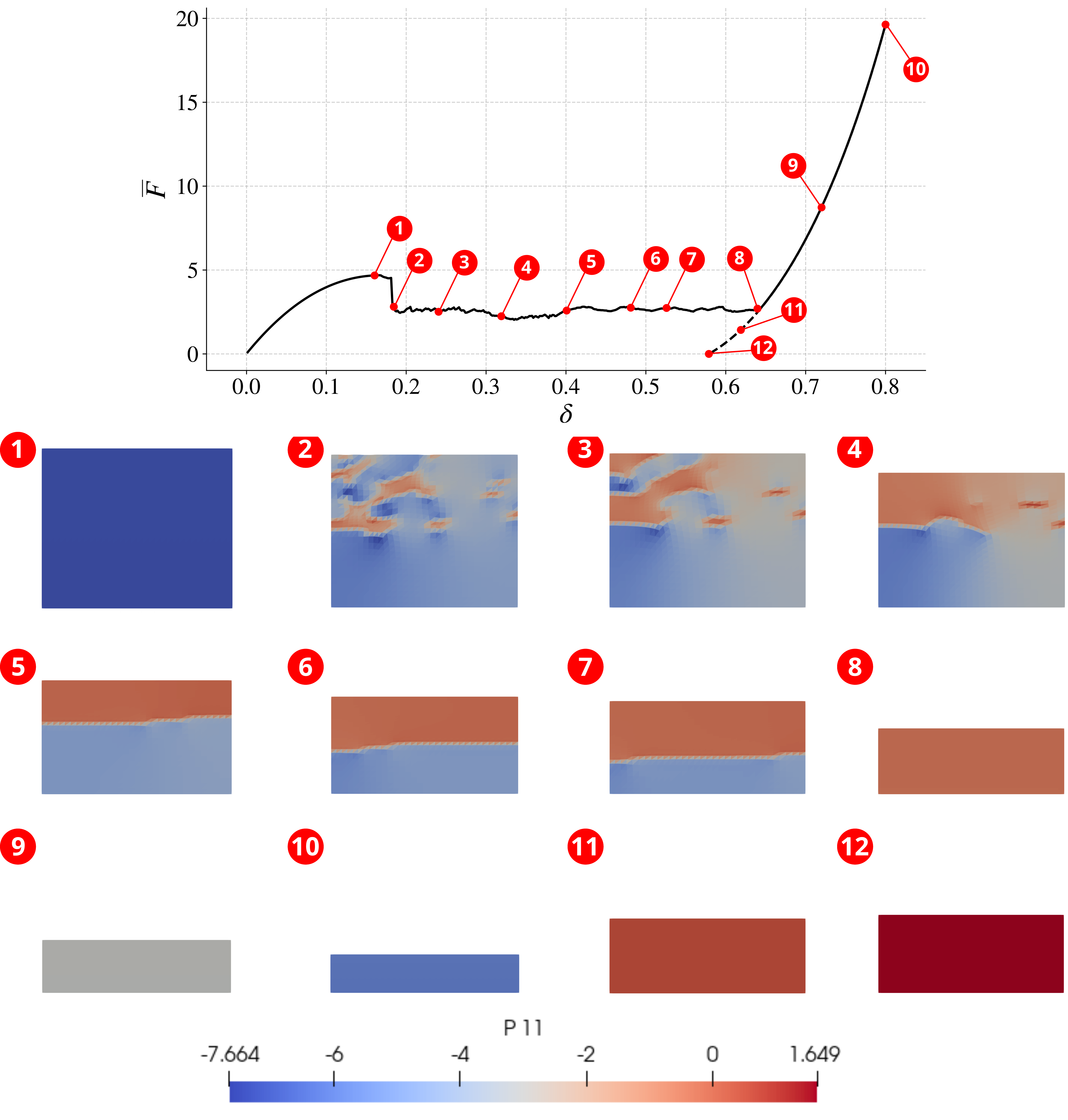}
    \caption{Evolution of the first Piola-Kirchhoff stress component \( P_{11} \) during cyclic loading for a bistable material. Snapshots correspond to the force–displacement stages marked in Figure~\ref{fig:bistable_cyclic}.}
    \label{fig:P11_Bistable}
\end{figure}

Fig.~\ref{fig:P11_Bistable} shows the progression of the first Piola-Kirchhoff stress component \( P_{11} \). Initially, the domain remains nearly stress-free, but as loading begins, compressive stresses concentrate near the top corners, aligning with the nucleation of densified zones. As the deformation front propagates downward, a band of highly compressive stress forms and becomes more uniform across the domain. During unloading, partial relaxation occurs; however, residual stress remains in \( P_{11} \) direction. Ideally, a fully relaxed state would correspond to hydrostatic compression at a new stretch level, lateral constraints prevent free contraction, inducing non-zero transverse tension and locking in residual stress-reflecting the system's bistable energy landscape.

\begin{figure}
    \centering
    \includegraphics[width=\linewidth]{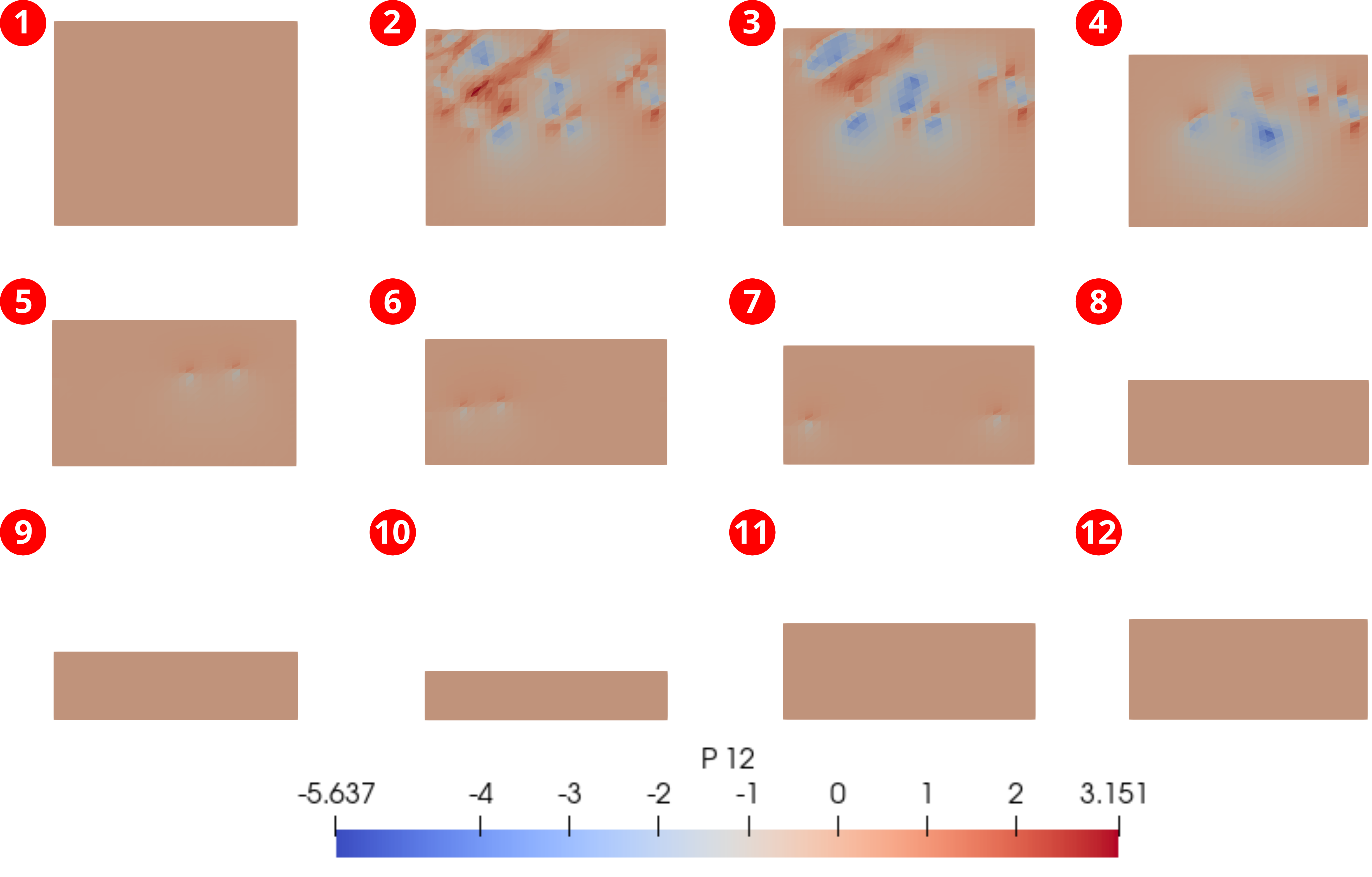}
    \caption{Contours of the shear stress component \( P_{12} \) during the same cyclic process. The asymmetric fields reflect internal shear reorganization during the loading and unloading paths.}
    \label{fig:P12_Bistable}
\end{figure}

The evolution of the shear stress component \( P_{12} \), depicted in Fig.~\ref{fig:P12_Bistable}, reveals localized high shear stress regions that arise during the loading phase. These stress heterogeneities are particularly visible at the interfaces of densified regions, indicating internal reorganization and reorientation of material microstructure. While the shear intensity diminishes during unloading, residual shear stresses remain in several parts of the domain, further evidencing path-dependence and the locking of deformation patterns.

\begin{figure}
    \centering
    \includegraphics[width=\linewidth]{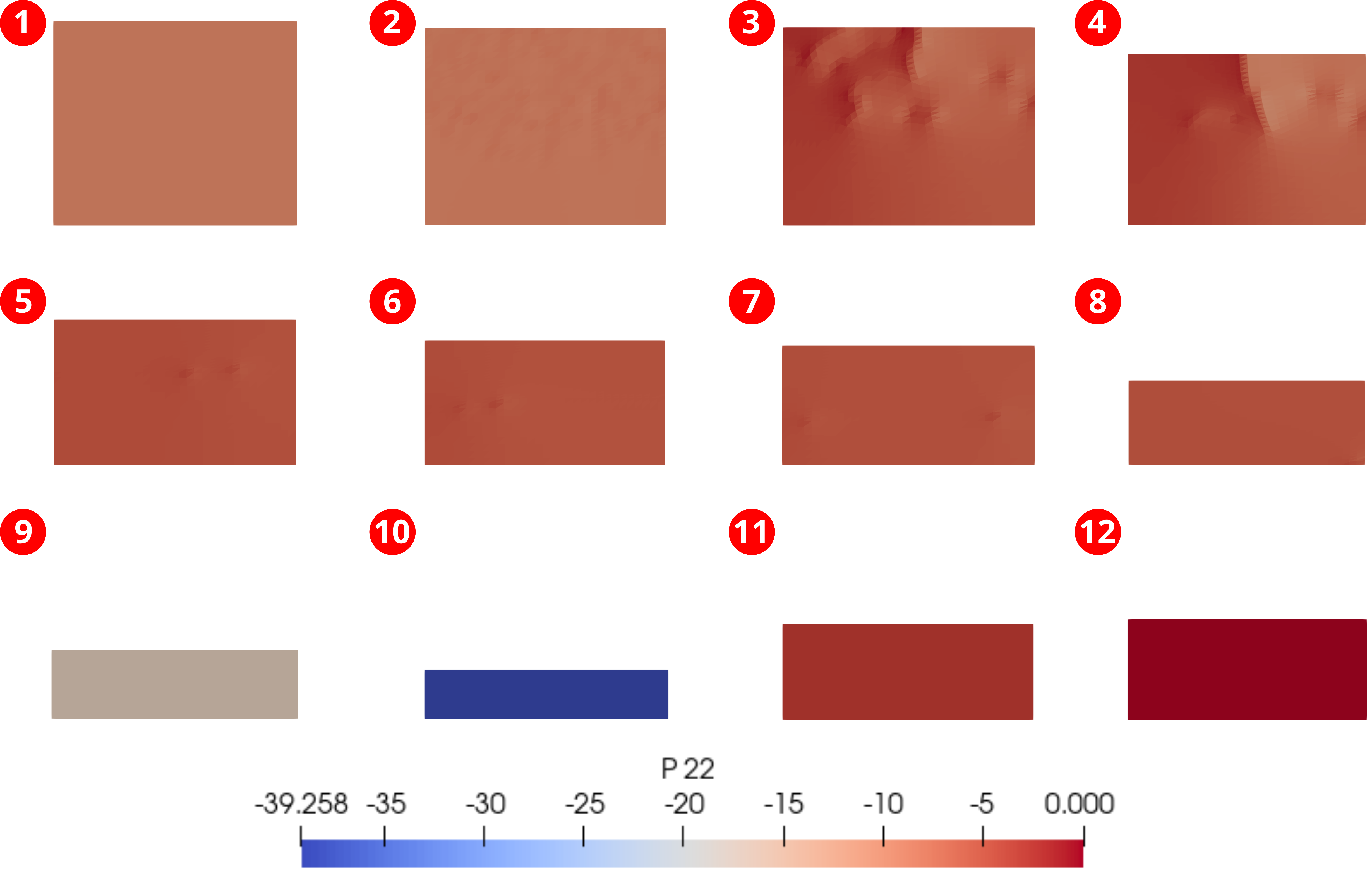}
    \caption{Contours of the first Piola-Kirchhoff stress component \( P_{22} \) during cyclic loading. Significant compressive stress builds up during loading and only partially relaxes during unloading.}
    \label{fig:P22_Bistable}
\end{figure}

Finally, Fig.~\ref{fig:P22_Bistable} illustrates the first Piola-Kirchhoff stress component \( P_{22} \), which dominates the deformation response due to the applied loading. Peak compressive stresses occur near mid-cycle at maximum load, while unloading leads to a fully stress-free state in \( P_{22} \) direction within the densified phase. 

Taken together, these stress contour plots provide a detailed view into the spatial evolution of internal forces during cyclic loading. They confirm the persistence of irreversible mechanical states post-unloading and underscore the mechanical memory inherent in bistable architected materials.

\newpage

\section{Volume ratio contour for increased artificial viscosity} \label{appendix: viscous_micro_buckling}

This section presents the evolution of the volume ratio fields over selected loading and unloading points along the force–displacement curve for both metastable and bistable systems under displacement-controlled cyclic compression. The objective is to examine how increasing the artificial viscosity parameter \( \eta \) influences the nature of compressive collapse and recovery, which is often interpreted as a surrogate for viscous micro-buckling behavior in architected materials.

\begin{figure}
    \centering
    \includegraphics[width=\linewidth]{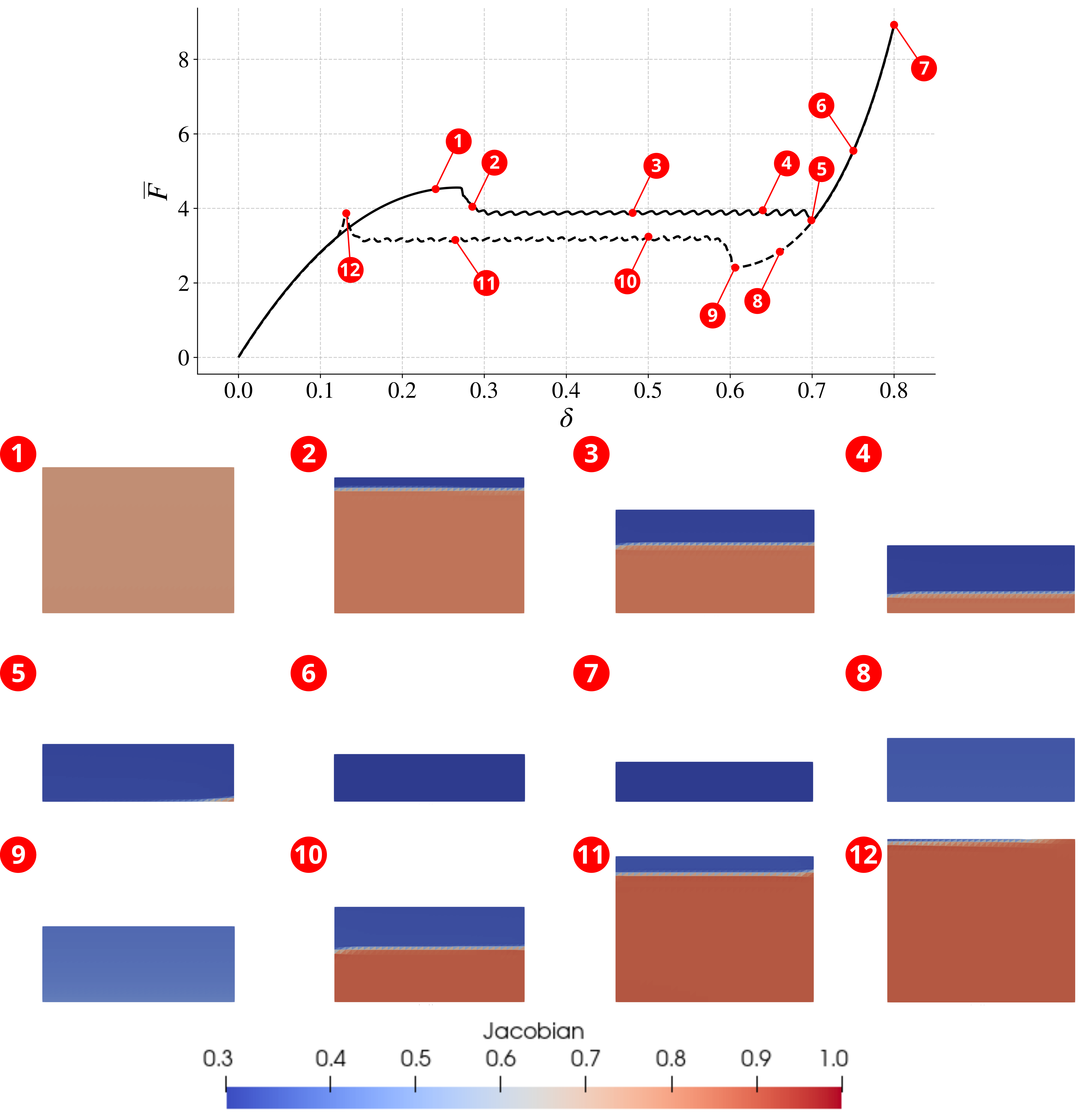}
    \caption{Jacobian field for metastable model with artificial viscosity \( \eta = 10.0 \).}
    \label{fig:LnU_Metastable_eta_10.0}
\end{figure}

Fig.~\ref{fig:LnU_Metastable_eta_10.0} illustrates the Jacobian field evolution for the metastable case with a relatively low viscosity value \( \eta = 10.0 \). The localized transition from undeformed to densified states is observed along a diffuse front, with a gradual volumetric collapse emerging over multiple time steps. The asymmetry between loading and unloading paths is captured by the persistence of densified zones, even after force removal.

\begin{figure}
    \centering
    \includegraphics[width=\linewidth]{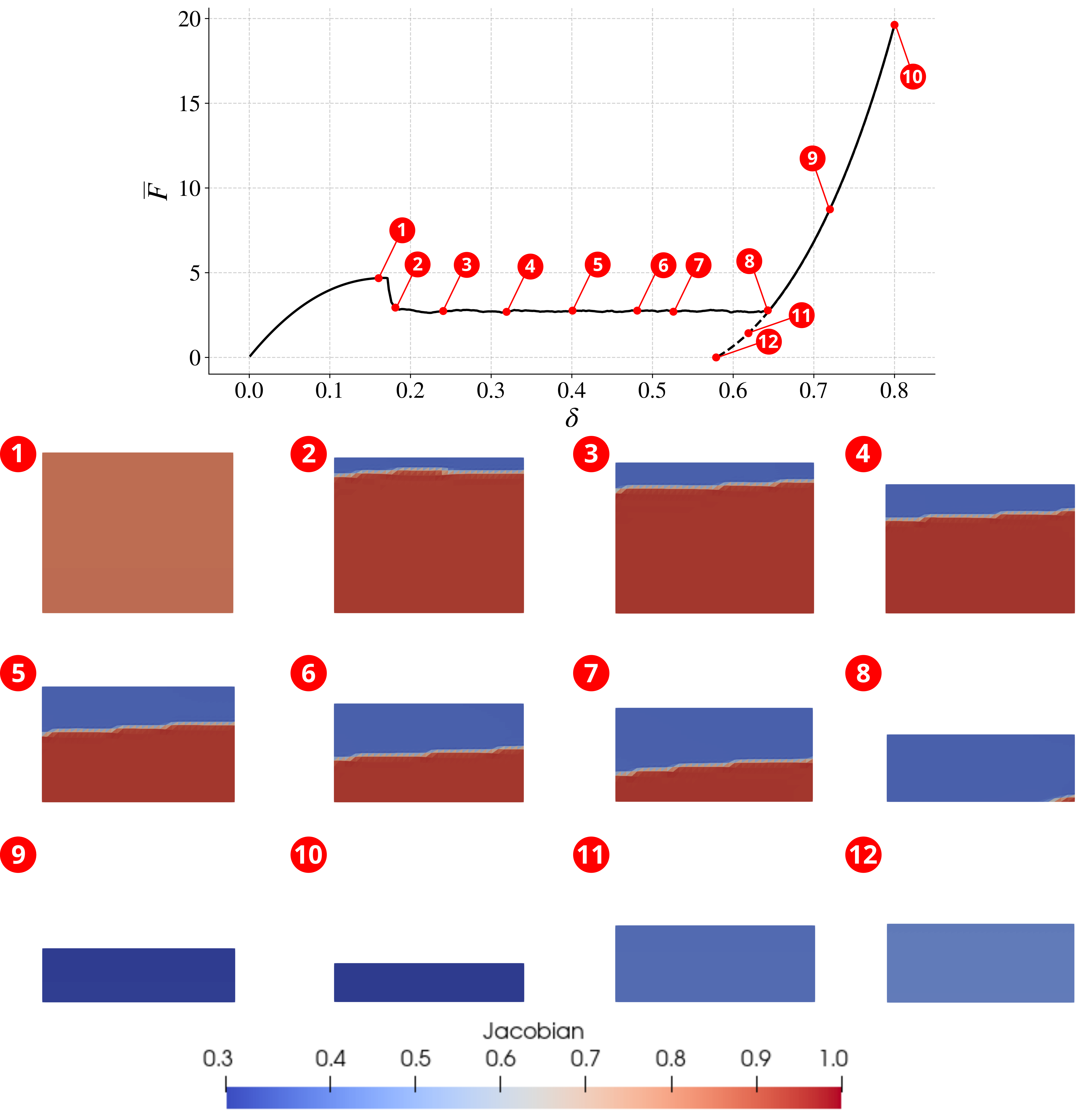}
    \caption{Jacobian field for bistable model with artificial viscosity \( \eta = 10.0 \).}
    \label{fig:LnU_Bistable_eta_10.0}
\end{figure}

The bistable counterpart, shown in Figure~\ref{fig:LnU_Bistable_eta_10.0}, displays a similar pattern of densified front propagation. However, due to the nature of the energy landscape, the post-collapse Jacobian values remain suppressed even during unloading, suggesting reduced recoverability. The collapse appears more abrupt, consistent with bistable phase transitions.

\begin{figure}
    \centering
    \includegraphics[width=\linewidth]{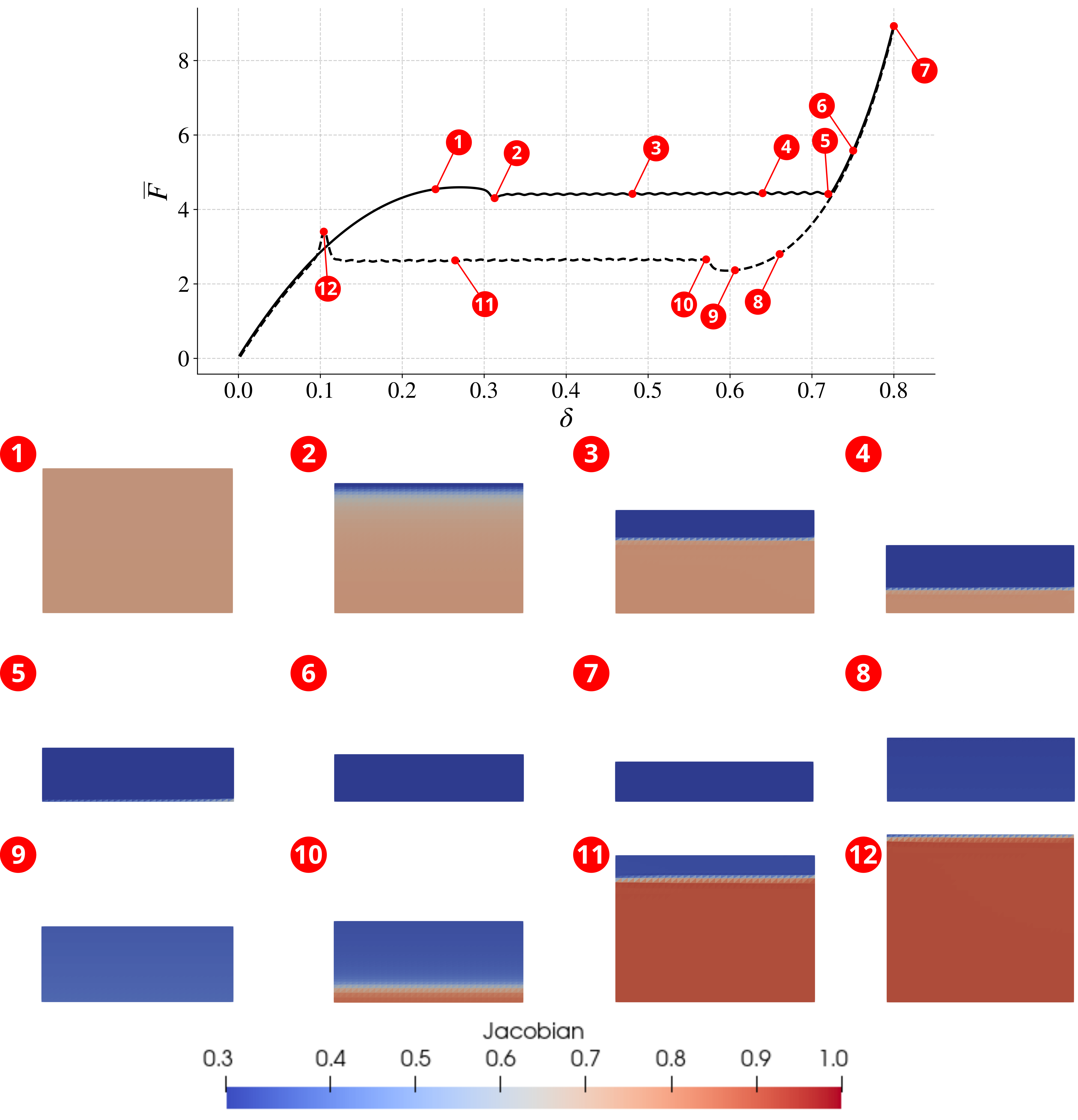}
    \caption{Jacobian field for metastable model with high artificial viscosity \( \eta = 50.0 \).}
    \label{fig:LnU_Metastable_eta_50.0}
\end{figure}

Increasing the artificial viscosity to \( \eta = 50.0 \) in the metastable configuration (Fig.~\ref{fig:LnU_Metastable_eta_50.0}) results in a more coherent and uniform progression of densification. The sharpness of front propagation is reduced, and volumetric deformation evolves more smoothly. Notably, partial recovery during unloading is observed, in contrast to lower \( \eta \) cases, reflecting the stabilizing role of viscosity.

\begin{figure}
    \centering
    \includegraphics[width=\linewidth]{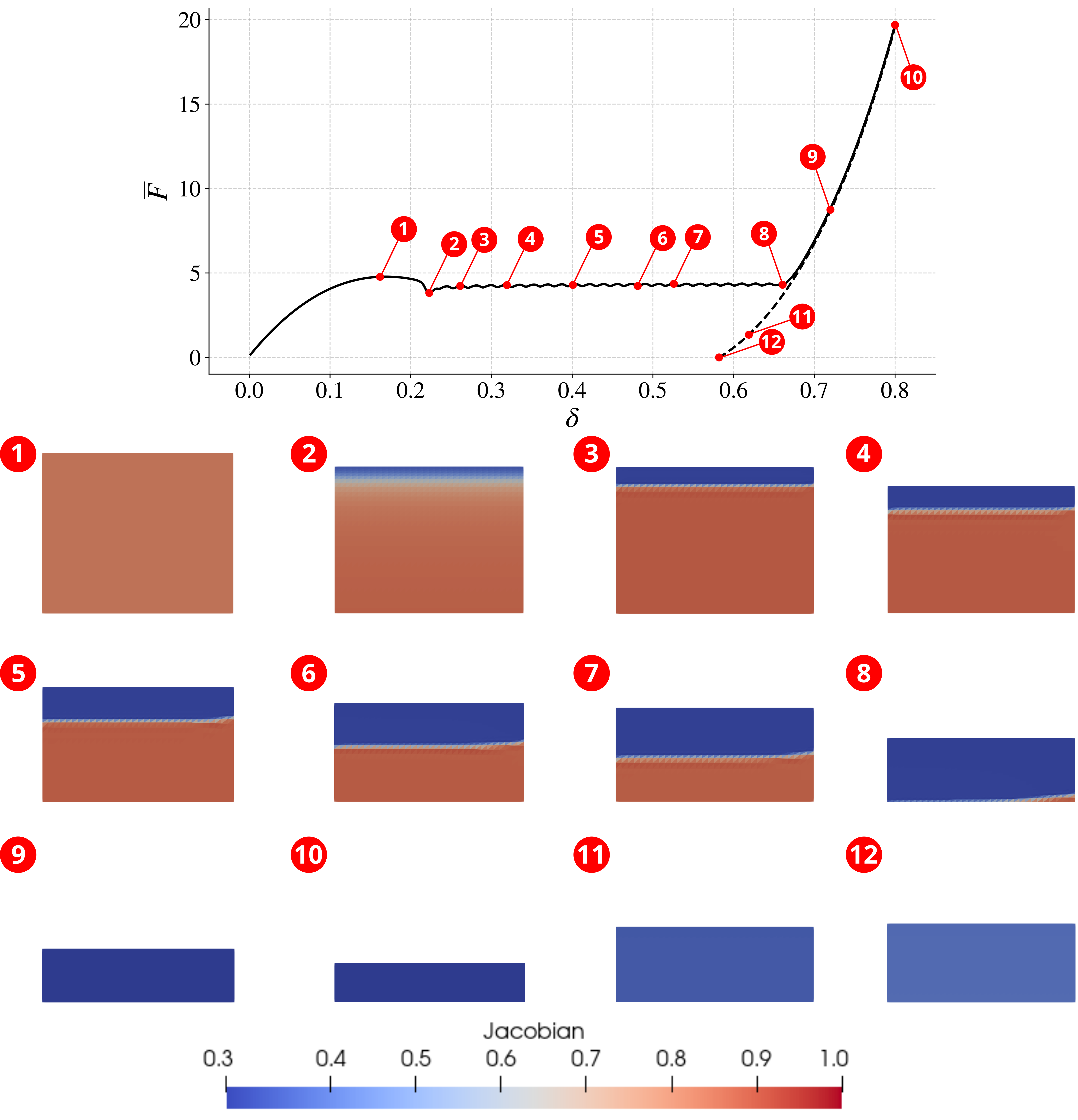}
    \caption{Jacobian field for bistable model with high artificial viscosity \( \eta = 120.0 \).}
    \label{fig:LnU_Bistable_eta_120.0}
\end{figure}

Finally, Fig.~\ref{fig:LnU_Bistable_eta_120.0} presents the high-viscosity bistable case (\( \eta = 120.0 \)), where the material response becomes nearly homogeneous post-collapse, with almost no recovery upon unloading. The high dissipation smooths out local instabilities, leading to uniform deformation zones with persistently low Jacobian values, effectively suppressing front propagation and indicating irreversible collapse.

Taken together, these results highlight the influence of artificial viscosity within this nonlocal model as a tuning parameter to regularize numerical instabilities and approximate physical behavior. High \( \eta \) suppresses abrupt transitions and promotes smoother, more dissipative responses.

\newpage

\section{Effect of Internal Length Scale and Mesh Resolution} \label{section:length_scale_study}

This section investigates the combined influence of internal length scale and mesh resolution on the deformation behavior of architected metastable materials subjected to compression. These materials are modeled using a non-(poly)convex energy landscape augmented by gradient regularization, which introduces a characteristic length scale \( l \) that governs the spatial extent over which deformation is smoothed. We aim to understand how this internal length influences the mechanical response—especially the sharpness and propagation of phase fronts—and how sensitive the predictions are to the underlying discretization of the domain.

The reference internal length \( l_0/H = 0.017 \) is defined based on the average element size of the baseline mesh (32 elements along the height), and all length scales are reported relative to the domain height (normalized to unity). A series of simulations were performed while systematically varying \( l \), with all other parameters held constant: viscosity \( \eta = 5.0 \), a 5\% vertical gradient in bulk modulus \( \kappa \), and a structured finite element mesh.

\begin{figure}[h!]
    \centering
    \includegraphics[width=0.9\linewidth]{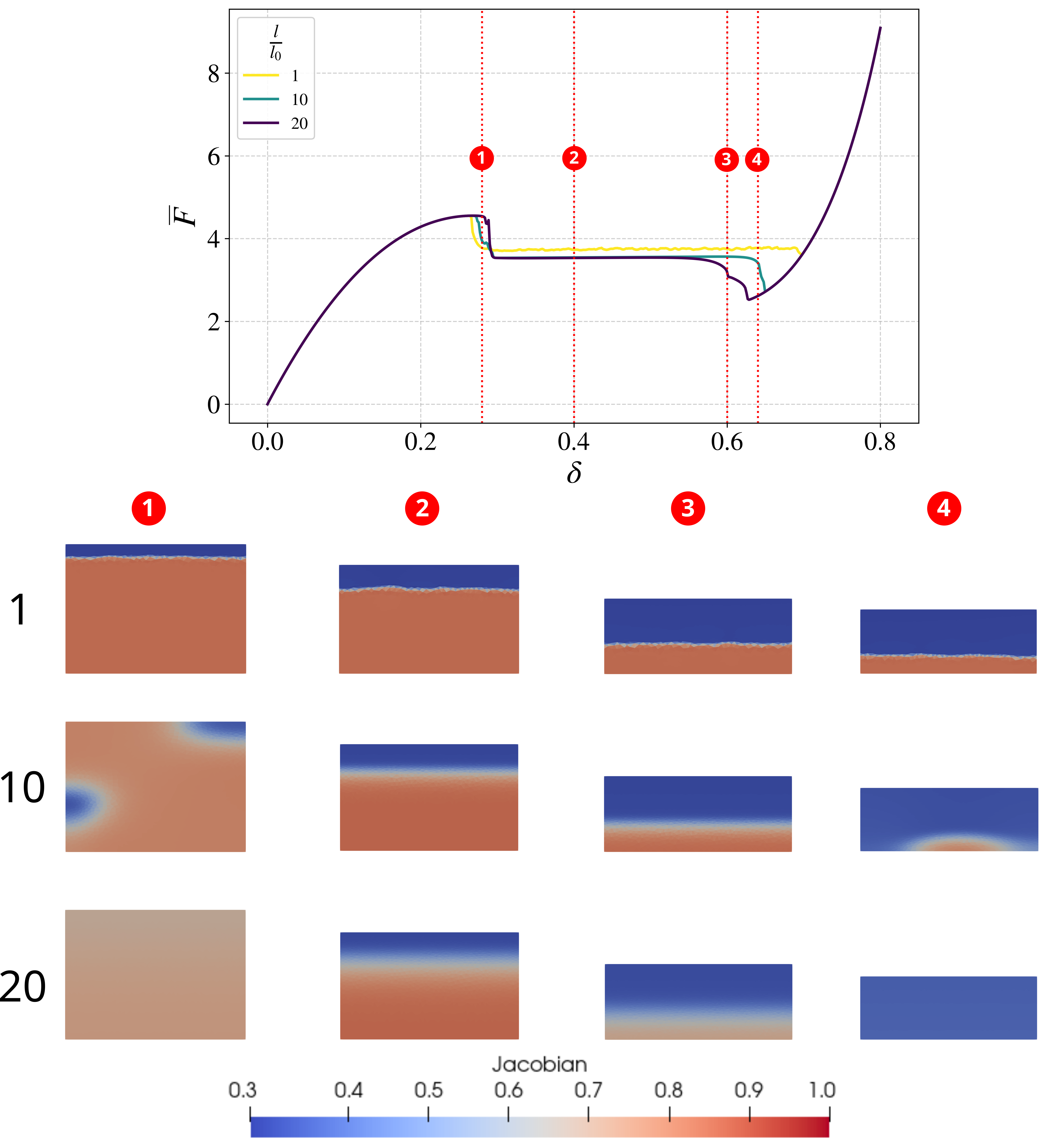}
    \caption{Influence of normalized internal length scale \( l/l_0 \in \{1, 10, 20\} \) on the normalized force–indentation response \( \bar{F}(\delta) \). The top plot shows the macroscopic force–displacement curves, while the subpanels illustrate spatial Jacobian fields \( J \) at indentation depths labeled (1)--(4). Increasing \( l \) broadens the deformation front and suppresses localized instabilities.}
    \label{fig:length_scale_variation}
\end{figure}

As shown in Fig.~\ref{fig:length_scale_variation}, the internal length scale has a significant effect on the observed force–displacement response and the morphology of deformation fronts. At a small internal length (\( l/l_0 = 1 \)), the system exhibits a pronounced force drop at \( \delta \approx 0.4 \), corresponding to a highly localized and abrupt densification event—a signature of an underlying mechanical instability. This sharp transition arises because the deformation is concentrated within a narrow band, leading to steep gradients and sudden energy release.

In contrast, for larger internal lengths (\( l/l_0 = 10, 20 \)), the response becomes significantly smoother. The onset of instability is delayed, and the force drop is more gradual, indicating a more diffuse and stable transition. This behavior is directly attributed to the role of the gradient term, which penalizes steep gradients in the deformation field and thus spreads the deformation over a wider region. The Jacobian contours shown below the force curves further illustrate this effect: for \( l/l_0 = 1 \), the transition front is sharp and discontinuous, whereas for \( l/l_0 = 20 \), the front spans a large portion of the domain and evolves smoothly with increasing indentation.

Physically, this demonstrates that the internal length scale \( l \) acts as a regularization parameter, controlling the width and smoothness of evolving phase boundaries. Larger values of \( l \) promote gradual transformations by diffusing localized strain, which in turn mitigates stress concentrations and suppresses snap-through-like behavior. This makes the model particularly well-suited for capturing realistic phase-front dynamics in soft architected systems with intrinsic microstructural length scales.

\begin{figure}[h!]
    \centering
    \includegraphics[width=\linewidth]{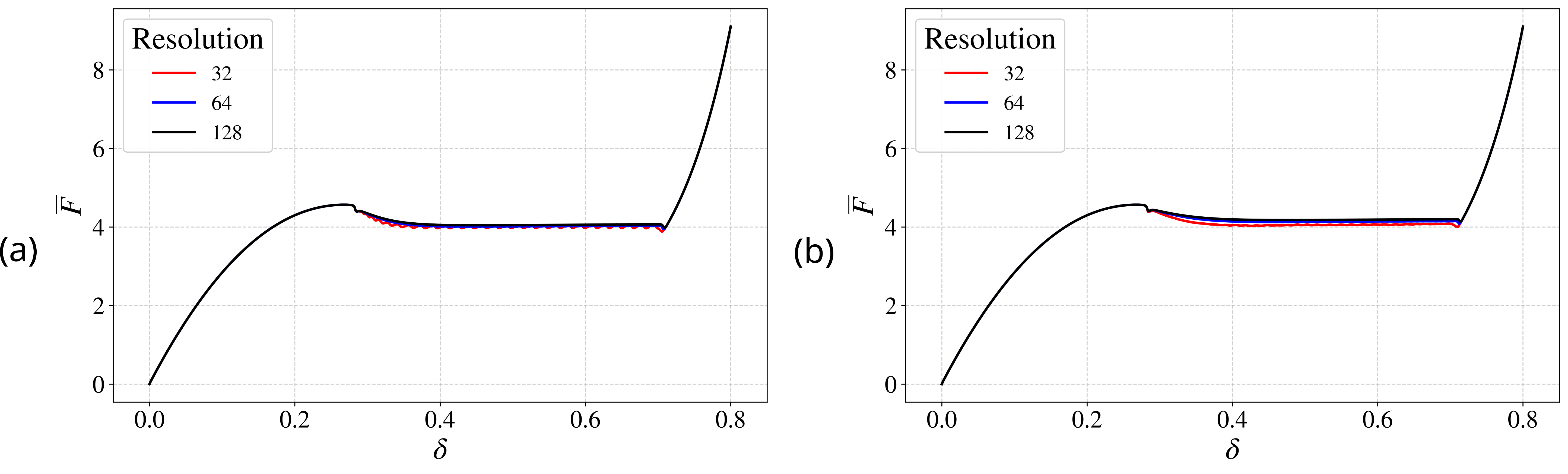}
    \caption{Mesh convergence study for the normalized force–indentation response \( \bar{F}(\delta) \) using (a) structured and (b) unstructured meshes with 32, 64, and 128 elements in the vertical direction. Simulations use fixed internal length \( l \), viscosity \( \eta = 20.0 \), and 5\% vertical variation in \( \kappa \). The excellent agreement between curves demonstrates mesh independence and robustness of the gradient-regularized formulation.}
    \label{fig:length_scale_mesh_convergence}
\end{figure}

To assess the numerical robustness of the model, a convergence study was conducted using both structured and unstructured meshes of increasing resolution. As shown in Fig.~\ref{fig:length_scale_mesh_convergence}, the normalized force–indentation curves for 32, 64, and 128 elements per height exhibit excellent agreement across both mesh types. This agreement holds not only in the elastic loading and densification, but also across the transition to the plateau stage, where instabilities would typically be most sensitive to discretization.

The tight collapse of the curves—particularly in the post-instability regime—confirms that the gradient regularization endows the model with strong mesh independence. Unlike local formulations that often suffer from mesh-induced localization or spurious front pinning, the inclusion of the internal length scale allows the model to converge reliably with relatively coarse discretizations. Furthermore, the mirrored deformation fields between structured and unstructured configurations validate the symmetry and reproducibility of the solution under ideal loading, suggesting that the formulation is robust with respect to both element arrangement and numerical perturbations.


\newpage
\FloatBarrier


\begin{thebibliography}{10}
\expandafter\ifx\csname url\endcsname\relax
  \def\url#1{\texttt{#1}}\fi
\expandafter\ifx\csname urlprefix\endcsname\relax\def\urlprefix{URL }\fi
\expandafter\ifx\csname href\endcsname\relax
  \def\href#1#2{#2} \def\path#1{#1}\fi

\bibitem{yang_review_2004}
W.~Yang, Z.-M. Li, W.~Shi, B.-H. Xie, M.-B. Yang, \href{https://link.springer.com/10.1023/B:JMSC.0000026928.93231.e0}{Review on auxetic materials}, Journal of Materials Science 39~(10) (2004) 3269--3279, publisher: Springer Science and Business Media {LLC}.
\newblock \href {https://doi.org/10.1023/b:jmsc.0000026928.93231.e0} {\path{doi:10.1023/b:jmsc.0000026928.93231.e0}}.
\newline\urlprefix\url{https://link.springer.com/10.1023/B:JMSC.0000026928.93231.e0}

\bibitem{faure_design_2017}
A.~Faure, G.~Michailidis, G.~Parry, N.~Vermaak, R.~Estevez, \href{http://link.springer.com/10.1007/s00158-017-1688-2}{Design of thermoelastic multi-material structures with graded interfaces using topology optimization}, Structural and Multidisciplinary Optimization 56~(4) (2017) 823--837, publisher: Springer Science and Business Media {LLC}.
\newblock \href {https://doi.org/10.1007/s00158-017-1688-2} {\path{doi:10.1007/s00158-017-1688-2}}.
\newline\urlprefix\url{http://link.springer.com/10.1007/s00158-017-1688-2}

\bibitem{kaminakis_design_2015}
N.~T. Kaminakis, G.~A. Drosopoulos, G.~E. Stavroulakis, \href{http://link.springer.com/10.1007/s00419-014-0970-7}{Design and verification of auxetic microstructures using topology optimization and homogenization}, Archive of Applied Mechanics 85~(9) (2015) 1289--1306, publisher: Springer Science and Business Media {LLC}.
\newblock \href {https://doi.org/10.1007/s00419-014-0970-7} {\path{doi:10.1007/s00419-014-0970-7}}.
\newline\urlprefix\url{http://link.springer.com/10.1007/s00419-014-0970-7}

\bibitem{bertoldi_negative_2010}
K.~Bertoldi, P.~M. Reis, S.~Willshaw, T.~Mullin, \href{https://onlinelibrary.wiley.com/doi/10.1002/adma.200901956}{Negative poisson's ratio behavior induced by an elastic instability}, Advanced Materials 22~(3) (2010) 361--366, publisher: Wiley.
\newblock \href {https://doi.org/10.1002/adma.200901956} {\path{doi:10.1002/adma.200901956}}.
\newline\urlprefix\url{https://onlinelibrary.wiley.com/doi/10.1002/adma.200901956}

\bibitem{lakes_deformation_1991}
R.~Lakes, \href{http://link.springer.com/10.1007/BF01130170}{Deformation mechanisms in negative poisson's ratio materials: structural aspects}, Journal of Materials Science 26~(9) (1991) 2287--2292, publisher: Springer Science and Business Media {LLC}.
\newblock \href {https://doi.org/10.1007/bf01130170} {\path{doi:10.1007/bf01130170}}.
\newline\urlprefix\url{http://link.springer.com/10.1007/BF01130170}

\bibitem{deshpande_isotropic_2000}
V.~Deshpande, N.~Fleck, \href{https://linkinghub.elsevier.com/retrieve/pii/S0022509699000824}{Isotropic constitutive models for metallic foams}, Journal of the Mechanics and Physics of Solids 48~(6) (2000) 1253--1283, publisher: Elsevier {BV}.
\newblock \href {https://doi.org/10.1016/s0022-5096(99)00082-4} {\path{doi:10.1016/s0022-5096(99)00082-4}}.
\newline\urlprefix\url{https://linkinghub.elsevier.com/retrieve/pii/S0022509699000824}

\bibitem{scarpa_dynamic_2002}
F.~Scarpa, J.~R. Yates, L.~G. Ciffo, S.~Patsias, \href{https://journals.sagepub.com/doi/10.1243/095440602321029382}{Dynamic crushing of auxetic open-cell polyurethane foam}, Proceedings of the Institution of Mechanical Engineers, Part C: Journal of Mechanical Engineering Science 216~(12) (2002) 1153--1156, publisher: {SAGE} Publications.
\newblock \href {https://doi.org/10.1243/095440602321029382} {\path{doi:10.1243/095440602321029382}}.
\newline\urlprefix\url{https://journals.sagepub.com/doi/10.1243/095440602321029382}

\bibitem{gaitanaros2012crushing}
S.~Gaitanaros, S.~Kyriakides, A.~M. Kraynik, On the crushing response of random open-cell foams, International Journal of Solids and Structures 49~(19-20) (2012) 2733--2743.

\bibitem{estrin_architectured_2019}
\href{https://link.springer.com/10.1007/978-3-030-11942-3}{Architectured materials in nature and engineering: Archimats}, {ISSN}: 0933-033X, 2196-2812 (2019).
\newblock \href {https://doi.org/10.1007/978-3-030-11942-3} {\path{doi:10.1007/978-3-030-11942-3}}.
\newline\urlprefix\url{https://link.springer.com/10.1007/978-3-030-11942-3}

\bibitem{kim_foam-like_2016}
O.~V. Kim, X.~Liang, R.~I. Litvinov, J.~W. Weisel, M.~S. Alber, P.~K. Purohit, \href{http://link.springer.com/10.1007/s10237-015-0683-z}{Foam-like compression behavior of fibrin networks} 15~(1)  213--228, publisher: Springer Science and Business Media {LLC}.
\newblock \href {https://doi.org/10.1007/s10237-015-0683-z} {\path{doi:10.1007/s10237-015-0683-z}}.
\newline\urlprefix\url{http://link.springer.com/10.1007/s10237-015-0683-z}

\bibitem{liang2017phase}
X.~Liang, I.~Chernysh, P.~K. Purohit, J.~W. Weisel, Phase transitions during compression and decompression of clots from platelet-poor plasma, platelet-rich plasma and whole blood, Acta biomaterialia 60 (2017) 275--290.

\bibitem{chen20243d}
E.~Chen, B.~Kim, N.~Bouklas, L.~J.~Bonassar, S.~Gaitanaros, 3d in-situ characterization reveals the instability-induced auxetic behavior of collagen scaffolds for tissue engineering, bioRxiv (2024) 2024--06.

\bibitem{ament_autonomous_2021}
S.~Ament, M.~Amsler, D.~R. Sutherland, M.-C. Chang, D.~Guevarra, A.~B. Connolly, J.~M. Gregoire, M.~O. Thompson, C.~P. Gomes, R.~B. Van~Dover, \href{https://www.science.org/doi/10.1126/sciadv.abg4930}{Autonomous materials synthesis via hierarchical active learning of nonequilibrium phase diagrams}, Science Advances 7~(51), publisher: American Association for the Advancement of Science ({AAAS}) (2021).
\newblock \href {https://doi.org/10.1126/sciadv.abg4930} {\path{doi:10.1126/sciadv.abg4930}}.
\newline\urlprefix\url{https://www.science.org/doi/10.1126/sciadv.abg4930}

\bibitem{hopkins_design_2016}
J.~B. Hopkins, L.~A. Shaw, T.~H. Weisgraber, G.~R. Farquar, C.~D. Harvey, C.~M. Spadaccini, \href{https://asmedigitalcollection.asme.org/mechanismsrobotics/article/doi/10.1115/1.4032248/383969/Design-of-Nonperiodic-Microarchitectured-Materials}{Design of nonperiodic microarchitectured materials that achieve graded thermal expansions}, Journal of Mechanisms and Robotics 8~(5), publisher: {ASME} International (2016).
\newblock \href {https://doi.org/10.1115/1.4032248} {\path{doi:10.1115/1.4032248}}.
\newline\urlprefix\url{https://asmedigitalcollection.asme.org/mechanismsrobotics/article/doi/10.1115/1.4032248/383969/Design-of-Nonperiodic-Microarchitectured-Materials}

\bibitem{korner_systematic_2015}
C.~Körner, Y.~Liebold-Ribeiro, \href{https://iopscience.iop.org/article/10.1088/0964-1726/24/2/025013}{A systematic approach to identify cellular auxetic materials}, Smart Materials and Structures 24~(2) (2015) 025013, publisher: {IOP} Publishing.
\newblock \href {https://doi.org/10.1088/0964-1726/24/2/025013} {\path{doi:10.1088/0964-1726/24/2/025013}}.
\newline\urlprefix\url{https://iopscience.iop.org/article/10.1088/0964-1726/24/2/025013}

\bibitem{lakes_foam_1987}
R.~Lakes, \href{https://www.science.org/doi/10.1126/science.235.4792.1038}{Foam structures with a negative poisson's ratio}, Science 235~(4792) (1987) 1038--1040, publisher: American Association for the Advancement of Science ({AAAS}).
\newblock \href {https://doi.org/10.1126/science.235.4792.1038} {\path{doi:10.1126/science.235.4792.1038}}.
\newline\urlprefix\url{https://www.science.org/doi/10.1126/science.235.4792.1038}

\bibitem{prall_properties_1997}
D.~Prall, R.~Lakes, \href{https://linkinghub.elsevier.com/retrieve/pii/S0020740396000252}{Properties of a chiral honeycomb with a poisson's ratio of — 1}, International Journal of Mechanical Sciences 39~(3) (1997) 305--314, publisher: Elsevier {BV}.
\newblock \href {https://doi.org/10.1016/s0020-7403(96)00025-2} {\path{doi:10.1016/s0020-7403(96)00025-2}}.
\newline\urlprefix\url{https://linkinghub.elsevier.com/retrieve/pii/S0020740396000252}

\bibitem{bertoldi_negative_2010-1}
K.~Bertoldi, P.~M. Reis, S.~Willshaw, T.~Mullin, \href{https://onlinelibrary.wiley.com/doi/10.1002/adma.200901956}{Negative poisson's ratio behavior induced by an elastic instability}, Advanced Materials 22~(3) (2010) 361--366, publisher: Wiley.
\newblock \href {https://doi.org/10.1002/adma.200901956} {\path{doi:10.1002/adma.200901956}}.
\newline\urlprefix\url{https://onlinelibrary.wiley.com/doi/10.1002/adma.200901956}

\bibitem{hughes_auxetic_2010}
T.~Hughes, A.~Marmier, K.~Evans, \href{https://linkinghub.elsevier.com/retrieve/pii/S0020768310000442}{Auxetic frameworks inspired by cubic crystals}, International Journal of Solids and Structures 47~(11) (2010) 1469--1476, publisher: Elsevier {BV}.
\newblock \href {https://doi.org/10.1016/j.ijsolstr.2010.02.002} {\path{doi:10.1016/j.ijsolstr.2010.02.002}}.
\newline\urlprefix\url{https://linkinghub.elsevier.com/retrieve/pii/S0020768310000442}

\bibitem{ganghoffer_frontiers_2023}
J.-F. Ganghoffer, A.~Wazne, H.~Reda, \href{https://linkinghub.elsevier.com/retrieve/pii/S0093641323000721}{Frontiers in homogenization methods towards generalized continua for architected materials}, Mechanics Research Communications 130 (2023) 104114, publisher: Elsevier {BV}.
\newblock \href {https://doi.org/10.1016/j.mechrescom.2023.104114} {\path{doi:10.1016/j.mechrescom.2023.104114}}.
\newline\urlprefix\url{https://linkinghub.elsevier.com/retrieve/pii/S0093641323000721}

\bibitem{liu_spatially_2024}
C.~Liu, M.~Pham, \href{https://onlinelibrary.wiley.com/doi/10.1002/adma.202305846}{Spatially programmable architected materials inspired by the metallurgical phase engineering}, Advanced Materials 36~(8), publisher: Wiley (2024).
\newblock \href {https://doi.org/10.1002/adma.202305846} {\path{doi:10.1002/adma.202305846}}.
\newline\urlprefix\url{https://onlinelibrary.wiley.com/doi/10.1002/adma.202305846}

\bibitem{pasternak_materials_2012}
E.~Pasternak, A.~Dyskin, \href{https://linkinghub.elsevier.com/retrieve/pii/S0020722511002369}{Materials and structures with macroscopic negative poisson’s ratio}, International Journal of Engineering Science 52 (2012) 103--114, publisher: Elsevier {BV}.
\newblock \href {https://doi.org/10.1016/j.ijengsci.2011.11.006} {\path{doi:10.1016/j.ijengsci.2011.11.006}}.
\newline\urlprefix\url{https://linkinghub.elsevier.com/retrieve/pii/S0020722511002369}

\bibitem{kotani_materials_2016}
M.~Kotani, S.~Ikeda, \href{https://www.tandfonline.com/doi/full/10.1080/14686996.2016.1180233}{Materials inspired by mathematics}, Science and Technology of Advanced Materials 17~(1) (2016) 253--259, publisher: Informa {UK} Limited.
\newblock \href {https://doi.org/10.1080/14686996.2016.1180233} {\path{doi:10.1080/14686996.2016.1180233}}.
\newline\urlprefix\url{https://www.tandfonline.com/doi/full/10.1080/14686996.2016.1180233}

\bibitem{alderson_elastic_2010}
A.~Alderson, K.~Alderson, D.~Attard, K.~Evans, R.~Gatt, J.~Grima, W.~Miller, N.~Ravirala, C.~Smith, K.~Zied, \href{https://linkinghub.elsevier.com/retrieve/pii/S0266353809002814}{Elastic constants of 3-, 4- and 6-connected chiral and anti-chiral honeycombs subject to uniaxial in-plane loading}, Composites Science and Technology 70~(7) (2010) 1042--1048, publisher: Elsevier {BV}.
\newblock \href {https://doi.org/10.1016/j.compscitech.2009.07.009} {\path{doi:10.1016/j.compscitech.2009.07.009}}.
\newline\urlprefix\url{https://linkinghub.elsevier.com/retrieve/pii/S0266353809002814}

\bibitem{dirrenberger_effective_2013}
J.~Dirrenberger, S.~Forest, D.~Jeulin, \href{http://link.springer.com/10.1007/s10999-012-9192-8}{Effective elastic properties of auxetic microstructures: anisotropy and structural applications}, International Journal of Mechanics and Materials in Design 9~(1) (2013) 21--33, publisher: Springer Science and Business Media {LLC}.
\newblock \href {https://doi.org/10.1007/s10999-012-9192-8} {\path{doi:10.1007/s10999-012-9192-8}}.
\newline\urlprefix\url{http://link.springer.com/10.1007/s10999-012-9192-8}

\bibitem{truskinovsky1996ericksen}
L.~Truskinovsky, G.~Zanzotto, Ericksen's bar revisited: Energy wiggles, Journal of the Mechanics and Physics of Solids 44~(8) (1996) 1371--1408.

\bibitem{gao2008multiple}
D.~Gao, R.~W. Ogden, Multiple solutions to non-convex variational problems with implications for phase transitions and numerical computation, The Quarterly Journal of Mechanics \& Applied Mathematics 61~(4) (2008) 497--522.

\bibitem{gao_multiple_2008}
D.~Y. Gao, R.~W. Ogden, \href{https://academic.oup.com/qjmam/article-lookup/doi/10.1093/qjmam/hbn014}{Multiple solutions to non-convex variational problems with implications for phase transitions and numerical computation}, The Quarterly Journal of Mechanics and Applied Mathematics 61~(4) (2008) 497--522, publisher: Oxford University Press ({OUP}).
\newblock \href {https://doi.org/10.1093/qjmam/hbn014} {\path{doi:10.1093/qjmam/hbn014}}.
\newline\urlprefix\url{https://academic.oup.com/qjmam/article-lookup/doi/10.1093/qjmam/hbn014}

\bibitem{ericksen1975equilibrium}
J.~L. Ericksen, Equilibrium of bars, Journal of elasticity 5 (1975) 191--201.

\bibitem{abeyaratne2006evolution}
R.~Abeyaratne, J.~K. Knowles, Evolution of phase transitions: a continuum theory, Cambridge University Press, 2006.

\bibitem{purohit2019compression}
P.~K. Purohit, Compression of fiber networks modeled as a phase transition, Mechanics and Physics of Solids at Micro-and Nano-Scales (2019) 131--155.

\bibitem{geers2010multi}
M.~G. Geers, V.~G. Kouznetsova, W.~Brekelmans, Multi-scale computational homogenization: Trends and challenges, Journal of computational and applied mathematics 234~(7) (2010) 2175--2182.

\bibitem{khajehtourian2021continuum}
R.~Khajehtourian, D.~M. Kochmann, A continuum description of substrate-free dissipative reconfigurable metamaterials, Journal of the Mechanics and Physics of Solids 147 (2021) 104217.

\bibitem{khajehtourian2021soft}
R.~Khajehtourian, D.~M. Kochmann, Soft adaptive mechanical metamaterials, Frontiers in Robotics and AI 8 (2021) 673478.

\bibitem{shan2015multistable}
S.~Shan, S.~H. Kang, J.~R. Raney, P.~Wang, L.~Fang, F.~Candido, J.~A. Lewis, K.~Bertoldi, Multistable architected materials for trapping elastic strain energy, Adv. Mater 27~(29) (2015) 4296--4301.

\bibitem{gong2005compressive}
L.~Gong, S.~Kyriakides, Compressive response of open cell foams part ii: Initiation and evolution of crushing, International Journal of Solids and Structures 42~(5-6) (2005) 1381--1399.

\bibitem{long_2020}
K.~N. Long, C.~M. Hamel, R.~Waymel, D.~S. Bolintineanu, E.~C. Quintana, S.~L. Kramer, \href{https://www.osti.gov/biblio/1673821}{Room temperature quasi-static characterization and constitutive model parametrization of flexible polyurethane foams of different densities loaded in different orientations}, Tech. rep., Sandia National Laboratories (SNL-NM), Albuquerque, NM (United States) (10 2020).
\newblock \href {https://doi.org/10.2172/1673821} {\path{doi:10.2172/1673821}}.
\newline\urlprefix\url{https://www.osti.gov/biblio/1673821}

\bibitem{restrepo_phase_2015}
D.~Restrepo, N.~D. Mankame, P.~D. Zavattieri, \href{https://linkinghub.elsevier.com/retrieve/pii/S2352431615000929}{Phase transforming cellular materials}, Extreme Mechanics Letters 4 (2015) 52--60, publisher: Elsevier {BV}.
\newblock \href {https://doi.org/10.1016/j.eml.2015.08.001} {\path{doi:10.1016/j.eml.2015.08.001}}.
\newline\urlprefix\url{https://linkinghub.elsevier.com/retrieve/pii/S2352431615000929}

\bibitem{jamalimehr2022rigidly}
A.~Jamalimehr, M.~Mirzajanzadeh, A.~Akbarzadeh, D.~Pasini, Rigidly flat-foldable class of lockable origami-inspired metamaterials with topological stiff states, Nature communications 13~(1) (2022) 1816.

\bibitem{deshpande2000high}
V.~Deshpande, N.~Fleck, High strain rate compressive behaviour of aluminium alloy foams, International journal of impact engineering 24~(3) (2000) 277--298.

\bibitem{deshpande2001multi}
V.~Deshpande, N.~Fleck, Multi-axial yield behaviour of polymer foams, Acta materialia 49~(10) (2001) 1859--1866.

\bibitem{chen2002size}
C.~Chen, N.~Fleck, Size effects in the constrained deformation of metallic foams, Journal of the Mechanics and Physics of Solids 50~(5) (2002) 955--977.

\bibitem{aifantis1984microstructural}
E.~C. Aifantis, On the microstructural origin of certain inelastic models (1984).

\bibitem{fleck1997strain}
N.~Fleck, J.~W. Hutchinson, Strain gradient plasticity, Advances in applied mechanics 33 (1997) 295--361.

\bibitem{toupin1962elastic}
R.~Toupin, Elastic materials with couple-stresses, Archive for rational mechanics and analysis 11~(1) (1962) 385--414.

\bibitem{mindlin1963microstructure}
R.~D. Mindlin, Microstructure in linear elasticity, Tech. rep. (1963).

\bibitem{mindlin1965second}
R.~D. Mindlin, Second gradient of strain and surface-tension in linear elasticity, International journal of solids and structures 1~(4) (1965) 417--438.

\bibitem{sagiyama_numerical_2018}
K.~Sagiyama, S.~Rudraraju, K.~Garikipati, \href{http://arxiv.org/abs/1701.04564}{A numerical study of branching and stability of solutions to three-dimensional martensitic phase transformations using gradient-regularized, non-convex, finite strain elasticity} (2018).
\newblock \href {http://arxiv.org/abs/1701.04564 [math]} {\path{arXiv:1701.04564 [math]}}, \href {https://doi.org/10.48550/arXiv.1701.04564} {\path{doi:10.48550/arXiv.1701.04564}}.
\newline\urlprefix\url{http://arxiv.org/abs/1701.04564}

\bibitem{bacigalupo_homogenization_2014}
A.~Bacigalupo, L.~Gambarotta, \href{https://linkinghub.elsevier.com/retrieve/pii/S0263822314002438}{Homogenization of periodic hexa- and tetrachiral cellular solids}, Composite Structures 116 (2014) 461--476, publisher: Elsevier {BV}.
\newblock \href {https://doi.org/10.1016/j.compstruct.2014.05.033} {\path{doi:10.1016/j.compstruct.2014.05.033}}.
\newline\urlprefix\url{https://linkinghub.elsevier.com/retrieve/pii/S0263822314002438}

\bibitem{dalaq_finite_2016}
A.~S. Dalaq, D.~W. Abueidda, R.~K. Abu Al-Rub, I.~M. Jasiuk, \href{https://linkinghub.elsevier.com/retrieve/pii/S0020768316000202}{Finite element prediction of effective elastic properties of interpenetrating phase composites with architectured 3d sheet reinforcements}, International Journal of Solids and Structures 83 (2016) 169--182, publisher: Elsevier {BV}.
\newblock \href {https://doi.org/10.1016/j.ijsolstr.2016.01.011} {\path{doi:10.1016/j.ijsolstr.2016.01.011}}.
\newline\urlprefix\url{https://linkinghub.elsevier.com/retrieve/pii/S0020768316000202}

\bibitem{iltchev_computational_2015}
A.~Iltchev, V.~Marcadon, S.~Kruch, S.~Forest, \href{https://linkinghub.elsevier.com/retrieve/pii/S0020740315000612}{Computational homogenisation of periodic cellular materials: Application to structural modelling}, International Journal of Mechanical Sciences 93 (2015) 240--255, publisher: Elsevier {BV}.
\newblock \href {https://doi.org/10.1016/j.ijmecsci.2015.02.007} {\path{doi:10.1016/j.ijmecsci.2015.02.007}}.
\newline\urlprefix\url{https://linkinghub.elsevier.com/retrieve/pii/S0020740315000612}

\bibitem{mousavi2025chain}
S.~M. Mousavi, J.~Mulderrig, B.~Talamini, N.~Bouklas, A chain stretch-based gradient-enhanced model for damage and fracture in elastomers, Computer Methods in Applied Mechanics and Engineering 444 (2025) 118103.

\bibitem{chen2002phase}
L.-Q. Chen, Phase-field models for microstructure evolution, Annual review of materials research 32~(1) (2002) 113--140.

\bibitem{borden2012phase}
M.~J. Borden, C.~V. Verhoosel, M.~A. Scott, T.~J. Hughes, C.~M. Landis, A phase-field description of dynamic brittle fracture, Computer Methods in Applied Mechanics and Engineering 217 (2012) 77--95.

\bibitem{li2020variational}
B.~Li, N.~Bouklas, A variational phase-field model for brittle fracture in polydisperse elastomer networks, International Journal of Solids and Structures 182 (2020) 193--204.

\bibitem{forest_continuum_2005}
S.~Forest, J.-S. Blazy, Y.~Chastel, F.~Moussy, \href{http://link.springer.com/10.1007/s10853-005-5041-6}{Continuum modeling of strain localization phenomena in metallic foams}, Journal of Materials Science 40~(22) (2005) 5903--5910, publisher: Springer Science and Business Media {LLC}.
\newblock \href {https://doi.org/10.1007/s10853-005-5041-6} {\path{doi:10.1007/s10853-005-5041-6}}.
\newline\urlprefix\url{http://link.springer.com/10.1007/s10853-005-5041-6}

\bibitem{combescure_generalized_nodate}
C.~Combescure, C.~Saint-Cyr, Generalized continuum media confronted to long and short wavelength instabilities in architected materials.

\bibitem{sperling2023enriched}
S.~Sperling, T.~Guo, R.~Peerlings, V.~Kouznetsova, M.~Geers, O.~Roko{\v{s}}, Enriched computational homogenization schemes applied to pattern-transforming elastomeric mechanical metamaterials, arXiv preprint arXiv:2307.10952 (2023).

\bibitem{guo2025reduced}
T.~Guo, V.~Kouznetsova, M.~Geers, K.~Veroy, O.~Roko{\v{s}}, Reduced-order modeling for second-order computational homogenization with applications to geometrically parameterized elastomeric metamaterials, International Journal for Numerical Methods in Engineering 126~(1) (2025) e7604.

\bibitem{maraghechi2024harvesting}
S.~Maraghechi, O.~Roko{\v{s}}, R.~Peerlings, M.~Geers, J.~Hoefnagels, Harvesting deformation modes for micromorphic homogenization from experiments on mechanical metamaterials, International Journal of Solids and Structures 301 (2024) 112916.

\bibitem{van2020newton}
S.~van Bree, O.~Roko{\v{s}}, R.~H. Peerlings, M.~Do{\v{s}}k{\'a}{\v{r}}, M.~G. Geers, A newton solver for micromorphic computational homogenization enabling multiscale buckling analysis of pattern-transforming metamaterials, Computer Methods in Applied Mechanics and Engineering 372 (2020) 113333.

\bibitem{brechet_architectured_2013}
Y.~Brechet, J.~Embury, \href{https://linkinghub.elsevier.com/retrieve/pii/S135964621200499X}{Architectured materials: Expanding materials space}, Scripta Materialia 68~(1) (2013) 1--3, publisher: Elsevier {BV}.
\newblock \href {https://doi.org/10.1016/j.scriptamat.2012.07.038} {\path{doi:10.1016/j.scriptamat.2012.07.038}}.
\newline\urlprefix\url{https://linkinghub.elsevier.com/retrieve/pii/S135964621200499X}

\bibitem{dirrenberger_homogenization_2011}
J.~Dirrenberger, S.~Forest, D.~Jeulin, C.~Colin, \href{https://linkinghub.elsevier.com/retrieve/pii/S1877705811004954}{Homogenization of periodic auxetic materials}, Procedia Engineering 10 (2011) 1847--1852, publisher: Elsevier {BV}.
\newblock \href {https://doi.org/10.1016/j.proeng.2011.04.307} {\path{doi:10.1016/j.proeng.2011.04.307}}.
\newline\urlprefix\url{https://linkinghub.elsevier.com/retrieve/pii/S1877705811004954}

\bibitem{xu_concurrent_2016}
B.~Xu, X.~Huang, S.~Zhou, Y.~Xie, \href{https://linkinghub.elsevier.com/retrieve/pii/S0263822316303804}{Concurrent topological design of composite thermoelastic macrostructure and microstructure with multi-phase material for maximum stiffness}, Composite Structures 150 (2016) 84--102, publisher: Elsevier {BV}.
\newblock \href {https://doi.org/10.1016/j.compstruct.2016.04.038} {\path{doi:10.1016/j.compstruct.2016.04.038}}.
\newline\urlprefix\url{https://linkinghub.elsevier.com/retrieve/pii/S0263822316303804}

\bibitem{vicente_concurrent_2016}
W.~Vicente, Z.~Zuo, R.~Pavanello, T.~Calixto, R.~Picelli, Y.~Xie, \href{https://linkinghub.elsevier.com/retrieve/pii/S0045782515004181}{Concurrent topology optimization for minimizing frequency responses of two-level hierarchical structures}, Computer Methods in Applied Mechanics and Engineering 301 (2016) 116--136, publisher: Elsevier {BV}.
\newblock \href {https://doi.org/10.1016/j.cma.2015.12.012} {\path{doi:10.1016/j.cma.2015.12.012}}.
\newline\urlprefix\url{https://linkinghub.elsevier.com/retrieve/pii/S0045782515004181}

\bibitem{allaire_shape_2002}
G.~Allaire, \href{http://link.springer.com/10.1007/978-1-4684-9286-6}{Shape Optimization by the Homogenization Method}, Applied Mathematical Sciences, Springer New York, 2002, {ISSN}: 0066-5452.
\newblock \href {https://doi.org/10.1007/978-1-4684-9286-6} {\path{doi:10.1007/978-1-4684-9286-6}}.
\newline\urlprefix\url{http://link.springer.com/10.1007/978-1-4684-9286-6}

\bibitem{guest_optimizing_2006}
J.~K. Guest, J.~H. Prévost, \href{https://linkinghub.elsevier.com/retrieve/pii/S0020768306000631}{Optimizing multifunctional materials: Design of microstructures for maximized stiffness and fluid permeability}, International Journal of Solids and Structures 43~(22) (2006) 7028--7047, publisher: Elsevier {BV}.
\newblock \href {https://doi.org/10.1016/j.ijsolstr.2006.03.001} {\path{doi:10.1016/j.ijsolstr.2006.03.001}}.
\newline\urlprefix\url{https://linkinghub.elsevier.com/retrieve/pii/S0020768306000631}

\bibitem{xu_design_2016}
S.~Xu, J.~Shen, S.~Zhou, X.~Huang, Y.~M. Xie, \href{https://linkinghub.elsevier.com/retrieve/pii/S0264127516300120}{Design of lattice structures with controlled anisotropy}, Materials \& Design 93 (2016) 443--447, publisher: Elsevier {BV}.
\newblock \href {https://doi.org/10.1016/j.matdes.2016.01.007} {\path{doi:10.1016/j.matdes.2016.01.007}}.
\newline\urlprefix\url{https://linkinghub.elsevier.com/retrieve/pii/S0264127516300120}

\bibitem{andreassen_design_2014}
E.~Andreassen, B.~S. Lazarov, O.~Sigmund, \href{https://linkinghub.elsevier.com/retrieve/pii/S0167663613002093}{Design of manufacturable 3d extremal elastic microstructure}, Mechanics of Materials 69~(1) (2014) 1--10, publisher: Elsevier {BV}.
\newblock \href {https://doi.org/10.1016/j.mechmat.2013.09.018} {\path{doi:10.1016/j.mechmat.2013.09.018}}.
\newline\urlprefix\url{https://linkinghub.elsevier.com/retrieve/pii/S0167663613002093}

\bibitem{ghaedizadeh_tuning_2016}
A.~Ghaedizadeh, J.~Shen, X.~Ren, Y.~Xie, \href{https://www.mdpi.com/1996-1944/9/1/54}{Tuning the performance of metallic auxetic metamaterials by using buckling and plasticity}, Materials 9~(1) (2016) 54, publisher: {MDPI} {AG}.
\newblock \href {https://doi.org/10.3390/ma9010054} {\path{doi:10.3390/ma9010054}}.
\newline\urlprefix\url{https://www.mdpi.com/1996-1944/9/1/54}

\bibitem{osanov_topology_2016}
M.~Osanov, J.~K. Guest, \href{https://www.annualreviews.org/doi/10.1146/annurev-matsci-070115-031826}{Topology optimization for architected materials design}, Annual Review of Materials Research 46~(1) (2016) 211--233, publisher: Annual Reviews.
\newblock \href {https://doi.org/10.1146/annurev-matsci-070115-031826} {\path{doi:10.1146/annurev-matsci-070115-031826}}.
\newline\urlprefix\url{https://www.annualreviews.org/doi/10.1146/annurev-matsci-070115-031826}

\bibitem{challis_design_2008}
V.~Challis, A.~Roberts, A.~Wilkins, \href{https://linkinghub.elsevier.com/retrieve/pii/S0020768308000929}{Design of three dimensional isotropic microstructures for maximized stiffness and conductivity}, International Journal of Solids and Structures 45~(14) (2008) 4130--4146, publisher: Elsevier {BV}.
\newblock \href {https://doi.org/10.1016/j.ijsolstr.2008.02.025} {\path{doi:10.1016/j.ijsolstr.2008.02.025}}.
\newline\urlprefix\url{https://linkinghub.elsevier.com/retrieve/pii/S0020768308000929}

\bibitem{kumar2020inverse}
S.~Kumar, S.~Tan, L.~Zheng, D.~M. Kochmann, Inverse-designed spinodoid metamaterials, npj Computational Materials 6~(1) (2020) 73.

\bibitem{zheng2023unifying}
L.~Zheng, K.~Karapiperis, S.~Kumar, D.~M. Kochmann, Unifying the design space and optimizing linear and nonlinear truss metamaterials by generative modeling, Nature Communications 14~(1) (2023) 7563.

\bibitem{thakolkaran2025experiment}
P.~Thakolkaran, M.~Espinal, S.~Dhulipala, S.~Kumar, C.~M. Portela, Experiment-informed finite-strain inverse design of spinodal metamaterials, Extreme Mechanics Letters 74 (2025) 102274.

\bibitem{kalina_neural_2024}
K.~A. Kalina, J.~Brummund, W.~Sun, M.~Kästner, \href{http://arxiv.org/abs/2410.03378}{Neural networks meet anisotropic hyperelasticity: A framework based on generalized structure tensors and isotropic tensor functions} (2024).
\newblock \href {http://arxiv.org/abs/2410.03378 [cs]} {\path{arXiv:2410.03378 [cs]}}, \href {https://doi.org/10.48550/arXiv.2410.03378} {\path{doi:10.48550/arXiv.2410.03378}}.
\newline\urlprefix\url{http://arxiv.org/abs/2410.03378}

\bibitem{huang_optimized_2024}
Z.~Huang, D.~Panozzo, D.~Zorin, \href{http://arxiv.org/abs/2310.08609}{Optimized shock-protecting microstructures}, {ACM} Transactions on Graphics 43~(6) (2024) 1--21.
\newblock \href {http://arxiv.org/abs/2310.08609 [math]} {\path{arXiv:2310.08609 [math]}}, \href {https://doi.org/10.1145/3687765} {\path{doi:10.1145/3687765}}.
\newline\urlprefix\url{http://arxiv.org/abs/2310.08609}

\bibitem{jones2025differentiable}
R.~E. Jones, A.~B. Tepole, J.~N. Fuhg, Differentiable neural network representation of multi-well, locally-convex potentials, arXiv preprint arXiv:2506.17242 (2025).

\bibitem{fuhg2024review}
J.~N. Fuhg, G.~A. Padmanabha, N.~Bouklas, B.~Bahmani, W.~Sun, N.~N. Vlassis, M.~Flaschel, P.~Carrara, L.~De~Lorenzis, A review on data-driven constitutive laws for solids, arXiv preprint arXiv:2405.03658 (2024).

\bibitem{holzapfel2002nonlinear}
G.~A. Holzapfel, Nonlinear solid mechanics: a continuum approach for engineering science (2002).

\bibitem{ball1976convexity}
J.~M. Ball, Convexity conditions and existence theorems in nonlinear elasticity, Archive for rational mechanics and Analysis 63 (1976) 337--403.

\bibitem{knowles1978failure}
J.~K. Knowles, E.~Sternberg, On the failure of ellipticity and the emergence of discontinuous deformation gradients in plane finite elastostatics, Journal of Elasticity 8~(4) (1978) 329--379.

\bibitem{lopez2007homogenization}
O.~Lopez-Pamies, P.~P. Casta{\~n}eda, Homogenization-based constitutive models for porous elastomers and implications for macroscopic instabilities: I—analysis, Journal of the Mechanics and Physics of Solids 55~(8) (2007) 1677--1701.

\bibitem{agoras2025effect}
M.~Agoras, F.~F. Fontenele, N.~Bouklas, The effect of fiber plasticity on domain formation in soft biological composites--part i: a bifurcation analysis, arXiv preprint arXiv:2507.07843 (2025).

\bibitem{furer2018macroscopic}
J.~Furer, P.~P. Casta{\~n}eda, Macroscopic instabilities and domain formation in neo-hookean laminates, Journal of the Mechanics and Physics of Solids 118 (2018) 98--114.

\bibitem{iordanidis2025effect}
D.~Iordanidis, F.~F. Fontenele, K.~Poulios, M.~Agoras, N.~Bouklas, The effect of fiber plasticity on domain formation in soft biological composites--part ii: An imperfection analysis, arXiv preprint arXiv:2507.14764 (2025).

\bibitem{ball1989fine}
J.~M. Ball, R.~D. James, Fine phase mixtures as minimizers of energy, in: Analysis and Continuum Mechanics: A Collection of Papers Dedicated to J. Serrin on His Sixtieth Birthday, Springer, 1989, pp. 647--686.

\bibitem{tekog2011size}
C.~Tekog, L.~Gibson, T.~Pardoen, P.~Onck, et~al., Size effects in foams: Experiments and modeling, Progress in Materials Science 56~(2) (2011) 109--138.

\bibitem{miehe2010phase}
C.~Miehe, M.~Hofacker, F.~Welschinger, A phase field model for rate-independent crack propagation: Robust algorithmic implementation based on operator splits, Computer Methods in Applied Mechanics and Engineering 199~(45-48) (2010) 2765--2778.

\bibitem{ambati2015review}
M.~Ambati, T.~Gerasimov, L.~De~Lorenzis, A review on phase-field models of brittle fracture and a new fast hybrid formulation, Computational Mechanics 55~(2) (2015) 383--405.

\bibitem{logg2012automated}
A.~Logg, K.-A. Mardal, G.~Wells, Automated solution of differential equations by the finite element method: The FEniCS book, Vol.~84, Springer Science \& Business Media, 2012.

\bibitem{FEniCSAlnaes2015}
M.~S. Aln{\ae}s, J.~Blechta, J.~Hake, A.~Johansson, B.~Kehlet, A.~Logg, C.~Richardson, J.~Ring, M.~E. Rognes, G.~N. Wells, The fenics project version 1.5, Archive of Numerical Software 3~(100) (2015).
\newblock \href {https://doi.org/10.11588/ans.2015.100.20553} {\path{doi:10.11588/ans.2015.100.20553}}.

\bibitem{UFL}
M.~S. Aln\ae{}s, A.~Logg, K.~B. \O{}lgaard, M.~E. Rognes, G.~N. Wells, \href{https://doi.org/10.1145/2566630}{Unified form language: A domain-specific language for weak formulations of partial differential equations}, ACM Trans. Math. Softw. 40~(2) (Mar. 2014).
\newblock \href {https://doi.org/10.1145/2566630} {\path{doi:10.1145/2566630}}.
\newline\urlprefix\url{https://doi.org/10.1145/2566630}

\bibitem{BOLINTINEANU2021}
D.~S. Bolintineanu, R.~Waymel, H.~Collis, K.~N. Long, E.~C. Quintana, S.~L. Kramer, \href{https://www.sciencedirect.com/science/article/pii/S2589152921001150}{Anisotropy evolution of elastomeric foams during uniaxial compression measured via in-situ x-ray computed tomography}, Materialia 18 (2021) 101112.
\newblock \href {https://doi.org/https://doi.org/10.1016/j.mtla.2021.101112} {\path{doi:https://doi.org/10.1016/j.mtla.2021.101112}}.
\newline\urlprefix\url{https://www.sciencedirect.com/science/article/pii/S2589152921001150}

\bibitem{long_2022}
K.~N. Long, C.~M. Hamel, \href{https://www.osti.gov/biblio/1870771}{Stabilized hyperfoam modeling of the general plastics ef4003 (3 pcf) flexible foam}, Tech. rep., Sandia National Laboratories (SNL-NM), Albuquerque, NM (United States) (05 2022).
\newblock \href {https://doi.org/10.2172/1870771} {\path{doi:10.2172/1870771}}.
\newline\urlprefix\url{https://www.osti.gov/biblio/1870771}

\bibitem{luan2022microscopic}
S.~Luan, A.~M. Kraynik, S.~Gaitanaros, Microscopic and macroscopic instabilities in elastomeric foams, Mechanics of Materials 164 (2022) 104124.

\bibitem{bertoldi2017flexible}
K.~Bertoldi, V.~Vitelli, J.~Christensen, M.~Van~Hecke, Flexible mechanical metamaterials, Nature Reviews Materials 2~(11) (2017) 1--11.

\bibitem{mousavi2025capturing}
S.~M. Mousavi, J.~Mulderrig, B.~Talamini, N.~Bouklas, Capturing the fractocohesive length scale through a gradient-enhanced damage model for elastomers, arXiv preprint arXiv:2509.00313 (2025).

\end{thebibliography}
\end{document}